\documentclass{aa}  
\usepackage{natbib}
\usepackage{xcolor}
\usepackage{graphicx}
\usepackage{txfonts}
\usepackage{hyperref}
\usepackage{float}
\usepackage{tablefootnote}
\usepackage{lipsum}
\usepackage{caption}
\usepackage{afterpage}
\usepackage{subcaption}
\usepackage{multirow}
\usepackage{lipsum}
\usepackage{arydshln}
\usepackage{comment}
\usepackage{oplotsymbl} % for pentagon symbol
\usepackage{lscape}

\begin{document}

   \title{{Radiation-hydrodynamics of star–disc collisions:\\ From system parameters to outflows and lightcurves}}

   \author{Taj Jankovi\v{c} \inst{1}
            \and
          Sergey Karpov \inst{1}
          \and
          Michal Zaja\v{c}ek \inst{2}
          \and 
          Vladim\'{\i}r Karas\inst{3}
          \and
          Marzena {\'S}niegowska\inst{3}
          }

   \institute{Institute of Physics of the Czech Academy of Sciences, Na Slovance 1999/2, 182~21 Prague, Czech Republic
    \and
     Department of Theoretical Physics and Astrophysics, Masaryk University, Kotlá\v{r}ská 267/2, 611~37 Brno, Czech Republic
     \and
     Astronomical Institute of the Czech Academy of Sciences, Bo\v{c}n\'{i} II 1401, 141~00 Prague, Czech Republic\\[3pt]
              \email{jankovic@fzu.cz}
             }

   \date{Received XX; accepted XX}

  \abstract
{ Quasi-periodic eruptions (QPEs) are nuclear transients producing bright, repeating soft X-ray flares superimposed on quiescent emission. A promising interpretation is that they are powered by star-disc collisions, in which a star crosses an accretion disc around a supermassive black hole, drives shocks, and launches dense outflows from which radiation emerges. }
{We present a systematic study of star-disc collisions, { linking} the physical parameters of the collision to the resulting outflows and emerging bolometric luminosities. }
{We perform three-dimensional local radiation-hydrodynamics simulations, varying the disc surface density { and vertical density profiles}, stellar velocity and radius, and { local collision angle}. {We focus on the regime where the star remains unperturbed by the collision.}}
{Variations of stellar velocity and accretion disc surface density leave the bow shock and outflow morphology largely unchanged. However, faster stars produce brighter flares, while denser discs mainly increase the flare duration. Increasing the stellar radius increases the momentum of the forward outflow, and produces brighter and longer flares. More centrally concentrated discs yield brighter and shorter flares because radiation escapes more efficiently through outer low-density layers. More oblique crossings reduce the momentum and luminosity asymmetry of two outflows, and lengthen the flares. We provide empirical scalings of the peak luminosity and flare duration with the individual system parameters and apply them to GSN 069. {{ The best candidate solutions} favour a star with a radius $\sim R_\odot$ on a retrograde orbit, colliding with a dense post-TDE disc with a vertically concentrated density profile. } }
{Our findings suggest that specific combinations of system parameters can reproduce characteristic flare amplitudes, durations, duty cycles, and `strong-weak' flare patterns observed in QPE sources.}

   \keywords{ Black hole physics - Radiation: dynamics -  Accretion, accretion discs}

\titlerunning{From system parameters to outflows and lightcurves}
   \maketitle
%
%-------------------------------------------------------------------

\section{Introduction}

\noindent
Quasiperiodic { eruptions} \citep[QPEs; see][for a review]{2026arXiv260715359W} are a recently discovered class of { repeating} soft X-ray nuclear transients associated with supermassive black holes (SMBHs), characterized by luminous flares recurring on timescales of hours to days \citep{Miniutti2019,Giustini2020,Arcodia2021,Chakraborty_2021ApJ...921L..40C,Quintin_2023A&A...675A.152Q,Nicholl_2024Natur.634..804N,chakraborty2025discoveryquasiperiodiceruptionstidal,Hernandez_2025NatAs...9..895H,Baldini_2026A&A...706L..15B}. Individual eruptions typically last from tens of minutes to a few hours and reach peak luminosities $\sim10^{41}$--$10^{43}\,{\rm erg\,s^{-1}}$, with flux enhancements of $\sim10$--$100$ above the quiescent level in the soft X-ray band. Their spectra are quasi-thermal with characteristic temperatures $\sim100$--$200~{\rm eV}$. Several sources show an alternating `long--short' pattern, where the recurrence time alternates between longer and shorter intervals from one flare to the next, \citep[e.g.][]{Miniutti2019,Giustini2020,Arcodia2021,Arcodia_2024A&A...690A..80A,Hernandez_2025NatAs...9..895H}.

The physical origin of QPEs remains an open question, and a wide range of models has been proposed. Broadly, these can be grouped into two classes. In the first class, the variability is generated by processes intrinsic to the accretion flow, including disc instabilities and related variability \citep[e.g.][]{Raj_2021ApJ...909...82R,pan_2022ApJ...928L..18P,pan_2023ApJ...952...32P,2020A&A...641A.167S,Sniegowska2023}, as well as precession-driven viewing-angle effects in super-Eddington flows \citep[e.g.][]{2024Natur.630..325P,Middleton_2025MNRAS.537.1688M}. The second class involves an orbiting secondary body on a tight orbit around the SMBH. In this case, QPEs can arise from mass transfer from the secondary through Roche-lobe overflow \citep[e.g.][]{Zalamea_2010MNRAS.409L..25Z,King_2020MNRAS.493L.120K,Metzger2022,Wang_2022ApJ...933..225W,Lu_2023MNRAS.524.6247L,Linial_2023ApJ...945...86L,Krolik_2022ApJ...941...24K}, or from repeated collisions of a star (or its unbound debris) or compact object with an accretion disc \citep[e.g.][]{Ivanov_1998ApJ...507..131I,Dai2010,Xian2021,Sukova2021,Linial2023,Linial_2025ApJ...991..147L,Franchini2023,Tagawa_2023MNRAS.526...69T,Yao_2025ApJ...978...91Y,vurm_2025ApJ...983...40V,Huang_2025ApJ...993..186H,Jankovic_2026A&A...710A.405J,Liu_2026ApJ..1006L..50L,2026A&A...709A.147M}. Another suggested QPE model involved a self-lensing SMBH binary system  { \citep{Ingram2021}}; however, this model is currently disfavoured due to the energy-dependence of the QPEs \citep{Miniutti2019} while lensing is achromatic. For a more general review on repeating nuclear transients and potential involved mechanisms, see also \citet{2024arXiv241104592S}.

The star--disc collision model is among the most promising scenarios because it links the recurrence time directly to the orbital motion of the colliding body while naturally producing the observed `long--short' pattern for a moderately eccentric and inclined stellar orbit \citep{Linial2023}. Furthermore, since the orbit of a colliding body (star, compact remnant) shrinks due to gravitational-wave losses, a fraction of the QPE-like systems with proper parameters could be detected via gravitational waves by space interferometers \citep{2024MNRAS.532.2143K}. In this model, a main-sequence star { presumably} intersects the disc twice per orbit; { however, depending on the viewing angle and the shocked material -- disc length-scales, only one eruption may be visible per orbit \citep{Yao_2025ApJ...978...91Y}}. Because the two crossings occur at different true anomalies, the time between one pair of consecutive crossings is longer than between the next pair, producing alternating long and short recurrence intervals, as it was observed in several QPE sources \citep[e.g.][]{Miniutti2019, Giustini2020,Arcodia2021,Arcodia_2024A&A...690A..80A}. Each crossing heats the intercepted disc gas through shocks, expelling dense, initially optically thick ejecta above and below the disc midplane. As these outflows expand under radiation pressure, their density drops until radiation can escape on timescales comparable to the observed flare durations \citep[e.g.][]{Linial2023,vurm_2025ApJ...983...40V,Huang_2025ApJ...993..186H}. The two ejecta components are intrinsically asymmetric \citep[e.g.][]{Jankovic_2026A&A...710A.405J,Liu_2026ApJ..1006L..50L}, providing a natural explanation for a `strong--weak' pattern, in which consecutive flares alternate in peak brightness, observed in several QPEs \citep[e.g.][]{Miniutti2023,Arcodia2021}. In addition, a fraction of the stellar envelope may be ablated and stretched into an elongated stream that can collide with the disc roughly half an orbit after the direct star--disc encounter. Owing to its geometry, such stream--disc interactions can persist longer than a direct star-disc collision and may become increasingly important for long-period QPEs \citep[e.g.][]{Linial_2025ApJ...991..147L,mummery_collisions_2025}.

{ The origin of the disc in this scenario is uncertain. It may be a faint AGN disc or a transient disc formed following a stellar tidal disruption event (TDE), as the bound stellar debris circularises around the SMBH. The latter possibility is supported by the direct association of several QPE sources with preceding TDEs \citep{Quintin_2023A&A...675A.152Q,Nicholl_2024Natur.634..804N,chakraborty2025discoveryquasiperiodiceruptionstidal,Miniutti2023,Chakraborty_2021ApJ...921L..40C,Baldini_2026A&A...706L..15B}, similarities between the host galaxies of QPEs and TDEs \citep{Wevers24,Gilbert25}, and by the compact disc properties inferred in some QPE systems, which are consistent with a transient post-TDE origin \citep{eRO2_UV,Nicholl_2024Natur.634..804N,chakraborty2025discoveryquasiperiodiceruptionstidal,mummery_fitted,Guolo_2025ApJ...985..146G,Guolo_2025ApJ...992..114G}. In addition, the decreasing eruption amplitudes over time in some QPE sources \citep[e.g. GSN 069 and eRO-QPE1;][]{Miniutti2023,2024ApJ...963L..47P} appear to be consistent with the TDE origin of the disc, which exhibits a gradual decrease in the accretion rate \citep{2026A&A...709A.147M}.}

Connecting the star-disc collisions model to observations requires hydrodynamics simulations because the flare properties are set by the coupled evolution of shocks, outflow geometry, and radiative transport in an expanding, optically thick medium. Radiation is generated in the shock during the disc crossing, but much of it is initially trapped and advected with the gas. The eventual luminosity and spectrum depend on how the shocked material flows around the star, breaks out of the disc surface, and expands before the radiation escapes. Early numerical studies explored star--disc interactions without radiation transport and outside the QPE-motivated regime \citep[e.g.][]{Syer1991,Zurek_1994ApJ...434...46Z,Armitage_1996ApJ...470..237A,1999A&A...352..452S,Dai2010,Sukova2021}. More recently, \citet{Yao_2025ApJ...978...91Y} performed hydrodynamics simulations of repeated collisions and quantified how cumulative heating drives mass ablation and modifies the stellar structure over many passages. \citet{vurm_2025ApJ...983...40V} carried out time-dependent 1D Monte Carlo radiation-hydrodynamics simulations to delineate the conditions under which the emerging emission matches QPE-like luminosities and temperatures. \citet{Huang_2025ApJ...993..186H} performed 2D multi-group radiation-hydrodynamics simulations of the broadband emission and showed that the spectral energy distribution (SED) evolution is roughly consistent with those seen in short-period QPEs. Frequency-integrated radiation-hydrodynamics simulations by \citet{Jankovic_2026A&A...710A.405J} and global hydrodynamics simulations of \citet{Liu_2026ApJ..1006L..50L} demonstrated that the intrinsic ejecta asymmetry can potentially explain the observed strong--weak pattern. \citet{Huang_2026ApJ..1005..235H} characterized the importance of numerical resolution and three-dimensional (3D) geometry, showing that resolving the bow-shock stand-off distance is essential and that oblique encounters can substantially reduce the luminosity asymmetry between the two ejecta components.

{A predictive star--disc collision model must connect the local physical conditions at each disc crossing to the observable properties of the resulting flare. The orbital configuration determines when collisions occur, but the peak luminosity, duration, radiated energy, spectral evolution, and brightness asymmetry of individual eruptions are set by the hydrodynamic response of the disc gas and by photon diffusion through the expanding ejecta. The relevant local parameters include the disc surface density, the relative size between the star and the disc height, {the magnitude of the relative velocity between the star and the disc gas, the angle that this velocity makes with the disc midplane,} and the disc vertical structure. Establishing how these parameters shape the outflows and their radiative signatures is therefore a necessary step toward constructing full QPE light-curve models from repeated star--disc crossings.} 

Several recent studies have explored parts of this problem. \citet{vurm_2025ApJ...983...40V} characterized the effect of disc surface density and collision speed in the quasi-spherical (1D) outflow limit, finding that the flare duration is set primarily by the photon diffusion time through the expanding ejecta, which increases with disc density and decreases with collision speed, whereas the characteristic luminosity depends most strongly on the shock energy thus increasing with the collision velocity and also weakly with disc surface density. Similar results were also obtained by \citet{Huang_2025ApJ...993..186H} through 2D radiation-hydrodynamics simulations. They also found that the decrease in the relative size between the star and the disc height prolongs the breakout/cooling emission and yields less asymmetric luminosities of the two outflows. \citet{Liu_2026ApJ..1006L..50L} and \citet{Huang_2026ApJ..1005..235H} {showed that more oblique collision geometries can reduce the luminosity asymmetry between the two outflows.} While existing studies have provided important information regarding the effect of system parameters on the outcome of collisions, there has not yet been a dedicated systematic 3D radiation-hydrodynamics study.

{In this work, we systematically investigate how the key physical parameters of the star-disc collisions model are imprinted on the outflow hydrodynamics and emerging bolometric flare luminosities. Using a suite of 3D radiation-hydrodynamics simulations in a local reference frame, we vary, one at a time relative to a fiducial model, the disc surface density, stellar velocity, stellar radius, vertical disc density profile, and {local collision angle}. We focus on the regime where the star itself remains unaffected by the passage. We characterize how the injected energy and momentum are redistributed within the disc gas for different parameters, and compute how these differences affect the flare properties such as the peak luminosity and flare duration.}

This paper is organized as follows. Section~\ref{sec:method} describes the physical setup of the star--disc collision model, the numerical method, and the initial conditions of our simulations. In Section~\ref{sec:results} we present the collision dynamics, shock heating, outflow properties, and the resulting lightcurves. In Section~\ref{sec:discussion} we discuss the implications of our findings and their connection to observed QPE phenomenology. Our main conclusions are summarized in Section~\ref{sec:conclusion}.

%-------------------------------------------------------------------
\section{Methodology}
\label{sec:method}
We performed 3D radiation-hydrodynamics simulations of the star-disc collision using the smoothed-particle hydrodynamics (SPH) code \texttt{Phantom} \citep{Price2018}. We adopt as our fiducial model the star--disc collision simulation presented in \citet{Jankovic_2026A&A...710A.405J}. In Section \ref{subsec:fiducial_sim} we briefly summarize the initial setup of the fiducial simulation, together with the key physics and approximations used in our simulations. In Section \ref{subsec:simulations} we present an overview of all the simulations. We simulate the collisions in a local Cartesian domain centered on the disc midplane, with the coordinate system defined by unit vectors $\hat{\boldsymbol e}_x$, $\hat{\boldsymbol e}_y$, and $\hat{\boldsymbol e}_{\rm z}$. 

\subsection{Fiducial simulation and method} \label{subsec:fiducial_sim}

The star with radius $R_\star$ is initially above the disc, with its center of mass positioned at $(0,0,5R_\star)$ and velocity $\boldsymbol v_\star=(0,0,-v_\star)$. The fiducial system parameters were chosen following \citet{Linial2023}, for an SMBH with mass $M_{\rm bh}=10^6\,M_\odot$ and a QPE period $P_{\rm QPE}=4\,$h, {which corresponds to a distance of {$r= 95R_{\rm g}$}, where $R_{\rm g}=GM_{\rm bh}c^{-2}$ is the gravitational radius, assuming two star-disc collisions per orbit. We adopt a steadily accreting, optically thick, radiation pressure-dominated $\alpha$-disc with viscosity parameter \(\alpha=0.1\) and relative Eddington mass accretion rate $\dot{m}=\dot{M}/\dot{M}_{\rm Edd}=0.1$, where $\dot{M}_{\rm Edd}=L_{\rm Edd}/(0.1c^2)$, $L_{\rm Edd}$ is the Eddington luminosity, and a characteristic radiative efficiency of 0.1 has been assumed} \citep{Shakura1973} \footnote{{ For this relative accretion rate, we have the bolometric accretion luminosity of $L_{\rm acc}=\dot{m}L_{\rm Edd}\simeq 1.26 \times 10^{43}\,(M_{\rm bh}/10^6\,M_{\odot})\,{\rm erg\,s^{-1}}$\, assuming the same radiative efficiency for the actual and the Eddington accretion rates. The numerical integration of the specific luminosity for a face-on view across the disc annuli from 6$r_{\rm g}$ up to $10^4\,r_{\rm g}$ gives the bolometric luminosity $L_{\rm bol}=2.7 \times 10^{43}\,{\rm erg\,s^{-1}}$, assuming optically thick thermal emission, which is consistent with $L_{\rm acc}$ within a factor of few. The soft X-ray luminosity (0.2-2 keV) is $3 \times 10^{40}\,{\rm erg\,s^{-1}}$, therefore any QPEs reaching $\sim 10^{41}\,{\rm erg\,s^{-1}}$ in the soft X-ray band will outshine the disc emission at this relative accretion rate. The thin disc emission is very soft, corresponding to 5.1 eV ($\sim 60\,000$ K) at the collision site at $r=95\,R_{\rm g}$ and {reaching its highest annulus temperature of} $\sim 21$ eV ($240\,000$ K) at $r=10\,R_{\rm g}$.}} We set $R_\star^{\rm f}=R_\odot$, $v_\star^{\rm f}=0.1c$, disc half-thickness $H=3R_\star$, and a uniform disc density $\rho_{\rm d}=3.7\times 10^{-8}\,{\rm g\,cm^{-3}}$. {Here and throughout, the superscript ‘${\rm f}$’ denotes the fiducial value.} The density normalization is chosen such that the mass of gas in the cylindrical column directly intercepted by the star is { $M_{\rm d}^{\rm f}=\pi \left (R_\star^{\rm f}\right)^2 H \rho_{\rm d}^{\rm f}=1.2\times 10^{-7}\,M_\odot$}. To prevent significant expansion of the disc due to its internal pressure over the course of the simulation, we set the specific internal energy to $u_{\rm d}=10^{-5}v_\star^2$.

To reduce computational costs, we simulate only a localized region of the accretion disc. The disc is constructed by randomly distributing $6\times 10^6$ SPH particles in a cuboidal domain centered at the origin, with horizontal dimensions $24R_\star$ and vertical extent $2H$. The particles are distributed according to a random uniform probability density function, and the particle mass is chosen to achieve a target density equal to $\rho_{\rm d}$. To smooth the initial particle distribution, we perform a relaxation process by evolving the system under periodic boundary conditions.

The gas pressure is modeled with an adiabatic equation of state with adiabatic index $\Gamma=5/3$, and radiation diffusion is treated in the flux-limited diffusion approximation (e.g., \citealt{Whitehouse_2004MNRAS.353.1078W,Whitehouse_2005MNRAS.364.1367W,Bate_2015MNRAS.449.2643B,Lau_2025A&A...699A.274L}{ )}, which assumes local thermodynamic equilibrium (LTE). The radiation diffusion flux is computed as
\begin{equation}\label{eq:Fdiff}
\boldsymbol F_{\rm diff}=-\lambda\,\frac{c}{\kappa_{\rm s}}\,\frac{\nabla e_{\rm rad}}{\rho},
\end{equation}
where $\rho$ is the gas density, $e_{\rm rad}$ is the radiation energy density. { We adopt a constant electron-scattering opacity $\kappa_{\rm s}=0.34\,{\rm cm^2\,g^{-1}}$, appropriate for fully
ionized, hydrogen-rich gas of approximately solar composition and commonly adopted in models of star--disc collision ejecta
\citep{vurm_2025ApJ...983...40V,Huang_2025ApJ...993..186H}}. The flux limiter $\lambda$ prevents unphysical, arbitrarily fast transport in optically thin gas. We adopt the \citet{Levermore_1981ApJ...248..321L} prescription, $\lambda=(2+K)/(6+3K+K^2)$ (with $\lambda\le 1/3$), where $K=|\nabla e_{\rm rad}|/(\kappa_{\rm s}\rho e_{\rm rad})$.

We neglect the SMBH gravitational field, Coriolis and centrifugal forces, and differential rotation of the disc, since the prompt shock formation and outflow launching occur on a timescale much shorter than the local orbital and shear timescales, and over a spatial scale much smaller than the orbital radius. {The star is modelled as a rigid sphere; in other words, the effects of collisions exerted upon the star atmosphere as well as the role of differential rotation of the accretion medium are not taken into account in this paper.} We also neglect the self-gravity of the gas and the gravity of the star, since the associated gravitational energies are small compared to the kinetic energy injected into the shocked gas during the early collision phase {and because the Bondi radius of the star is much smaller than the effective area of the direct star-disc collision (see Section 4.3 in \citealt{Jankovic_2026A&A...710A.405J} for a more extended discussion)}. We note, however, that the disc self-gravity may still affect the pre-collision vertical structure of the disc by modifying its geometrical thickness and density distribution; exploring such initial disc models is beyond the scope of this work. We further neglect a pre-existing coronal or ambient medium above the disc, as well as the effects of the star's own radiation pressure and wind on the disc gas. The interaction between the star and the gas is treated as elastic collisions with individual SPH particles. When a gas particle reaches the stellar surface, its velocity component normal to the surface is reversed while the tangential component is preserved. We keep the stellar velocity fixed, neglecting the loss of stellar momentum during the collision, which is justified because { the stellar mass $M_\star \sim M_\odot$} is much larger than the intercepted disc mass $M_{\rm d}$. { More precisely, the hydrodynamical-drag timescale due to star-disc collisions with the relative velocity $v_{\rm rel}$ can be estimated as the $e$-folding timescale for the specific angular momentum of an orbiting star \citep{2000ApJ...536..663N,2001A&A...376..686K,Sniegowska_2026arXiv260513799S},}
\begin{equation}
   \tau_{\rm drag}=\frac{M_{\star}v_{\star}}{|F_{\rm drag}|}\simeq \frac{M_{\star}v_{\star}}{\pi R_{\star}^2\rho_{\rm d}v_{\rm rel}^2}=\frac{M_{\star}}{\pi R_{\star}^2\rho_{\rm d}v_{\star}}\,, 
\end{equation}
where the last equality is valid for a perpendicular crossing. For fiducial values, we get $\tau_{\rm drag}\approx 37(M_{\star}/1\,M_{\odot})(R_{\star}/R_{\odot})^{-2}(\rho_{\rm d}/3.7\times 10^{-8}\,{\rm g\,cm^{-3}})^{-1}(v_{\star}/0.1c)^{-1}\, \text{years}\, \gg 2P_{\rm QPE}$, {and it is} much longer than the orbital period of the star.
{Furthermore, the gravitational merger timescale for a star of $1\,M_{\odot}$ at $95\,R_{\rm g}$ is { even larger} than the orbital timescale, $\tau_{\rm merge}\sim 250\,000\,(M_{\rm bh}/10^6\,M_{\odot})^{-2} (M_{\star}/1\,M_{\odot})^{-1} (r/95\,R_{\rm g})^4$ years, hence we can safely neglect orbital decay due to gravitational-wave losses. At the distance of $95\,R_{\rm g}$ we can also neglect general relativistic effects on the emerging radiation.} These assumptions are discussed in greater detail in \citet{Jankovic_2026A&A...710A.405J}.

We adopt the same numerical resolution as in \citet{Jankovic_2026A&A...710A.405J}. Specifically, the initial star--disc configuration results in a smoothing length $h_{\rm sl}\approx (m/\rho)^{1/3}\lesssim 0.05R_\star$, where $m$ is the particle mass, ensuring $h_{\rm sl}$ remains much smaller than the size of the star and disc. %We have verified convergence by repeating the fiducial simulation with $h_{\rm sl}\approx 0.01R_\star$ and $h_{\rm sl}\approx 0.005R_\star$, finding no qualitative differences.

\subsection{Range of parameters}\label{subsec:simulations}

We performed a suite of simulations exploring how the collision dynamics and emerging radiation depend on: i) stellar velocity $v_\star$, ii) disc surface density $\Sigma_{\rm d}$, iii) stellar radius $R_\star$, iv) vertical disc density profile, {and v) the local collision angle $i$}. In each run, we vary {only one parameter at a time} relative to the fiducial model, while keeping the numerical method and remaining physical parameters fixed, in order to isolate the effect of each parameter. The explored parameter ranges are summarized in Table~\ref{tab:parameters}.

i) We consider $v_\star/v_{\star}^{\rm f}=0.5,\ 0.75,\ 1,\ 1.25,$ and $1.5$. This range is chosen to explore the range of velocities constrained by the orbital periods of the star given by the time elapsed between consecutive flares observed in QPEs, assuming a fixed SMBH mass $M_{\rm bh}= 10^6\,M_\odot$.

ii) We consider $\Sigma_{\rm d}/\Sigma_{\rm d}^{\rm f}=1,\ 3,\ 10,\ 30,$ and $100$. In practice, this is implemented by scaling the disc density normalization while keeping the remaining disc geometry fixed. This wide range is motivated by the large theoretical uncertainty in the local disc surface density at the crossing radius, spanning from Shakura--Sunyaev radiation-pressure dominated discs with $\Sigma_{\rm d}\sim 10^{4}\,{\rm g}\,{\rm cm^{-2}}$ to denser disc configurations obtained from models assuming a disc formed in a stellar tidal disruption event with $\Sigma_{\rm d} \sim 10^{6}\,{\rm g}\,{\rm cm^{-2}}$, assuming $P_{\rm QPE}=4\,$h, $M_{\rm bh}= 10^6\,M_\odot$, and $R_\star=R_\odot$ (e.g., \citealt{Linial_2025ApJ...991..147L,mummery_collisions_2025}).

iii) We consider $R_\star/R_\star^{\rm f}=0.5,\  1,\ 2,\ 3,$ and $4$. The adopted range is intended to span modestly smaller effective radii, corresponding to collisions of smaller stars or ones which lost mass in previous collisions, to larger radii corresponding to larger stars or stars which puff up due to previous interactions by a factor of a few as found by \citet{Yao_2025ApJ...978...91Y}. We note, that changing $R_\star$ effectively explores gas dynamics of different $H/R_\star$, since for the same values of $H/R_\star$ (while keeping $\Sigma_{\rm d}$ fixed), the gas dynamics of star-disc collisions remain largely unchanged due to approximate scale invariance (e.g. \citealt{Yao_2025ApJ...978...91Y}). For our adopted values, $H/R_\star=6,\ 3,\ 3/2,\ 1,$ and $3/4$, respectively. For $R_\star/R_\star^{\rm f}=0.5$ we decrease the lateral extent of the disc to $4.5R_\odot$ in order to keep the same numerical resolution, i.e. the same ratio between $h_{\rm sl}$ and $R_\star$.

iv) We consider different disc vertical density profiles by varying the standard deviation $\sigma/R_\odot = \infty,\ 2.5,\ 2,\ 1.5,$ and $1$ of the Gaussian profile, $\rho_{\rm d}(z)\propto e^{-z^2/2\sigma^2}$, where $\sigma/R_\odot=\infty$ denotes the uniform density profile of the fiducial simulation. For these simulations, the density normalization is adjusted such that $M_{\rm d}$ is the same as in the fiducial model. We restrict to $\sigma/R_\odot \geq 1$ because for $\sigma/R_\odot \lesssim 1$ the disc becomes so thin that the effective ratio $H/R_\star$ is altered, and the variation of $\sigma/R_\odot$ is no longer cleanly single-parameter. This range of $\sigma$ corresponds to disc midplane densities $\rho_{\rm mid}/\rho_{\rm mid}^{\rm f}=1$, $1.3$, $1.4$, $1.7$, $2.4$, where $\rho_{\rm mid}^{\rm f}=3.7\times 10^{-8}\,{\rm g\,cm^{-3}} $ is the disc midplane density for the fiducial simulation. 

v) We consider $i=90^\circ,\ 75^\circ,\ 60^\circ,\ 45^\circ,$ and $30^\circ$, where $i=90^\circ$ corresponds to a local perpendicular crossing (the fiducial geometry). {Here $i$ denotes the local collision angle, defined as the angle that the relative velocity between the star and the disc gas makes with the disc midplane. This angle should not be confused with the global inclination of the stellar orbit. In general, $i$ depends on the global orbital geometry and on the velocities of both the star and the disc gas at the crossing point. Varying $i$ therefore explores different local collision geometries that can arise from different combinations of stellar orbit and disc rotation.} We restrict the range to $i\ge 30^\circ$ since decreasing $i$ increases the path length through dense gas, which can significantly increase the computational time. 

We emphasize that the simulations in this parameter survey should be interpreted primarily as controlled variations of individual local collision parameters. Their purpose is to characterize how each parameter affects the outflow hydrodynamics and the resulting flare properties, rather than to provide a set of fully self-consistent global QPE models. Constructing such models would require coupling the local collision conditions to the global orbital evolution and to the evolution of the stellar and disc properties between successive crossings, which is beyond the scope of this work. For example, at fixed \(M_{\rm bh}\), choosing a different fiducial collision radius would change \(v_\star\), $\Sigma_{\rm d}$ for a fixed accretion rate, and the tidal constraint on $R_\star$. However, since the local hydrodynamic evolution remains approximately self-similar when varying \(v_\star\), a different choice of the fiducial collision radius would not qualitatively affect the main results discussed below. In this sense, the largest \(R_\star\) simulations should be viewed mainly as controlled variations of the effective obstacle size (or \(H/R_\star\)), rather than as literal \(1\,M_\odot\) stars at the exact fiducial collision radius \(\sim 100R_{\rm g}\), where stars with \(R_\star\gtrsim 2R_\odot\) would likely be tidally disrupted by the SMBH.

\begin{table}
    \caption{Parameter ranges explored in this work.}
    \label{tab:parameters}
    \centering
    \begin{tabular}{l l l}
        \hline\hline
        Parameter & Fiducial value & Values w.r.t. fiducial \\
        \hline\hline
        $v_\star$ & $0.1\,c$ & $0.5$, $0.75$, $1$, $1.25$, $1.5$\\
        $\Sigma_{\rm d}$ & $1.6\times 10^{4}\,{\rm g\,cm^{-2}}$ & $1$, $3$, $10$, $30$, $100$\\
        $R_\star$ & $1\,R_\odot$ & $0.5$, $1$, $2$, $3$, $4$\\
        $\sigma/R_\odot$ & uniform & $\infty$, $2.5$, $2$, $1.5$, $1$\\
        $i$ & $90^\circ$ & $90^\circ$, $75^\circ$, $60^\circ$, $45^\circ$, $30^\circ$\\
        \hline
    \end{tabular}
    \tablefoot{Only one parameter is varied at a time, while all other parameters are held fixed. }
\end{table}

\section{Results}
\label{sec:results}

\subsection{Collision dynamics}\label{subsec:coll_dyn}

\begin{figure*}[t!]
   \centering
   \includegraphics[width=\textwidth]{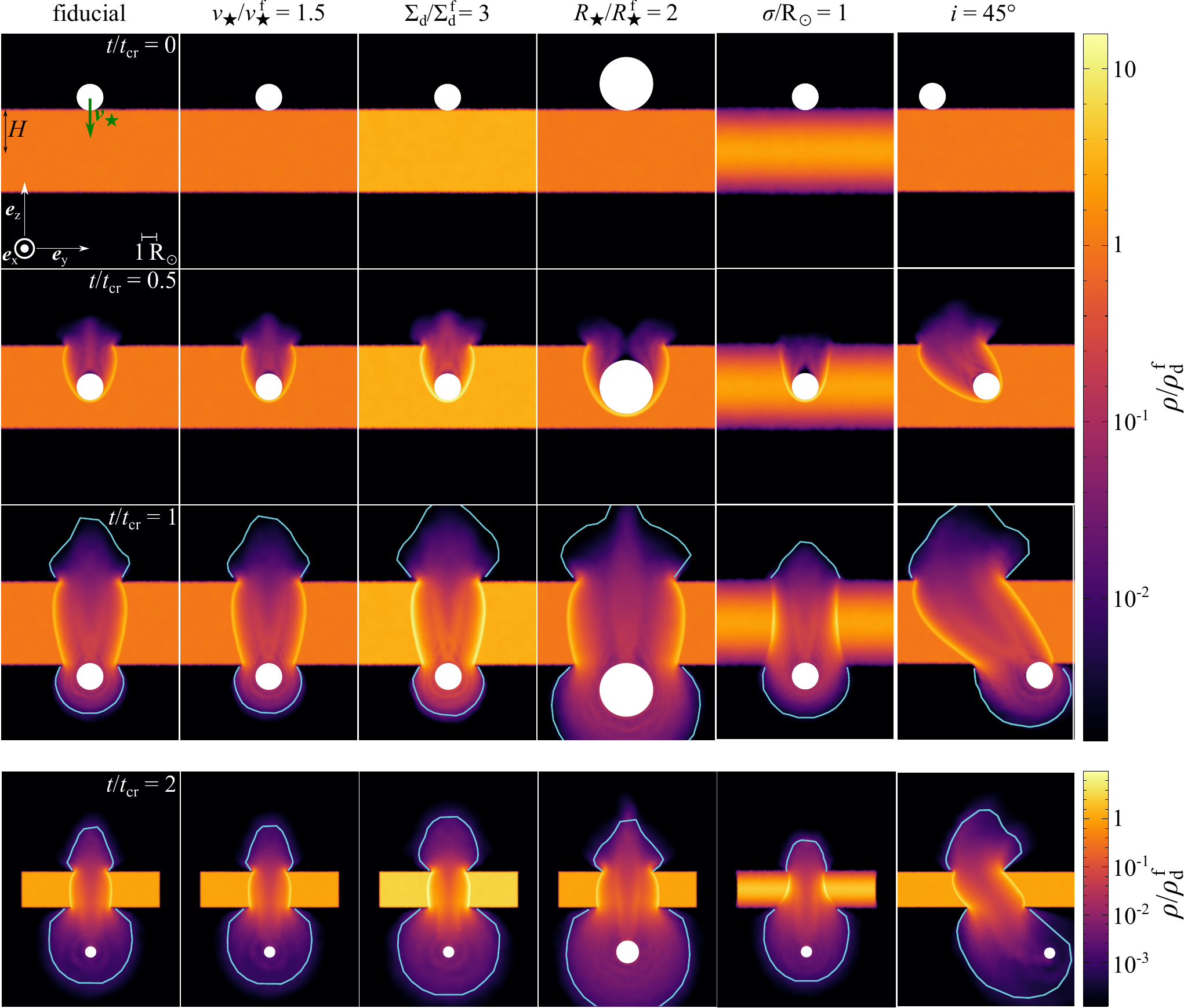}
   \caption{Gas density slices in the $yz$-plane at $x=0$ at different times for the following simulations: fiducial (first column), $v_\star/v_\star^{\rm f}=1.5$  (second column), $\Sigma_{\rm d}/\Sigma_{\rm d}^{\rm f}=3$  (third column), $R_\star/R_\star^{\rm f}=2$  (fourth column), $\sigma/R_\odot=1$  (fifth column), $i=45^\circ$  (sixth column). Snapshots in the first, second, third, and fourth row, correspond to $t/t_{\rm cr}=0$, $t/t_{\rm cr}=0.5$, $t/t_{\rm cr}=1$, $t/t_{\rm cr}=2$, respectively. In the last row, we zoom out and also change colorbar limits. The white circle denotes the star. Solid blue lines in the third and fourth row correspond to the photosphere surface evaluated according to the method described in Section \ref{subsec:lcs}.}
   	\label{fig:qpe_sim_density}
\end{figure*}

\begin{figure*}[h] 
   \centering
   \includegraphics[width=\textwidth]{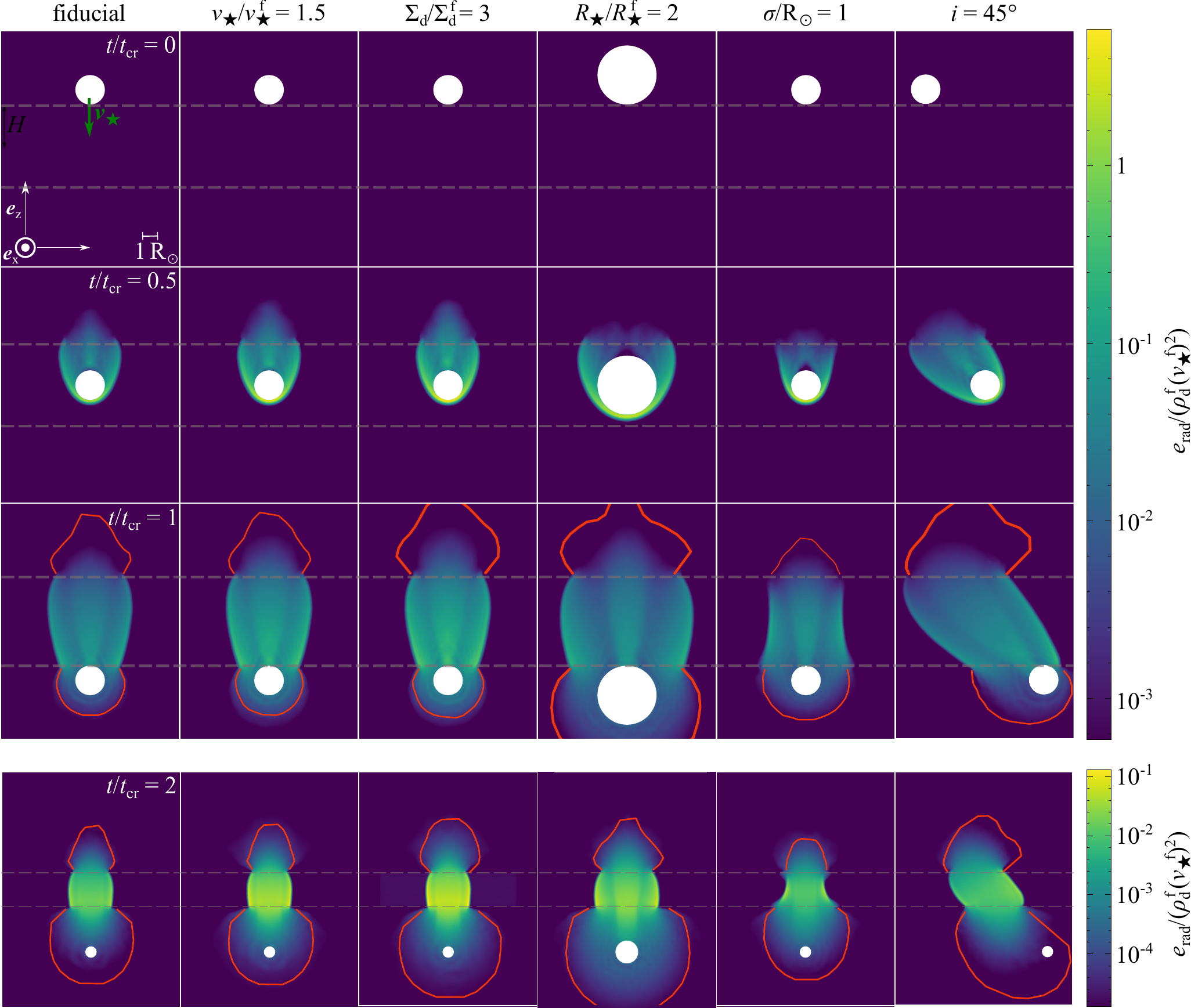}
  \caption{Radiation energy density slices in the $yz$-plane at $x=0$ at different times for the following simulations: fiducial (first column), $v_\star/v_\star^{\rm f}=1.5$  (second column), $\Sigma_{\rm d}/\Sigma_{\rm d}^{\rm f}=3$  (third column), $R_\star/R_\star^{\rm f}=2$  (fourth column), $\sigma/R_\odot=1$  (fifth column), $i=45^\circ$  (sixth column). Snapshots in the first, second, third, and fourth row, correspond to $t/t_{\rm cr}=0$, $t/t_{\rm cr}=0.5$, $t/t_{\rm cr}=1$, $t/t_{\rm cr}=2$, respectively.  In the last row, we zoom out and also change colorbar limits. The white circle denotes the star, while the dashed gray lines mark the outer edges of the accretion disc. Solid red lines in the third and fourth row correspond to the photosphere surface evaluated according to the method described in Section \ref{subsec:lcs}.}
  
   	\label{fig:qpe_sim_erad}
\end{figure*}

The dynamics of the star--disc interaction unfolds in several stages, driven by the formation of shocks, the redistribution of momentum and energy, and the subsequent ejection of gas. Figures~\ref{fig:qpe_sim_density} and \ref{fig:qpe_sim_erad} show $\rho$ and $e_{\rm rad}$, respectively, contained in slices in the $yz$-plane at $x=0$. Here we show only one representative simulation for each parameter variation, while snapshots from all simulations are shown in Section \ref{app:all_sims} (Figures \ref{fig:qpe_sim_density_all} and \ref{fig:qpe_sim_erad_all}). We express time in the units of the crossing time of the star through the disc $t_{\rm cr}=2(H+R_\star) (v_\star\sin i)^{-1}$. 

Focusing on the fiducial simulation (first column), at $t/t_{\rm cr}= 0$ (first row), the star collides supersonically with the disc, forming a radiation-pressure-dominated bow shock. This leads to a sharp increase in $\rho$ and temperature in the shocked layer and a corresponding rise in $e_{\rm rad}$. Soon, a quasi-steady state is established in which the inflow of material into the shock front is balanced by the outflow of gas escaping laterally around the star. As the shock front reaches the upper disc boundary, it triggers a shock breakout, where trapped radiation escapes as the post-shock gas expands into a low-pressure region. Gas flowing around the star subsequently converges in its wake, forming a secondary shock (visible at $t/t_{\rm cr}\approx 0.5$; second row), which re-heats the gas and produces a distinct ejecta component. At later times, the bow shock reaches the lower disc boundary, triggering a forward shock breakout and launching a quasi-spherical outflow along the forward direction (third and fourth rows). 

Varying $v_\star$ (second column) and $\Sigma_{\rm d}$ (third column) primarily changes the normalization of the thermodynamic variables, while leaving the qualitative gas dynamics largely unchanged. In particular, the morphology of the bow shock, lateral expansion, wake convergence, and the resulting outflow structure are similar across these suites when comparing simulations at the same $t/t_{\rm cr}$. The main difference is that larger $v_\star$ produces higher $e_{\rm rad}$ because the shock is stronger and the post-shock temperatures are higher. Likewise, increasing $\Sigma_{\rm d}$ increases the gas density throughout the interaction and results in higher $e_{\rm rad}$.

Varying $R_\star$ (fourth column) changes the lateral extent of the interaction region. In simulations with larger $R_\star$, the bow shock has a larger cross section and pushes disc gas to larger cylindrical radii, creating a larger evacuated cavity around the stellar trajectory. At the same time, a larger star intercepts and shocks a larger amount of gas. The asymmetry between the two outflows increases with $R_\star$, in the sense that the forward outflow becomes more dominant in mass.

The vertical structure of the disc (fifth column) mainly affects the shape of the shock front. For more centrally concentrated discs, the bow shock becomes less collimated near the disc edges as $\sigma$ decreases. This occurs because the shock front propagates faster through the lower-density outer disc layers. The outflow asymmetry does not vary strongly with decreasing $\sigma$.

For smaller $i$, the evacuated cavity carved by the star becomes elongated, producing an approximately elliptical shape in projection. As $i$ decreases, the star follows a longer path through the disc and shocks more gas, leading to a larger shocked mass. Additionally, the shock front and the resulting outflows are no longer axisymmetric around the $z$-axis. The mass asymmetry between the forward and backward outflows decreases with decreasing $i$, and the corresponding asymmetry in $e_{\rm rad}$ also becomes weaker. This reflects the longer interaction time, which allows a larger fraction of the shocked gas to flow around the star and reduces the relative importance of the quasi-spherical forward shock breakout compared to the perpendicular-crossing case. This is similar to the effect of decreasing $R_\star$ at fixed $H$, as discussed above.

\subsection{Shock heating}\label{subsec:shock}

\begin{figure}
	\centering  
        \begin{minipage}[b]{\linewidth}
        \includegraphics[width=\textwidth]{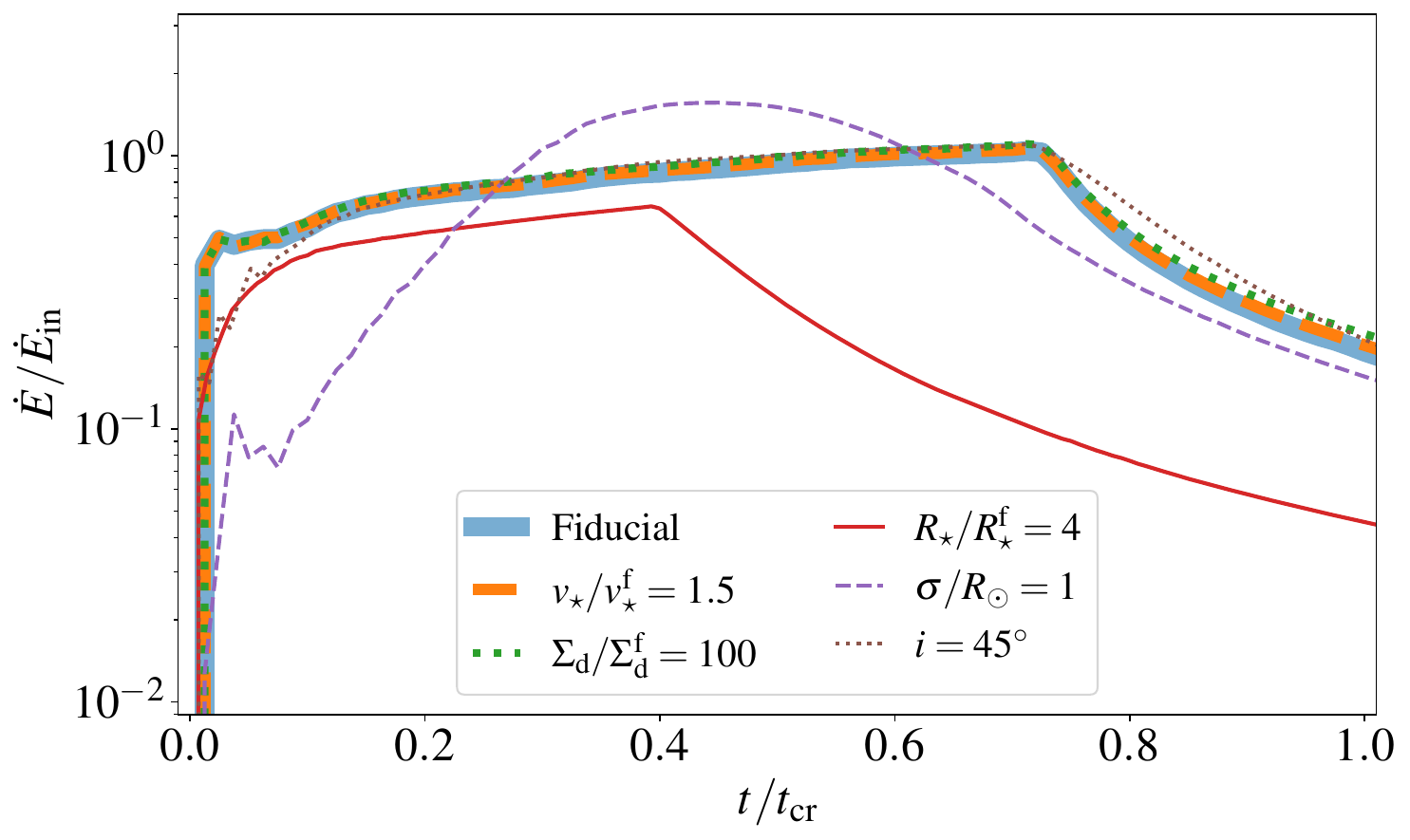}
	\end{minipage} 
	\caption{Shock heating rate $\dot{E}$ as a function of time for different simulations. The blue, orange, and green curves overlap because the dimensionless $\dot{E}$ remains nearly self-similar when varying \(v_\star\) or \(\Sigma_{\rm d}\).}
	\label{fig:dotE}
\end{figure}

Figure~\ref{fig:dotE} shows the shock heating rate $\dot{E}$ as a function of time for six representative simulations. The evolution of $\dot{E}$ for all the simulations is shown in Figure \ref{fig:dotE_all}. { {We obtain $\dot{E}$ directly from the simulations as the irreversible heating term, which captures shock dissipation in \textsc{Phantom}.}} The values are normalized to the shock heating rate for the gas inside the column through which the star moves 
\begin{align}
    \label{eq:dotE}
    \nonumber
    \dot{E}_{\rm in}&=\dot{M}_{\rm in}\,\Delta u\\
    &=5.6\times10^{42}\,{\rm erg\,s^{-1}} \left(\frac{v_\star }{0.1c}\right)^3 \left(\frac{\rho}{3.6\times 10^{-8}\, {\rm g\, cm^{-3}}}\right) \left(\frac{R_\star}{R_\odot}\right)^2,
\end{align}
where the mass inflow rate into the shock is $\dot{M}_{\rm in}=\pi R_\star^2 \rho v_\star$. { The quantity} $\Delta u=2(\Gamma+1)^{-2}v_\star^2 \approx 0.37\,v_\star^2$ corresponds to the characteristic increase in the specific internal energy of the gas due to the collision in the rest frame of the star, obtained from the Rankine--Hugoniot jump conditions assuming a strong shock and radiation-pressure-dominated gas with $\Gamma=4/3$.

In the fiducial simulation (solid blue line), $\dot{E}$ rises sharply as soon as the star makes first contact with the disc at $t=0$. While the star remains inside the disc, $\dot{E}$ continues to increase because the shock front is still laterally detaching and expanding around the star, and has not yet reached a perfect steady state. $\dot{E}/\dot{E}_{\rm in}$ starts to decrease at $t/t_{\rm cr}\approx 0.75$, when the shock front reaches the lower disc edge. However, it does not vanish entirely because the shock front continues to propagate laterally through the disc. 

Simulations with different $v_\star$ ({ dashed} orange line), $\Sigma_{\rm d}$ ({ dotted} green line), and $i$ ({ dotted} brown line) exhibit a very similar evolution to the fiducial case when expressed in $\dot{E}/\dot{E}_{\rm in}$ and compared at the same $t/t_{\rm cr}$, indicating that the shock structure and energy conversion efficiency remain broadly self-similar across these suites. Additionally, {simulations with smaller local collision angles} show a more gradual rise and fall of $\dot{E}$ because the star enters and leaves the disc more gradually.

{For larger $R_\star$ (solid red line), $\dot{E}/\dot{E}_{\rm in}$ is lower compared to the fiducial case at the same $t/t_{\rm cr}$ because the shock front has not yet expanded laterally as much, which reduces the contribution to $\dot{E}$ from the lateral shock-heated regions. Additionally, $\dot{E}/\dot{E}_{\rm in}$ begins to decrease earlier, at $t/t_{\rm cr}\approx 0.4$, because for larger $R_\star$, the shock front is thicker and reaches the disc boundary at an earlier $t/t_{\rm cr}$.}

For more centrally concentrated discs ({ dashed} purple line), $\dot{E}$ directly reflects the stratification of the disc. Because the gas density is lower near the disc surfaces, $\dot{E}$ is lower when the star enters the disc and again when it exits compared to the uniform-density fiducial simulation. By contrast, as the star crosses the disc center, where the density is highest, $\dot{E}$ increases correspondingly. This trend becomes more pronounced as $\sigma$ decreases and the density becomes higher in the midplane while remaining low near the surfaces (at fixed $\Sigma_{\rm d}$).

\subsection{Outflow}\label{subsec:outflow}

\begin{figure}[h!]
\centering
\includegraphics[width=0.94\linewidth]{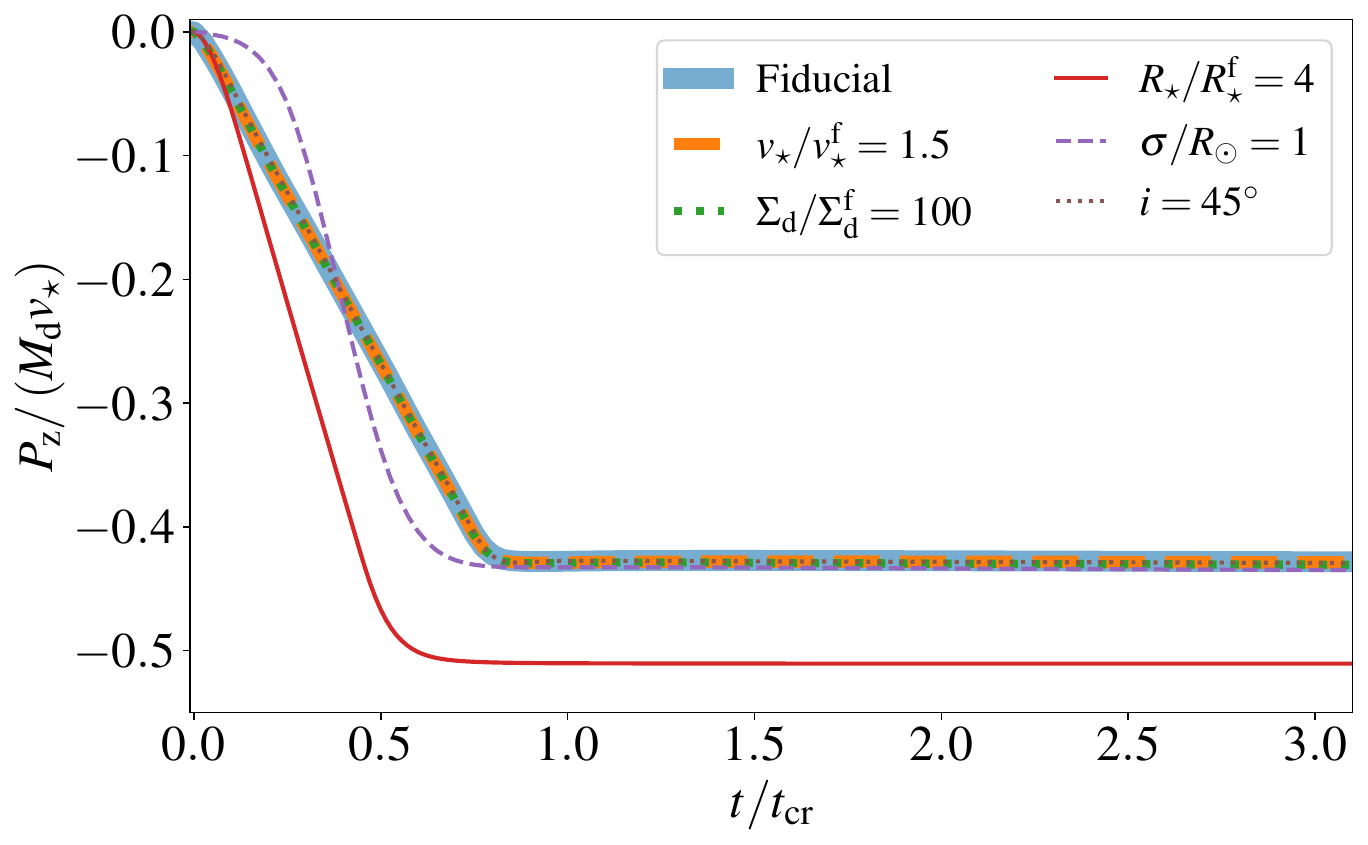}
\includegraphics[width=0.94\linewidth]{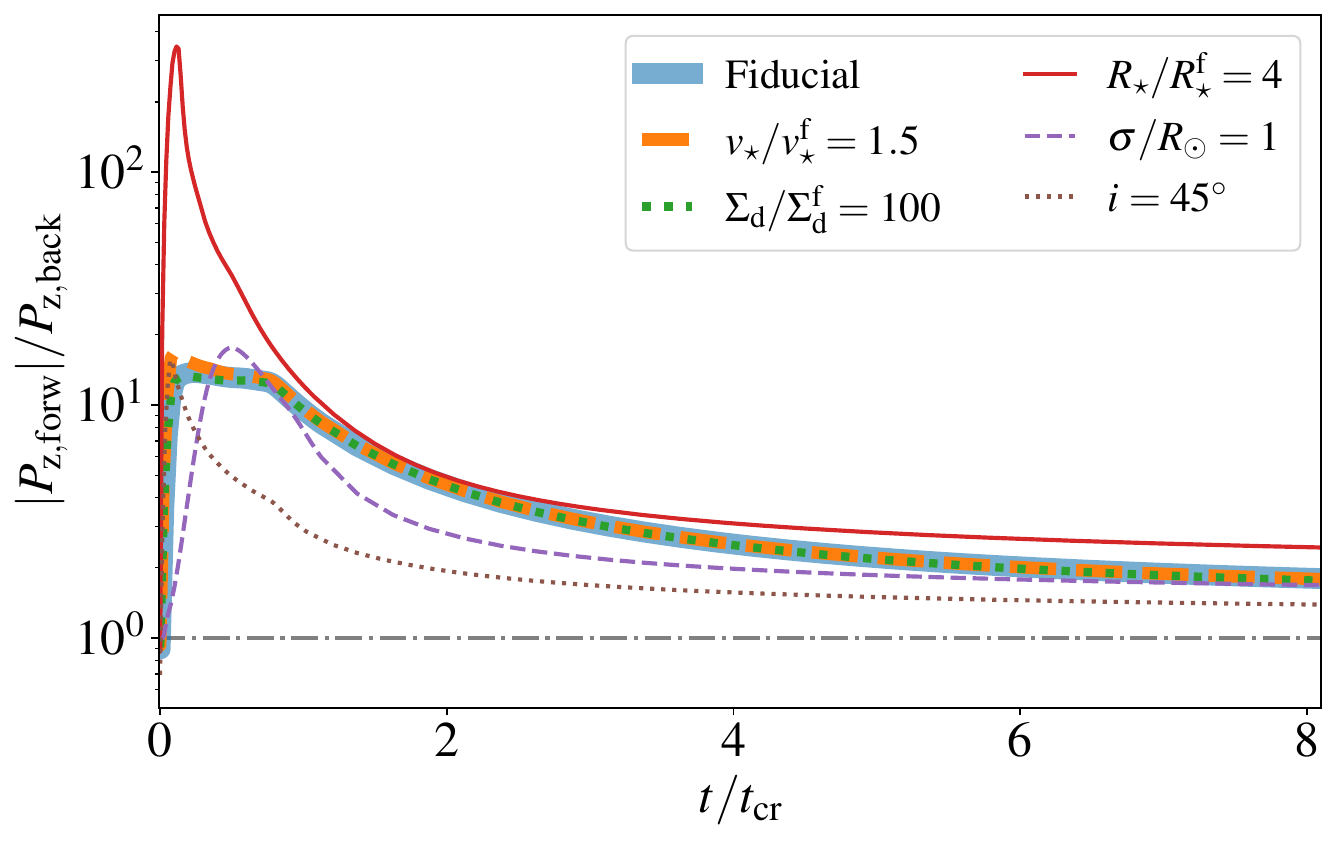}
\includegraphics[width=0.98\linewidth]{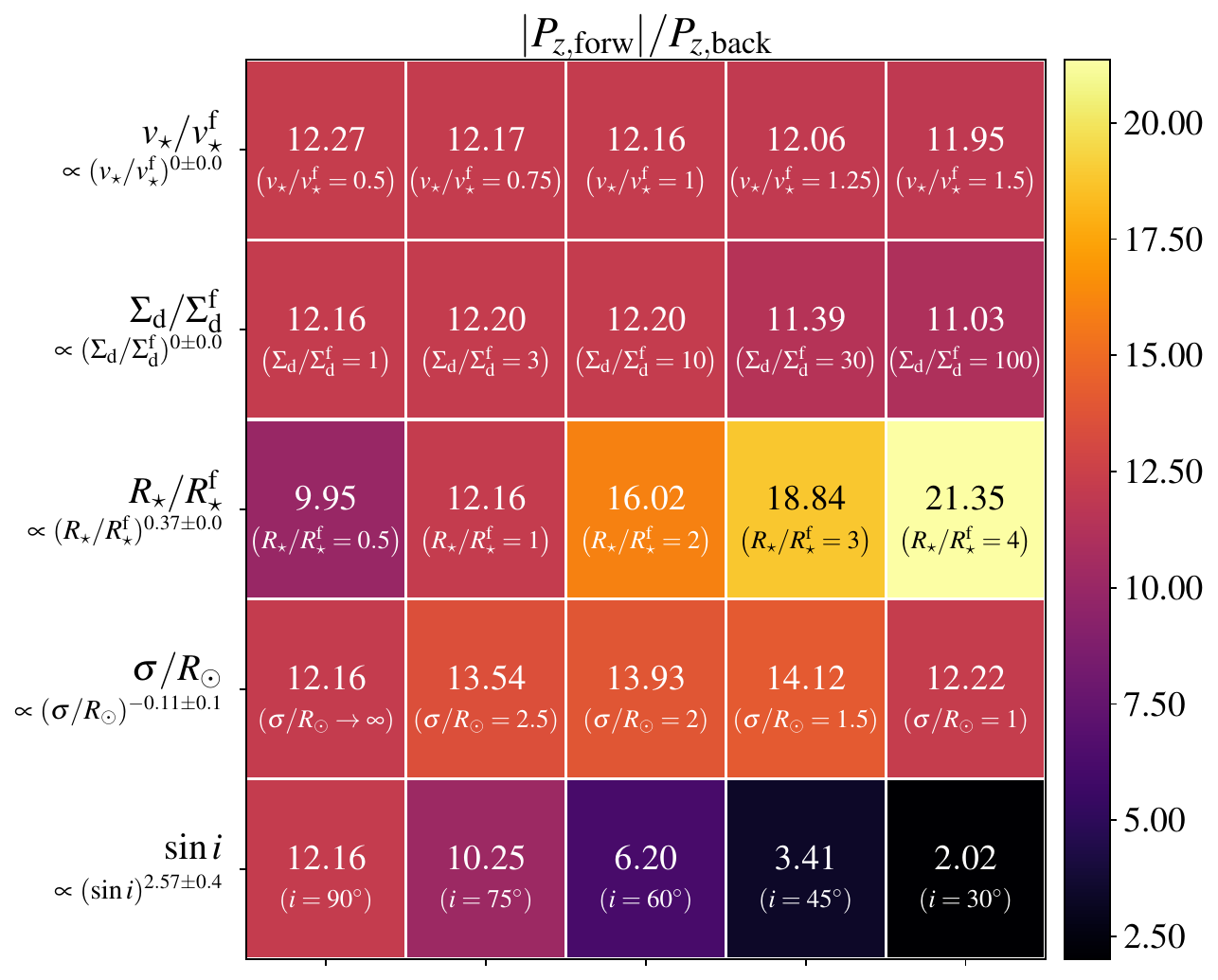}
\caption{Top: Momentum  $P_{\rm z}^{\rm }$ along $e_{\rm z}$ of the
shocked gas for different simulations. The blue, orange, green, and brown curves overlap
because the dimensionless redistribution of momentum remains nearly
self-similar { across these simulations}.
{ Middle: Ratio of the magnitudes of the forward and backward
vertical momenta, $|P_{z,\rm forw}|/P_{z,\rm back}$.
Bottom: Heatmap of $|P_{z,\rm forw}|/P_{z,\rm back}$ evaluated at the time when half of the star emerges from the disc. Each row corresponds to a different
varied parameter, while each column corresponds to a different parameter
value, shown in parentheses. Values of the momentum ratio are shown in
individual cells, which are also colored according to their values. On
the left, we show the best-fitting power-law dependence and its
uncertainty for each parameter.}}
\label{fig:Pz}
\end{figure}

Figure~\ref{fig:Pz} (top) shows the evolution of the vertical momentum
$P_{\rm z}$ for gas that has been affected by the shock by the end of
our simulation for six representative simulations. We calculate it as
the sum of momenta along $\hat{\boldsymbol e}_{\rm z}$ of gas that is
either outside the disc or with a specific internal energy $u>2u_{\rm d}$
in the last snapshot, where $u_{\rm d}$ is the value in the initial disc.
We normalize $P_{\rm z}$ by $M_{\rm d}v_\star$.

While the star is inside the disc, it continuously injects momentum into the surrounding gas, and $P_{\rm z}$ decreases as gas is accelerated predominantly along $-\hat{\boldsymbol e}_{\rm z}$. After the star exits the disc, the momentum asymptotes to a constant value, reflecting that the momentum of the shocked gas remains constant once the injection stops. { For all the simulations, except for those with different $R_\star$ (solid red line), the evolution of $P_{\rm z}/(M_{\rm d}v_\star)$ remains very similar to the fiducial case, with an asymptotic value of $P_{\rm z}/(M_{\rm d}v_\star)\simeq-0.43$. For more centrally concentrated discs (dashed purple line),  the transition to the asymptotic value is steeper, due to the increased density near the disc midplane and the correspondingly more pronounced momentum injection in that region.} { Increasing $R_\star$ (solid red line) produces a more negative asymptotic net momentum.}

\begin{figure}
	\centering
    \includegraphics[width=0.99\linewidth]{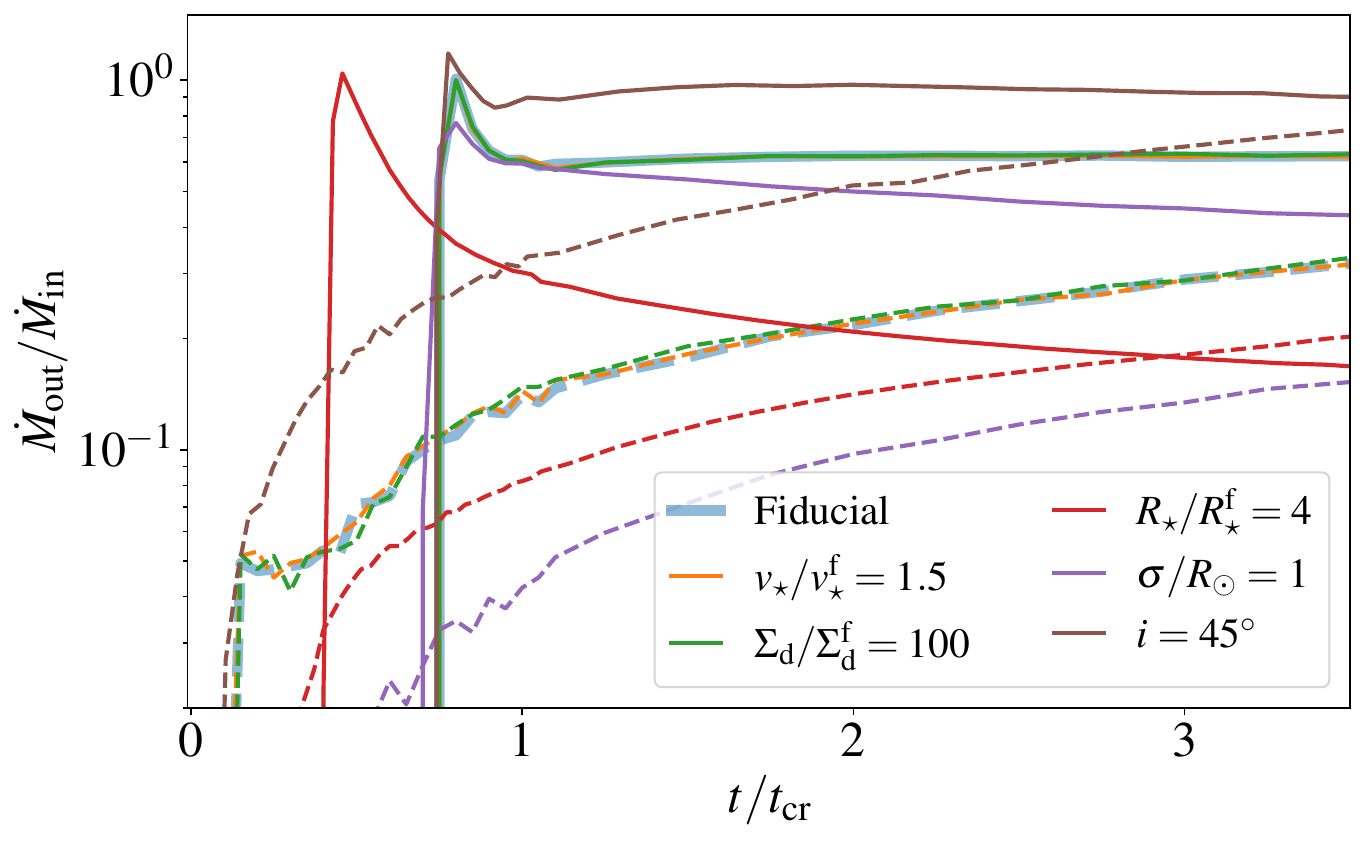}
    \caption{The mass outflow rate $\dot{M}_{\rm out}$ ejected along the forward (solid lines) and backward (dashed lines) direction for different simulations. The blue, orange, and green curves overlap because the dimensionless $\dot{M}_{\rm out}$ remains nearly self-similar when varying \(v_\star\) or \(\Sigma_{\rm d}\).}
	\label{fig:dotM_Mej}
\end{figure}

{ In Figure~\ref{fig:Pz} (middle), we show the momentum asymmetry, calculated as the ratio of the momentum magnitudes along  $\hat{\boldsymbol e}_{\rm z}$ of the forward and backward-moving gas, $|P_{z,\rm forw}|/P_{z,\rm back}$. The momentum is initially strongly
asymmetric between the two directions, with the forward component
dominating because the forward-moving gas is on average faster and because it contains more mass. At later times, $|P_{z,\rm forw}|/P_{z,\rm back}$
gradually decreases and approaches unity (horizontal dash-dotted line) as the masses of the forward and backward outflows
increase at increasingly similar rates (see Figure~\ref{fig:dotM_Mej}).}

{  In Figure~\ref{fig:Pz} (bottom), we show a heatmap of $|P_{z,\rm forw}|/P_{z,\rm back}$ evaluated at the time, when half of the star has emerged from the disc and the total $P_{\rm z}$ has reached its asymptotic value (see the top panel of Figure~\ref{fig:Pz}). On the left, we show the best-fitting power-law dependence and its uncertainty for each varied parameter, obtained through the least-squares method while holding all other parameters fixed at their fiducial values.}

{ Varying $v_\star$ (first row), $\Sigma_{\rm d}$ (second row) or vertical disc density profile (fourth row) produces
only a very small change relative to the fiducial case. Increasing $R_\star$ (third row) leads to a strong increase in the
momentum asymmetry with
$|P_{z,\rm forw}|/P_{z,\rm back}\propto R_\star^{0.37}$.} This is driven primarily by the fact that increasing
$R_\star$ decreases $H/R_\star$ and therefore shortens the time during
which shocked gas can flow around the star and reach the ingress side
of the disc. As a result, a smaller fraction of the shock-heated gas is
redirected into the backward outflow, and { the momentum distribution}
remains more strongly biased toward the forward direction.

{ The momentum asymmetry decreases with decreasing $i$ (fifth row). Specifically, we find
$|P_{z,\rm forw}|/P_{z,\rm back}\propto(\sin i)^{2.57}$.}
This is because in more grazing encounters, the injected momentum is redistributed more symmetrically between the two sides of the disc. In the limiting case $i\to0$, where the relative motion is nearly parallel to the disc midplane, {$|P_{z,\rm forw}|/P_{z,\rm back}$ would approach unity.}
{ Despite this decreasing asymmetry, the asymptotic total $P_z$ remains nearly unchanged with $i$ (dotted brown line in the top panel of Figure~\ref{fig:Pz}). More grazing collisions affect a larger amount of disc gas, increasing the magnitudes of both momentum components. Their difference, which determines the total $P_z$, can therefore remain approximately constant while their ratio approaches unity.}

Figure~\ref{fig:dotM_Mej} shows the mass outflow rate $\dot{M}_{\rm out}$ ejected along the forward (solid lines) and backward (dashed lines) direction {calculated from the gas mass crossing the lower and upper disc boundaries, respectively, per unit time. Values} are normalized to the mass inflow rate into the shock $\dot{M}_{\rm in}=\pi R_\star^2 \rho_{\rm d} v_\star$ for six representative simulations. The evolution of $\dot{M}$ for all the simulations is shown in Figure \ref{fig:dotM_Mej_all}. In the fiducial simulation (blue lines), the forward $\dot{M}_{\rm out}$ shows an early spike associated with the forward shock breakout{. At later times, most of the forward outflow emerges from an annular region where the laterally expanding shock front intersects the lower disc surface. The backward outflow is similarly supplied mainly by gas escaping near the intersection of the shock front with the upper disc surface. The forward $\dot{M}_{\rm out}$ then remains approximately constant, while the backward $\dot{M}_{\rm out}$ gradually increases and approaches it {as the conditions on the two sides of the interaction structure become increasingly symmetric}. This delayed rise occurs because gas contributing to the backward outflow must first flow around the star and can remain in the shocked cavity before escaping through the upper disc surface. Gas shocked deeper in the disc has a longer escape time, causing the backward mass flux to build up gradually.}\footnote{The constant forward $\dot{M}_{\rm out}$ can be understood from the scaling $\dot{M}_{\rm out}\approx A\rho v_{\rm z}$, where $A$, $\rho$, and $v_{\rm z}$ are the effective outflow area, density, and vertical velocity of the gas crossing the ejection surface \citep{Jankovic_2026A&A...710A.405J}. At late times, $A\propto r_{\perp,{\rm sh}}\propto t^{1/2}$, while $v_{\rm z}\propto v_{\rm sh}\propto t^{-1/2}$, and $\rho$ remains an approximately constant multiple of $\rho_{\rm d}$, leading to an approximately constant $\dot{M}_{\rm out}$. The backward $\dot{M}_{\rm out}$ instead increases because, after passing through the shock, matter accumulates in the cavity before escaping, with a time delay that is longer for gas deeper inside the disc.}

The simulations with different $v_\star$ (orange lines) and $\Sigma_{\rm d}$ (green lines) show a very similar behaviour to the fiducial case when compared at the same $t/t_{\rm cr}$ and expressed in $\dot{M}_{\rm in}$. In collisions with lower $i$ (brown lines), $\dot{M}_{\rm out}$ is higher than in the fiducial case for both outflows, due to the longer path length through the disc and the larger amount of shocked gas. Additionally, the asymmetry between the forward and backward $\dot{M}_{\rm out}$ becomes smaller, due to the more symmetric redistribution of the injected momentum (see Figure \ref{fig:Pz}).

For larger $R_\star$ (red lines) and lower $\sigma$ (purple lines), the late-time forward $\dot{M}_{\rm out}$ does not remain constant but instead decreases with time. This indicates that, in these cases, the late-time forward outflow is not dominated by a ring-like interface of approximately fixed width at the disc edge.

\subsection{Observable properties}\label{subsec:lcs}

\subsubsection{ Outflow energetics}
\label{sec:outflow_energy}

{ When the star collides with the disc, it injects energy into the gas.
This energy is initially redistributed into thermal and radiation energy
by shock heating and is then rapidly transferred into bulk kinetic energy
as the shocked gas expands. The increase in kinetic energy is driven both
by pressure forces in the optically thick gas and by the direct mechanical
pushing of the gas by the star. The detailed exchange between these energy components is discussed in \citet{Jankovic_2026A&A...710A.405J}.}

\begin{figure}%[H] 
   \centering
    \includegraphics[width=0.99\linewidth]{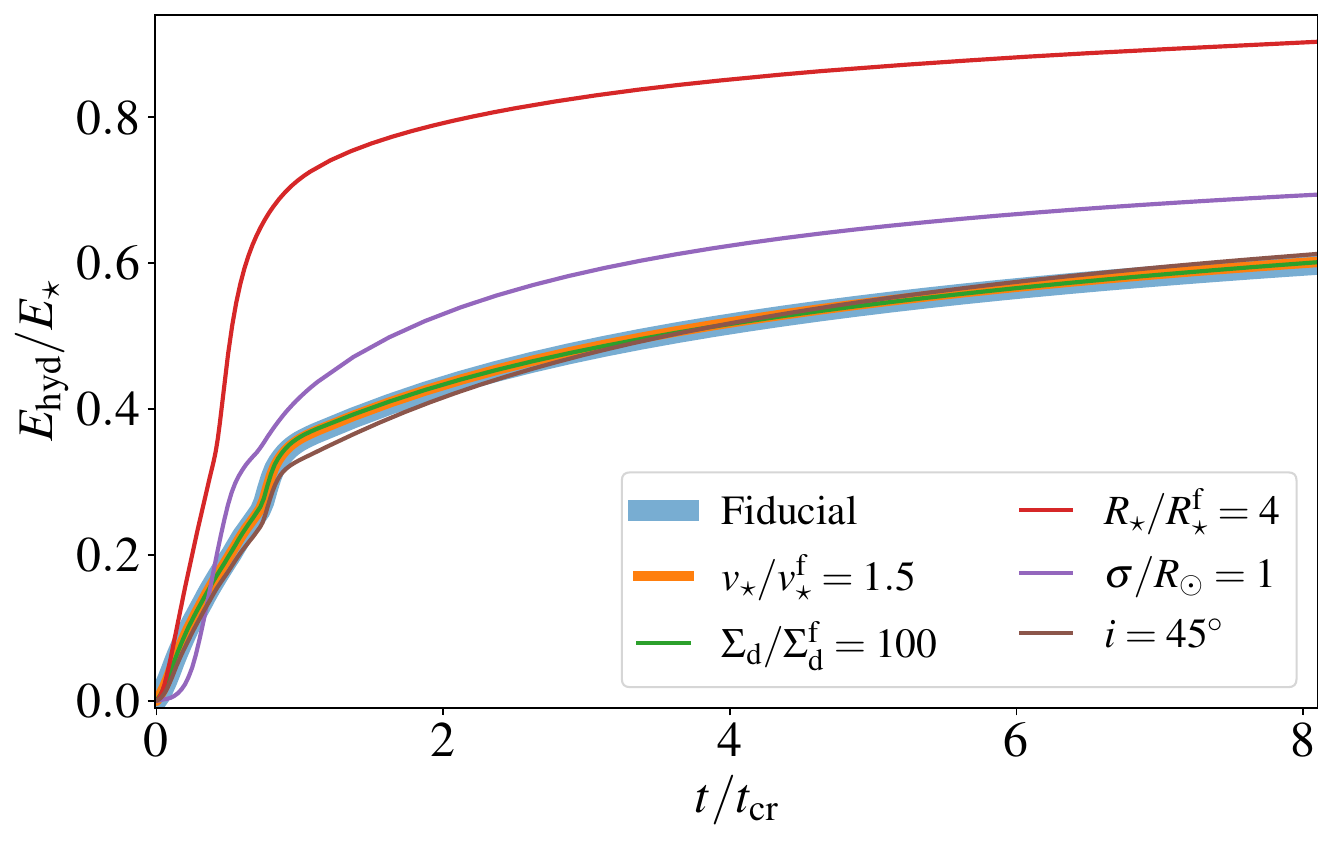}
     \includegraphics[width=0.99\linewidth]{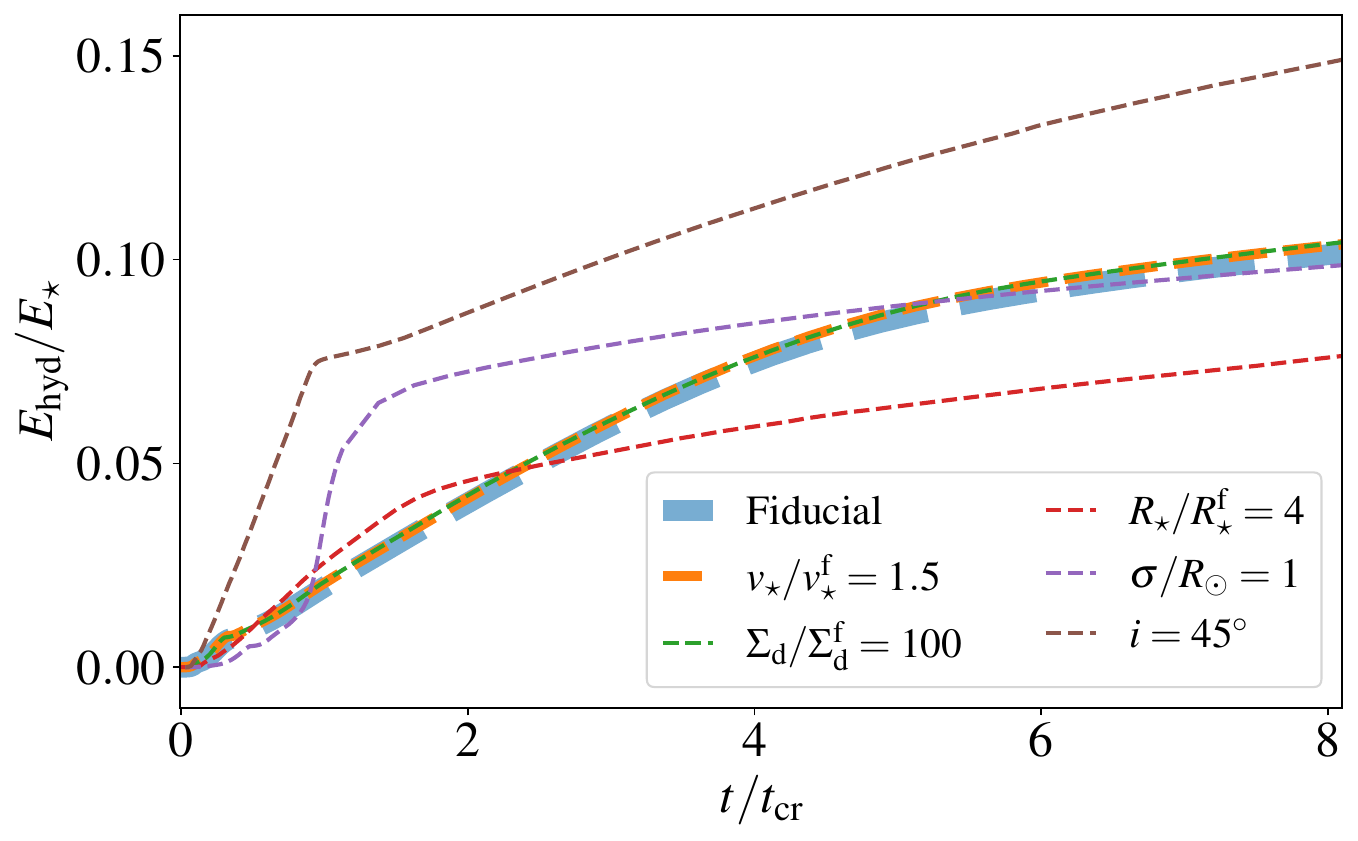}
\caption{ Hydrodynamic energy $E_{\rm hyd}$ of the forward-moving (top) and
backward-moving (bottom) gas, corresponding to gas with $v_z<0$ and $v_z>0$, respectively, evaluated at different times. The blue, orange, and green curves
overlap because the dimensionless redistribution of energy remains
nearly self-similar when varying $v_\star$ or $\Sigma_{\rm d}$.}
   	\label{fig:Eoutflow}
\end{figure}

{ In Figure~\ref{fig:Eoutflow}, we show the evolution of the hydrodynamic energy $E_{\rm hyd}$, obtained as the sum of the kinetic and thermal energies, of the forward-moving (top panel) and backward-moving (bottom panel) gas. {We normalize the energies by} 
\begin{equation}
    E_{\star}
    =\frac{1}{2}M_{\rm d}v_\star^2 
    =1.1\times10^{45}\,{\rm erg}
    \left(\frac{M_{\rm d}}{1.2\times10^{-7}\,M_\odot}\right)
    \left(\frac{v_\star}{0.1c}\right)^2 .
\end{equation}
While the star is crossing the disc and injects energy into the shocked gas, $E_{\rm hyd}$ rises rapidly in both components, although more strongly in the forward-moving gas. After the star leaves the disc, the total energy of the gas, which includes its kinetic, radiation, and thermal energy, remains
approximately constant at $\sim E_\star$  (see also Figure~11 of \citealt{Jankovic_2026A&A...710A.405J}). Nevertheless, $E_{\rm hyd}$ continues to increase more gradually as radiation energy is converted into bulk kinetic energy through radiation pressure during the expansion.  The forward-moving gas carries more hydrodynamic energy in all of the simulations. The asymmetry between the two components is strongest for the larger star (red lines). Conversely, the $i=45^\circ$ collision (brown lines) exhibits the smallest asymmetry, consistent with a more symmetric redistribution of the injected energy. }

{ The hydrodynamic energy carried by the outflows sets an upper limit on the total energy that could be radiated in a secondary interaction. For example, if an outflow collides with ambient gas, part of this energy may be converted into radiation through additional shocks. In optically thin, magnetized regions, such interactions could also accelerate nonthermal particles. The resulting thermal or nonthermal emission may
be observable. However, modelling these processes is beyond the scope
of this work.}

\subsubsection{ Light curve calculation and fiducial simulation}

\begin{figure}
   \centering
    \includegraphics[width=0.49\textwidth]{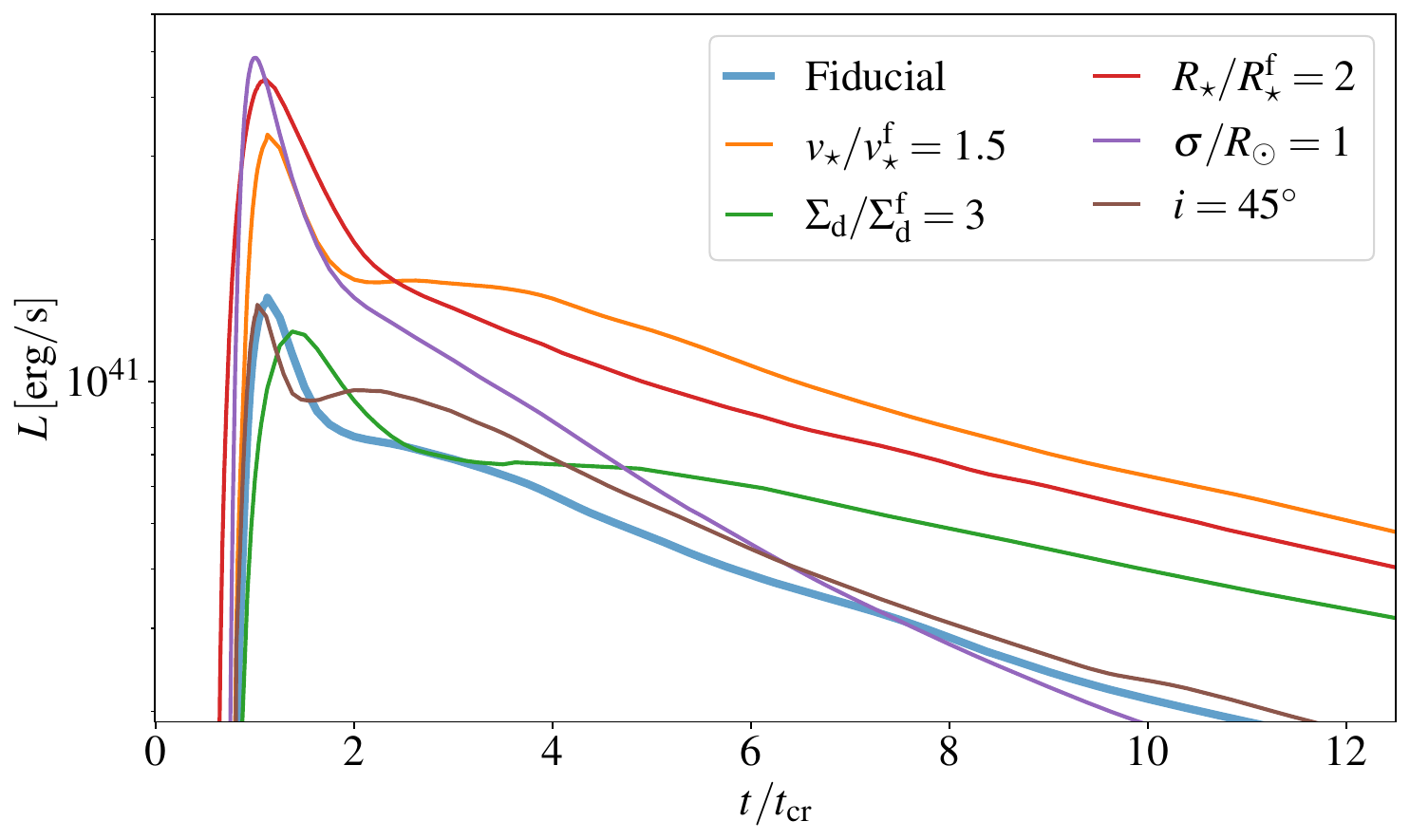}
     \includegraphics[width=0.49\textwidth]{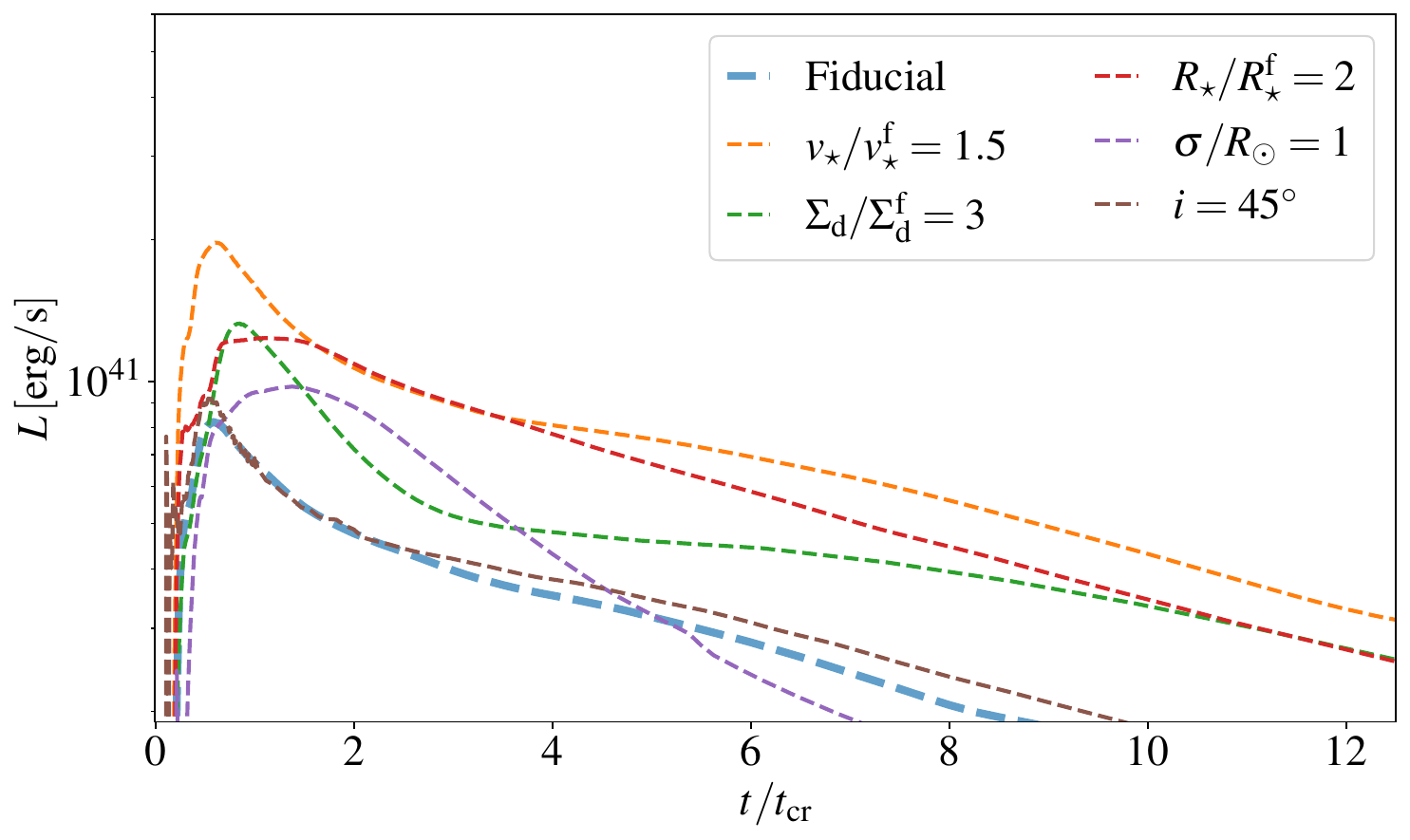}
   \caption{The time evolution of $L$ of the forward (top) and backward (bottom) outflow for different simulations. The dotted lines denote the extrapolated late-time decay.}
   	\label{fig:L_plot}
\end{figure}

Figure~\ref{fig:L_plot} shows the bolometric luminosity $L$ of the forward (top panel; solid lines) and backward (bottom panel; dashed lines) outflows as a function of time for six represenative simulations. We compute the emerging luminosity by integrating the sum of the diffusion and advective radiation fluxes through the photosphere. We determine the photosphere separately for each outflow by considering a hemisphere of rays centered on the breakout region at the disc surface, and locating along each ray the photospheric radius $R_{\rm ph}$ where the optical depth $\tau=\int_{R_{\rm ph}}^\infty \rho\kappa_{\rm s}\,{\rm d}r$ drops to unity. The diffusion flux is obtained from Equation~(\ref{eq:Fdiff}), while for the advective flux we follow \citet{Piro_2020ApJ...894....2P} and account for the motion of the photosphere, $\boldsymbol F_{\rm adv}^{\rm ph}=e_{\rm rad}(\boldsymbol v-\boldsymbol v_{\rm ph})$, where $\boldsymbol v$ is the gas velocity and $\boldsymbol v_{\rm ph}$ is the velocity of the receding photosphere, estimated from the change of $\boldsymbol R_{\rm ph}$ between snapshots. This ensures that only the net radiation energy crossing the moving photosphere contributes to $L$. The emerging luminosity is set by the shock heating rate $\dot{E}$ while the star is passing through the disc, which is then reduced by adiabatic losses during expansion, meaning that $L\sim \eta\,\dot{E}$, where $\eta<1$ accounts for adiabatic losses \citep[e.g.][]{Linial2023}.

In the fiducial simulation (blue lines in Figure \ref{fig:L_plot}), forward and backward lightcurves rise rapidly to $L=1.5\times 10^{41}\,{\rm erg\,s^{-1}}$ and $L=8.2\times 10^{40}\,{\rm erg\,s^{-1}}$ (see left panels of Figure \ref{fig:Lpeak_duration}), followed by a more gradual decline. The forward outflow is brighter than the backward one, reflecting a combination of asymmetric energy injection into the two ejecta components (asymmetry in $\dot{E}$) and differences in how efficiently each component converts the injected energy into escaping radiation (asymmetry in $\eta$). In the fiducial case, the forward outflow is brighter mainly due to the higher $\dot{E}$ of the forward ejected gas as explained in \citet{Jankovic_2026A&A...710A.405J}. The forward lightcurve also exhibits a break to a shallower decline at $t/t_{\rm cr}\approx 1.7$ corresponding to the transition from emission emerging from the forward breakout ejecta to emission emerging from gas that remains trapped longer in the cavity. {We note that the fiducial peak bolometric luminosities lie near the lower end of the observed QPE-candidate X-ray luminosity range. Specifically, they are comparable to the peak luminosity of XMMSL1~J024916.6-041244 $\sim10^{41}\,{\rm erg\,s^{-1}}$ \citep{Chakraborty_2021ApJ...921L..40C}, but are below the luminosities of brighter QPE sources such as GSN~069 $\sim10^{42}\,{\rm erg\,s^{-1}}$ \citep{Miniutti2019}. This motivates the parameter variations explored below.}

\subsubsection{ Dependence on the collision parameters}

\begin{figure*}[!t]
   \centering
    \includegraphics[width=0.49\textwidth]{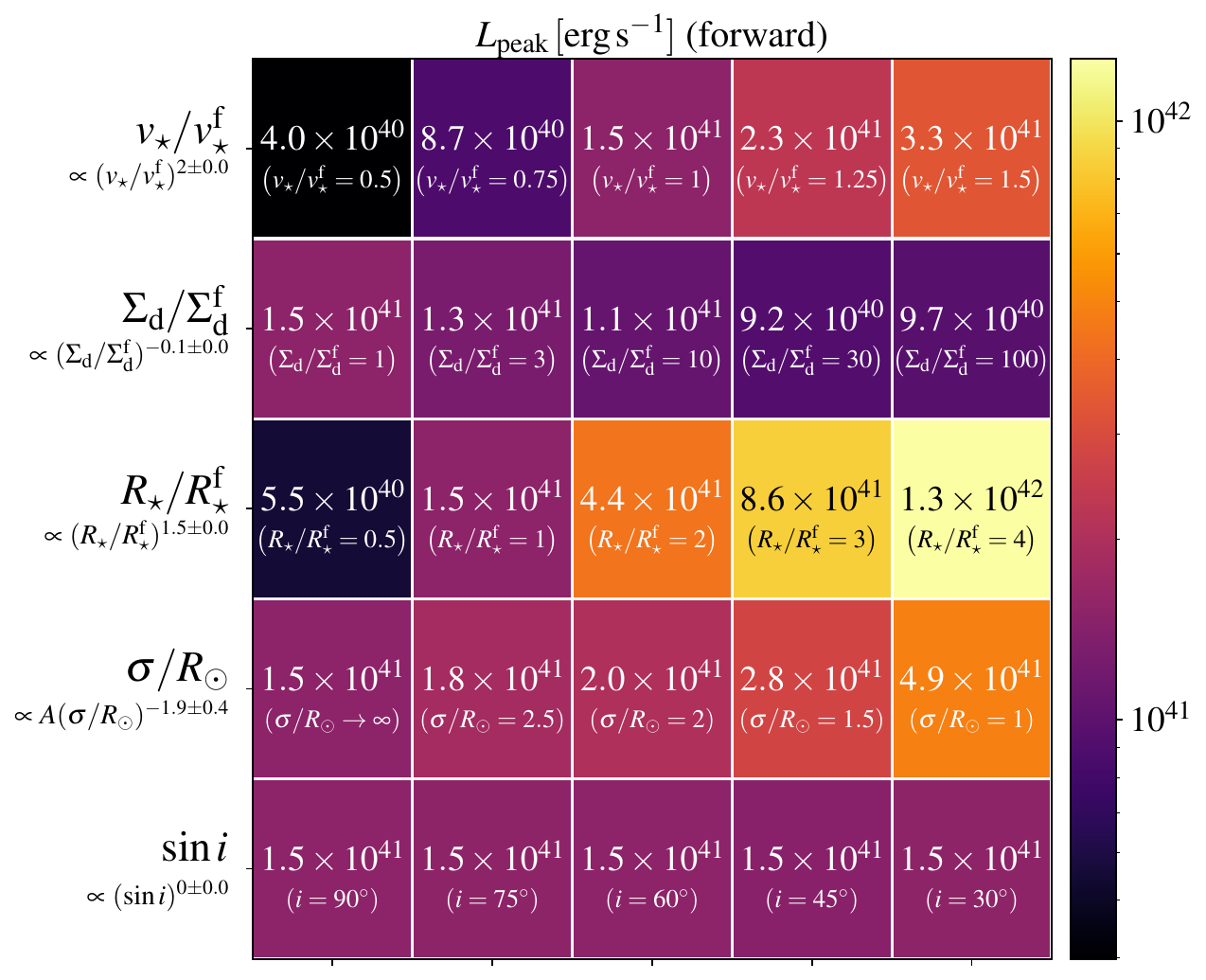}
    \includegraphics[width=0.49\textwidth]{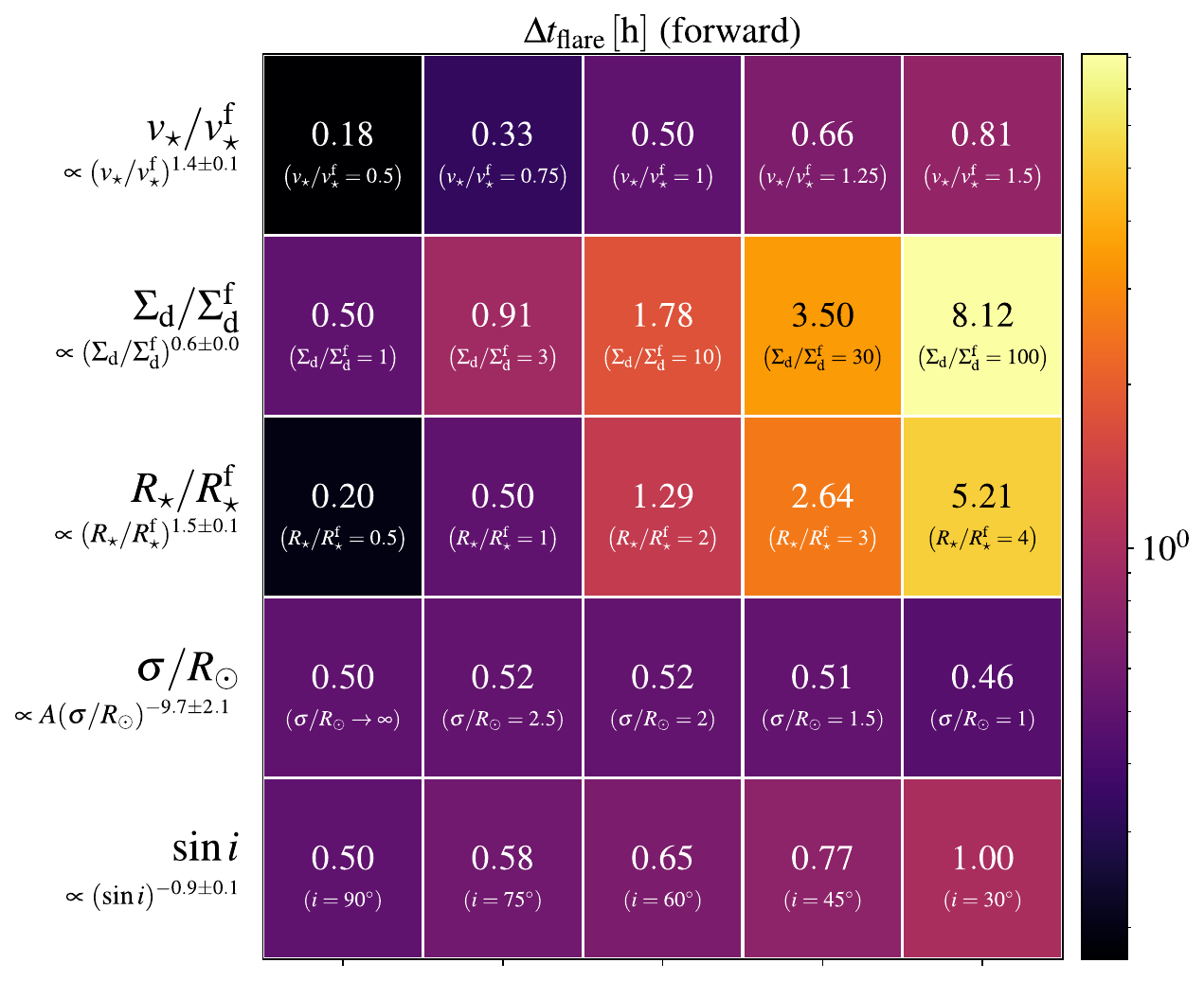}
     \includegraphics[width=0.49\textwidth]{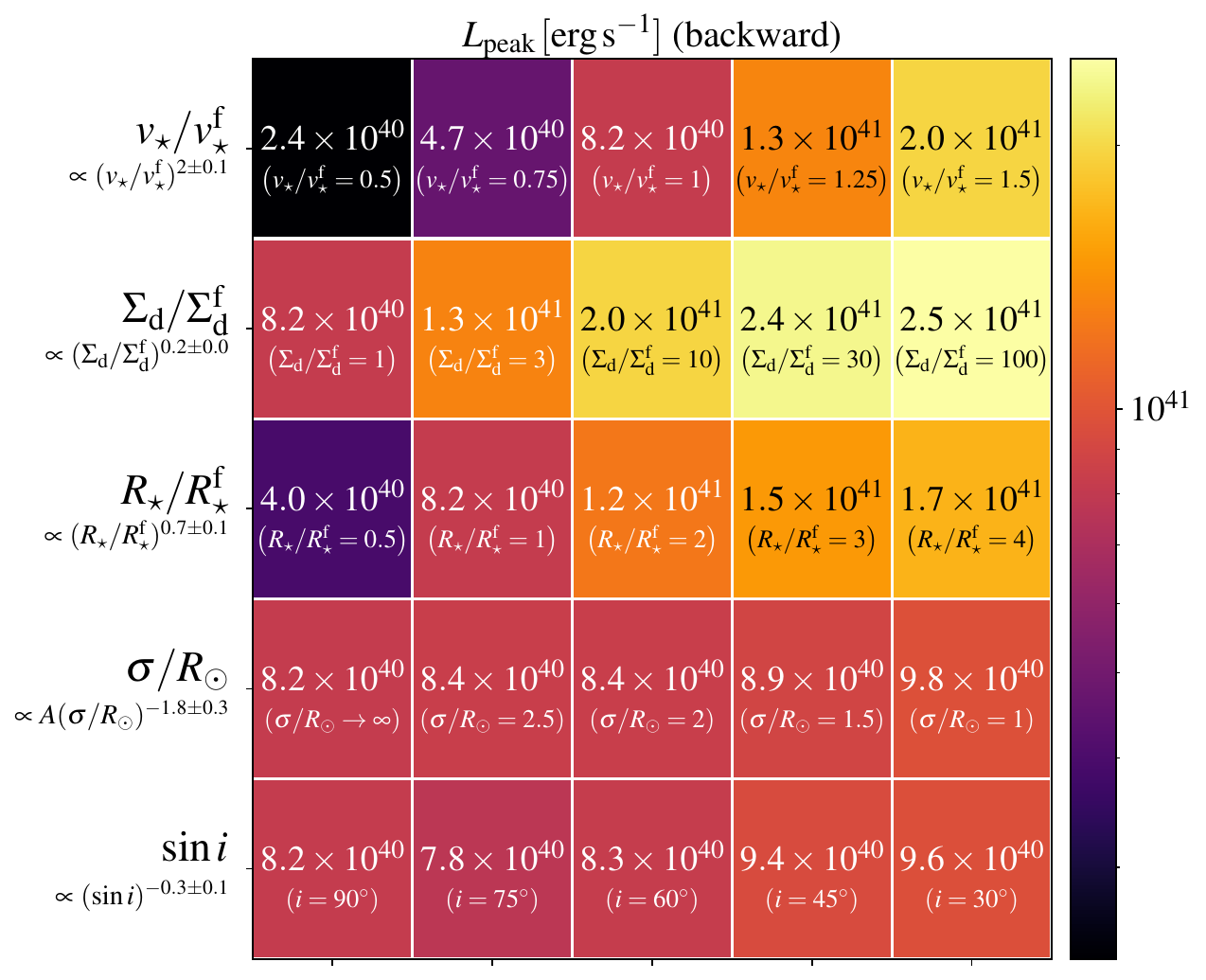}
    \includegraphics[width=0.49\textwidth]{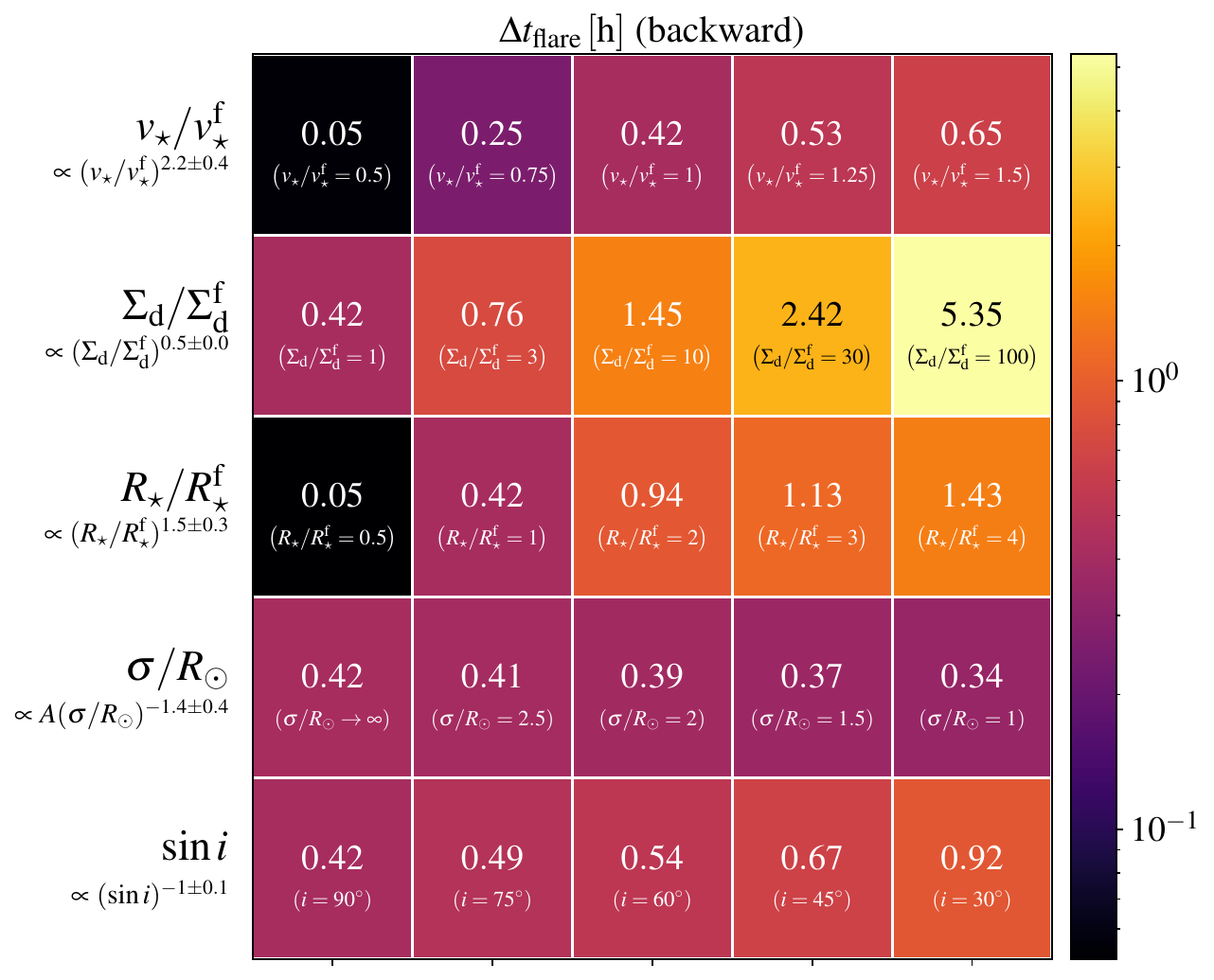}
   \caption{Heatmaps of the peak luminosity $L_{\rm peak}$ (first columns) and flare durations $\Delta t$ (right columns) for the forward (top) and backward (bottom) outflows for all simulations. Each row corresponds to a different varied parameter, while each column corresponds to a different parameter value, shown in parentheses. Values of $L_{\rm peak}$ and $\Delta t$ are shown in individual cells, which are also colored according to their values. On the left, we show the best-fitting power-law dependence and its uncertainty for each parameter. For backward lightcurves in which the prominent early spike associated with shock breakout is brighter than the rest of the lightcurve { (see Figure \ref{fig:L_plot_back_all})}, we report $L_{\rm peak}$ of the broader secondary peak and calculate $\Delta t$ using this peak as the reference value for the extrapolation.}
   	\label{fig:Lpeak_duration}
\end{figure*}

In Figure~\ref{fig:Lpeak_duration}, we show heatmaps of the peak luminosity $L_{\rm peak}$ (left) and flare duration $\Delta t$ (right) for the forward (top) and backward (bottom) outflows for all simulations. { The labels on the left side of each heatmap show the best-fitting power-law dependence and its uncertainty for each varied parameter, obtained through the least-squares method while holding all other parameters fixed at their fiducial values.}
\footnote{We define $\Delta t$ as the time interval over which the luminosity remains above a fixed threshold with $L>L_{\rm th} =2\times10^{40}\,{\rm erg\,s^{-1}}$. We chose $L_{\rm th}$ as a threshold that is low enough to be crossed by both the forward and backward lightcurves for all simulations. It is also of order $\sim 1\%$ of the bolometric peak luminosity for GSN~069 \citet{Miniutti2023}, making it suitable for the observational comparison discussed in Section \ref{sec:discussion}. If the lightcurve does not fall below $L_{\rm th}$ within the simulated time, we extrapolate the late-time post-peak decline by fitting $L(t)=L_{\rm peak}[(t-t_{\rm peak})/D+1]^{-\xi}$, where $t_{\rm peak}$ and $L_{\rm peak}$ are fixed by the selected peak, while $D$ and $\xi$ are fitted parameters (dotted lines in Figure~\ref{fig:L_plot}); this form is motivated by \citet{vurm_2025ApJ...983...40V} and \citet{Huang_2025ApJ...993..186H}. The parameter dependences are fitted as power laws in logarithmic space, except for the simulations with different $\sigma$, where we use a linear-space offset power law $C+A(R_\odot/\sigma)^\alpha$, because a single power law in $\sigma$ does not provide a good description. For the simulations with different $R_\star$, we performed additional runs with a larger lateral disc extent to more accurately determine $\Delta t$, since in the original simulations the shock front reached the lateral disc edges before the lightcurve entered a monotonic decline. We also tested different $L_{\rm th}$ values and found that the qualitative trends are unchanged. The exact fitted exponents, mainly for $v_\star$ and $R_\star$, vary somewhat with $L_{\rm th}$, and we therefore treat the fitted duration scalings as approximate.} All the lightcurves are shown in Figures \ref{fig:L_plot_forw_all} and \ref{fig:L_plot_back_all}.

Increasing $v_\star$ (orange lines in Figure \ref{fig:L_plot}) increases the amplitude of flares, reflecting the larger $\dot{E}$ (see Equation \ref{eq:dotE}), while the forward/backward luminosity asymmetry remains similar. Specifically, we find $L_{\rm peak} \propto v_\star^2$ for both outflows (see left panels of Figure \ref{fig:Lpeak_duration}). Additionally, the flare duration increases with $v_\star$, following approximately $\Delta t \propto v_\star^{1.4}$ and $\Delta t \propto v_\star^{2.2}$ for the forward and backward outflow, respectively (see right panels of Figure \ref{fig:Lpeak_duration}), as more energy is injected into the disc by a faster-moving star. However, $L$ decreases faster { in the lightcurves} with increasing $v_\star$ since adiabatic losses are larger due to a faster expanding outflow.

By contrast, increasing $\Sigma_{\rm d}$ (green lines in Figure \ref{fig:L_plot}) produces longer flares, reflecting the longer diffusion time through the denser ejecta. Specifically, we find $\Delta t \propto \Sigma_{\rm d}^{0.6}$ and $\Delta t \propto \Sigma_{\rm d}^{0.5}$ for the forward and backward outflow, respectively (see right panels of Figure \ref{fig:Lpeak_duration}).  As $\Sigma_{\rm d}$ increases beyond $\Sigma_{\rm d}/\Sigma_{\rm d}^{\rm f}\sim 3$, the forward peak luminosity decreases, while the backward peak continues to rise.\footnote{{\citet{Huang_2025ApJ...993..186H} found that the peak luminosity increases with $\Sigma_{\rm d}$. However, their trends are not directly comparable to ours, since the vertical density structure and opacity treatment are different. Additionally, they attributed the disc surface density dependence of the luminosity to opacity variations in the ejecta, and noted that the luminosity may become less sensitive to $\Sigma_{\rm d}$ at sufficiently high surface density.}} This behaviour indicates a competition between the larger $\dot{E}$ at higher $\Sigma_{\rm d}$ and stronger adiabatic losses in the more optically thick ejecta, such that the luminosity asymmetry can weaken and even reverse as $\Sigma_{\rm d}$ increases. Nevertheless, the dependence of $L_{\rm peak}$ is very weak with $L_{\rm peak}\propto \Sigma_{\rm d}^{-0.1}$ and $L_{\rm peak}\propto \Sigma_{\rm d}^{0.2}$ for the forward and backward outflow, respectively (see left panels of Figure \ref{fig:Lpeak_duration}).

{Increasing $R_\star$ (red lines in Figure \ref{fig:L_plot}) produces brighter flares in both directions, consistent with the larger $\dot{E}$ (see Figure \ref{fig:dotE}). Specifically, we find $L_{\rm peak}\propto R_\star^{1.5}$ and $L_{\rm peak}\propto R_\star^{0.7}$  for the forward and backward outflow, respectively (see left panels of Figure \ref{fig:Lpeak_duration}). Although a larger $R_\star$ ejects more mass, the peak luminosities do not show a turnover at increasing $R_\star$, contrary to the effect of increasing $\Sigma_{\rm d}$ explained above. This is because for larger $R_\star$, the outflows occupy a larger volume, which counteracts the effect of the larger ejected mass and reduces the adiabatic losses. In addition, the backward lightcurve develops a brief early spike, associated with a backward-directed shock breakout from the upper disc surface.}\footnote{The spike becomes more pronounced at larger $R_\star$ because the backward breakout occurs when the star is only partially embedded in the disc, while the gas near the shock cap is already strongly compressed and heated. The first escaping radiation therefore originates from the hotter layers closest to the tip of the star and is correspondingly brighter.} The flare duration increases with $R_\star$, following approximately $\Delta t\propto R_\star^{1.5}$ for both outflows (see right panels of Figure \ref{fig:Lpeak_duration}). This is because increasing $R_\star$ increases the amount of disc gas affected by the collision and therefore the diffusion time, resulting in a longer flare duration.

For more centrally concentrated discs (purple lines in Figure~\ref{fig:L_plot}), $L_{\rm peak}$ increases as $\sigma/R_\odot$ decreases at a similar rate for the two outflows (see left panels of Figure~\ref{fig:Lpeak_duration}). This is because the radiation energy density gradient becomes larger, since a larger fraction of the shock energy is deposited in the dense central layers, while the lower-density outer disc layers are only weakly heated. This enhances the radiative flux through the outer layers after breakout (see Equation~\ref{eq:Fdiff}). The radiation therefore escapes more promptly and suffers less adiabatic losses, leading to a higher peak luminosity. This also leads to slightly shorter flare durations for more centrally concentrated discs (right panels of Figure~\ref{fig:Lpeak_duration}).

{For collisions with smaller $i$ (brown lines in Figure~\ref{fig:L_plot}), the peak luminosity of the forward outflow remains comparable to the fiducial case. This is because the forward peak is dominated by radiation emerging from the forward shock-breakout ejecta. The heating rate and the subsequent adiabatic losses of this component do not change strongly with $i$, since the post-breakout expansion rate of the photosphere remains similar across different $i$ \citep[as also found by][]{Huang_2026ApJ..1005..235H}. For the backward outflow, $L_{\rm peak}$ increases with decreasing $i$, following $L_{\rm peak}\propto(\sin i)^{-0.8}$ (see left panels of Figure~\ref{fig:Lpeak_duration}). This reflects the longer interaction time in more oblique crossings, which allows a larger fraction of the shocked gas to flow around the star and be redirected toward the backward outflow before escaping. As a result, more radiation energy is carried by the backward ejecta, while the more gradual expansion reduces the adiabatic losses of this radiation before it escapes. The flares also become longer in physical units for smaller $i$, because the encounter becomes increasingly gradual and the shock heating, as well as the supply of shocked gas to the outflows, is distributed over a longer time interval. This trend is similar for the two outflows, with $\Delta t \propto (\sin i)^{-0.9}$ and $\Delta t \propto (\sin i)^{-1}$ for the forward and backward outflow, respectively (see right panels of Figure~\ref{fig:Lpeak_duration}). In addition, $L_{\rm back}$ shows a brief early spike associated with a backward-directed shock breakout from the upper disc surface, similar to the effect of increasing $R_\star$ described above. The forward lightcurve develops a bump following $L_{\rm peak}$, which develops into a prominent secondary peak for $i\lesssim 60^\circ$, as the amount of radiation emerging from gas that remains trapped longer in the cavity increases.}

These trends are broadly consistent with, and complementary to, recent studies of QPE-motivated star--disc collisions. \citet{Huang_2025ApJ...993..186H} and \citet{vurm_2025ApJ...983...40V} found that the observable emission depends sensitively on the collision conditions, with brighter flares produced by faster or larger stars. Additionally, \citet{Liu_2026ApJ..1006L..50L} and \citet{Huang_2026ApJ..1005..235H} found that the collision geometry can strongly affect the relative strength of the forward and backward ejecta, with more in-plane and co-rotating stellar orbits { with respect to the disc} producing more similar flares. However, a direct quantitative comparison is difficult, because these studies adopt different dimensionalities, radiation and opacity treatments, and dynamical setups.

\section{Discussion} \label{sec:discussion}

In the following, we connect our simulation parameter study to observed QPEs. Specifically, we focus on GSN~069, the best-characterised source with a clear `strong--weak' alternation and well-constrained flare properties \citep{Miniutti2019,Miniutti2023,Miniutti_2023A&A...674L...1M}. Our goal is not to perform a detailed parameter inference for this system, but rather to test whether the star--disc collision model can reproduce the observed peak luminosities, recurrence time, flare duration, and alternating flare amplitudes within the systematic uncertainties of the comparison. We therefore construct a simple model from the one-parameter trends measured in the heatmaps of peak luminosities and flare durations obtained from simulation (see Figure \ref{fig:Lpeak_duration}) and use it to identify the parameter space compatible with GSN~069.

{For GSN~069, we adopt a bolometric peak luminosity of $L_{\rm GSN}=3\times10^{42}\,{\rm erg\,s^{-1}}$, a characteristic flare duration of $\Delta t_{\rm GSN} =1\,{\rm h}$, a mean QPE period of $P_{\rm QPE}^{\rm GSN}=9\,{\rm h}$, a peak luminosity ratio between consecutive flares of $R_{\rm GSN}=1.5$, and SMBH mass $M_{\rm bh}=10^6\,M_\odot$.} {Assuming a circular stellar orbit and two star--disc crossings per orbit, the observed QPE period fixes the Keplerian velocity at the disc crossing radius, giving $v_{\rm K}/v_{\star}^{\rm f}=(P_{\rm QPE}^{\rm f}/P_{\rm QPE}^{\rm GSN})^{1/3}\simeq0.76$, where $P_{\rm QPE}^{\rm f}=4\,{\rm h}$ is the fiducial QPE period (see Section~\ref{subsec:fiducial_sim}). In the local simulations, however, the relevant velocity is the relative velocity between the star and the rotating disc gas. We therefore identify $i$ with the local collision angle, defined as the angle that this relative velocity makes with the disc midplane. For the simple case of a circular stellar orbit crossing a circular Keplerian disc with equal stellar and disc speeds, this local collision angle is related to the global orbital inclination $I$ by $I=180^\circ-2i$, and the corresponding relative velocity is $v_{\rm rel}/v_{\star}^{\rm f}=2\cos i\,(P_{\rm QPE}^{\rm f}/P_{\rm QPE}^{\rm GSN})^{1/3}$. Under this mapping, $i=45^\circ$ ($I=90^\circ$) corresponds to a globally polar orbit, while $i>45^\circ$ ($I<90^\circ$) and $i<45^\circ$ ($I>90^\circ$) correspond to prograde and retrograde global orbits, respectively. This mapping applies for $i<90^\circ$, since the $i=90^\circ$ limit would correspond to a co-moving circular orbit. The $i=90^\circ$ simulation is therefore treated as a controlled limiting case in the local disc frame, against which more oblique impacts can be compared. }

We then use the one-parameter trends obtained from the simulations to sample the parameter space more finely. For each quantity $Q\in\{L_{\rm peak}^{\rm forw},L_{\rm peak}^{\rm back},\Delta t_{\rm forw},\Delta t_{\rm back}\}$, we use the fitted parameter dependences shown in Figure~\ref{fig:Lpeak_duration} to calculate $Q(v_\star,\Sigma_{\rm d},R_\star,\sigma,i)=Q_{\rm f} f_v(v_\star) f_\Sigma(\Sigma_{\rm d}) f_R(R_\star) f_\sigma(\sigma) f_i(i)$, where $Q_{\rm f}$ is the fiducial value and each factor $f$ describes how $Q$ changes when one parameter is varied relative to its fiducial value. We then sample $10^6$ random combinations of $(\Sigma_{\rm d},R_\star,\sigma,i)$, with $v_\star$ computed from the sampled value of $i$ using the expression above. For each sampled parameter combination, we compute $L_{\rm peak}^{\rm forw}$, $L_{\rm peak}^{\rm back}$, $\Delta t_{\rm forw}$, and $\Delta t_{\rm back}$ and compare them with $L_{\rm GSN}$, $\Delta t_{\rm GSN}$, and $R_{\rm GSN}$. To rank the sampled parameter combination, we compute a $\chi^2$ score based on the difference between computed values and adopted GSN~069 values.

The candidate solutions separate into two broad regions. These regions in the parameter space are also visible in Figure \ref{fig:corner}. {Both regions favour retrograde orbits since $\sin i\lesssim \sqrt{2}/2$ ($I\gtrsim 90^\circ$}).\footnote{{The preference for retrograde global orbits was also found by \citet{Tagawa_2023MNRAS.526...69T}, based on an analytic model for the prompt emission from bow-shock breakout. The comparison should, however, be interpreted cautiously, since their model does not include radiation from shock-heated gas that flows around the star, remains trapped longer in the disc, and escapes later through the photosphere. Their solutions also favour more massive stars than our candidate models, although such stars may undergo stronger ablation in repeated retrograde collisions, potentially limiting their lifetime \citep{Sniegowska_2026arXiv260513799S}.}} The first region has $\Sigma_{\rm d}/\Sigma_{\rm d}^{\rm f}\gtrsim1$, with the best fitting examples often having $\Sigma_{\rm d}/\Sigma_{\rm d}^{\rm f}\gtrsim10$, $\sigma/R_\odot\lesssim0.5$, $R_\star/R_{\rm f}\sim0.8$, and $\sin i\sim0.4$. These models reproduce the adopted GSN~069 flare properties with good agreement, but require an accretion disc with a vertical density profile highly concentrated toward the disc midplane. One representative solution in this region has $\Sigma_{\rm d}/\Sigma_{\rm d}^{\rm f}=45.8$, $R_\star/R_{\rm f}=0.8$, $\sigma/R_\odot=0.3$, and $\sin i=0.3$. This solution gives $L_{\rm peak}^{\rm forw}=3.1\times10^{42}\,{\rm erg\,s^{-1}}$, $L_{\rm peak}^{\rm forw}/L_{\rm peak}^{\rm back}=1.5$, $\Delta t_{\rm forw}=1.0\,{\rm h}$, $\Delta t_{\rm back}=0.7\,{\rm h}$, and duty cycles of $0.1$ for both flares. Such $\Sigma_{\rm d}/\Sigma_{\rm d}^{\rm f}$ are high for a standard thin disc, but can be reached in dense post-TDE discs (e.g. \citealt{mummery_collisions_2025}). This region also requires a vertically concentrated density profile, with $\sigma/R_\odot\sim0.4$. At the collision radius inferred for GSN~069, $r\simeq365R_\odot$, this corresponds to an aspect ratio $\sigma/r\sim10^{-3}$, where $r$ is the distance from the SMBH. Such small aspect ratios could occur in TDE discs at late times, when the accretion rate is lower and the disc becomes thinner (e.g. \citealt{Zanazzi_2019MNRAS.487.4965Z}). { A TDE origin of the disc is also supported by \citet{Guolo_2025ApJ...992..114G} and \citet{Guolo_2025ApJ...985..146G}  who modelled the evolution of the TDE disc in GSN~069. They found that the disc was already sufficiently extended to encompass the inferred collision radius before the appearance of the QPEs. Additionally, they found a surface density of $\Sigma_{\rm d}\sim 10^3-10^4\,{\rm g\,cm^{-2}}$ at this radius, broadly consistent with our result.}

The second region has $\Sigma_{\rm d}/\Sigma_{\rm d}^{\rm f}\sim0.1$, $\sigma/R_\odot\sim1$, $R_\star/R_{\rm f}\sim1$, and $\sin i\sim0.1$. These models are more directly compatible with a standard thin disc vertical structure at that radius. However, they reproduce the adopted GSN~069 flare properties less accurately, and they require a very grazing collision geometry. One representative sampled solution in this region has $\Sigma_{\rm d}/\Sigma_{\rm d}^{\rm f}=0.1$, $R_\star/R_{\rm f}=1.0$, $\sigma/R_\odot=0.9$, and $\sin i=0.1$. This solution gives $L_{\rm peak}^{\rm forw}=1.6\times10^{42}\,{\rm erg\,s^{-1}}$, $L_{\rm peak}^{\rm forw}/L_{\rm peak}^{\rm back}=2.0$, $\Delta t_{\rm forw}=1.0\,{\rm h}$, $\Delta t_{\rm back}=2.0\,{\rm h}$, and duty cycles of $0.1$ and $0.2$ for the forward and backward flares, respectively.

{We also repeated this comparison for the later phase of GSN~069, detected after the reappearance of QPEs in the XMM12 observation \citep{Miniutti_2023A&A...674L...1M}. For this epoch, we adopt $L_{\rm GSN}=5\times10^{42}\,{\rm erg\,s^{-1}}$, $\Delta t_{\rm GSN}=1\,{\rm h}$, $P_{\rm QPE}^{\rm GSN}=7.4\,{\rm h}$, and $R_{\rm GSN}=2$. We find that the second region, with lower disc surface density, becomes disfavoured, while the first region, with a dense and vertically concentrated disc, remains viable. This strengthens the preference for the post-TDE disc interpretation, consistent with observational evidence for TDE-like activity in GSN~069 \citep{Shu2018,Sheng_2021ApJ...920L..25S}.}

Although the comparison favours a dense post-TDE disc, this inference relies on several simplifying assumptions. Due to this, we interpret these solutions as physically motivated regions of parameter space that can reproduce the observed GSN~069 flare properties, rather than as unique best-fitting models. First, our model assumes separable one parameter scalings, and does not capture possible coupling between $\Sigma_{\rm d}$, $R_\star$, $\sigma$, and $i$. Secondly, the luminosities are bolometric and integrated over hemispheres around the forward and backward outflows, whereas a real observer views the system from a particular line of sight and in a specific X-ray band. Viewing angle effects, band-limited emission, and spectral evolution can all affect flare amplitudes and their duration. Additionally, they could make the observed flares more symmetric than the bolometric light curves shown here, with a slower rise to peak, which is more similar to the flare profiles seen in GSN~069 (e.g. \citealt{Metzger2022}). We also note, that simple effective temperature estimates from our photospheric radii are typically lower than the blackbody temperatures inferred from observed QPE flares, $kT\sim100$ to $200,{\rm eV}$, likely because the grey LTE treatment does not capture effects such as photon starvation and Comptonization, which can harden the emerging spectrum (e.g. \citealt{Linial2023,vurm_2025ApJ...983...40V}). The local simulations also neglect disc rotation and shearing, as well as changes in the collision geometry between successive passages. These effects are likely to be especially important for the solutions with low $\sin i$, when the star travels a longer path through the disc. These limitations are discussed in more detail in \citet{Jankovic_2026A&A...710A.405J}. Additionally, the outcome of the collision could be affected by warping of the disc, which could modulate both recurrence time and burst energy \citep{Chen_2026arXiv260524905C}.

\section{Conclusions} \label{sec:conclusion}

{Repeated collisions between a star and an accretion disc are a promising model for QPEs. In this study, we performed a systematic suite of 3D local radiation-hydrodynamics simulations of star--disc collisions, focusing on the regime where the star remains effectively unperturbed by the collision. We varied the main system parameters relative to the fiducial model: stellar velocity, disc surface density, stellar radius, vertical disc density profile, and {the local collision angle}. The stellar radius is varied at fixed disc scale height, allowing us to explore the impact of the relative size of the star and the disc. Our goal was to determine how these physical parameters are imprinted on the outflow hydrodynamics and the resulting bolometric flare properties. Our main conclusions are as follows.}

{(i) Across the parameter space explored, the collision dynamics remains similar. The star drives a bow shock through the disc, deflecting a fraction of it around the star, and two outflows emerge on opposite sides of the disc. The forward outflow carries more momentum and is more energetic than the backward outflow, but the strength of this asymmetry depends on the system parameters.}

{(ii) Varying the stellar velocity mainly changes the thermodynamic normalization of the shocked gas rather than the overall outflow morphology, with the bow-shock structure and the asymmetry between the momenta of the forward and backward outflows remaining similar to the fiducial case. Collisions with faster stars produce brighter flares in both outflows, mainly reflecting the larger shock heating rate. Faster collisions also produce longer flares, because the larger injected energy keeps the luminosity longer above the adopted threshold, although their luminosity decreases more rapidly after peak due to stronger adiabatic losses in the faster-expanding outflow.}

{(iii) Varying the disc surface density also leaves the overall bow shock and outflow morphology largely unchanged. Denser discs increase the amount of shocked gas and therefore the mass and momentum carried by the outflows, while the momentum asymmetry between the forward and backward outflows remains similar to the fiducial case. The resulting flares become longer in denser discs, since the ejecta become more optically thick and radiation remains trapped longer. The peak luminosity depends only weakly on the disc surface density, because the larger shock heating rate in denser discs competes with stronger adiabatic losses in the more optically thick outflow.}

{(iv) Increasing the stellar radius causes the star to interact with a larger amount of disc gas. As a result, both outflows carry more mass and momentum, with the momentum of the forward outflow increasing more strongly than that of the backward outflow. Larger stars also produce brighter flares in both directions, mainly because the shock heating rate is larger. The forward flare becomes increasingly dominated by shock breakout radiation, while the backward lightcurve can develop a brief early spike from shock breakout along the backward direction. The flare durations also increase with stellar radius, since a larger amount of shocked gas increases the diffusion time.}

{(v) Varying the vertical disc density profile mainly affects the thermodynamic structure of the shocked gas and the efficiency of radiative escape, with the momentum asymmetry between the forward and backward outflows largely unaffected. In collisions with more centrally concentrated discs, the flares are brighter in both outflows and have slightly shorter durations. This is because a larger fraction of the shock energy is deposited in the dense central layers, while the lower density outer layers are only weakly heated. The resulting larger radiation energy density gradient enhances the radiative flux through the outer layers after shock breakout, allowing radiation to escape more promptly and with weaker adiabatic losses.}

{(vi) Decreasing the {local collision angle} breaks the approximate axial symmetry of the shock front and the resulting outflows. In more oblique collisions, the star travels a longer path through the disc, allowing a larger fraction of the shocked gas to flow around the star and be redirected backward. This reduces the momentum asymmetry between the two outflows and increases the luminosity of the backward outflow, while the forward outflow peak luminosity remains close to the fiducial value because it is still dominated by shock breakout radiation. The flares become longer, since the shock heating and the supply of shocked gas to the outflows are distributed over a longer time interval.}

{(vii) We provide empirical scalings for the peak luminosity and flare duration as functions of the individual system parameters. These scalings allow the simulation results to be translated into approximate predictions for different local star--disc collision conditions and provide an important step toward connecting local radiation-hydrodynamic simulations to observed QPE systems.}

{(viii) We apply the empirical scalings for the peak luminosity and flare duration to GSN~069, and identify regions of parameter space where the star--disc collision model can reproduce the observed peak luminosity, flare duration, duty cycle, and strong and weak luminosity pattern within factors of order unity. {The candidate solutions favour a star with radius $\sim R_\odot$ on a retrograde orbit, colliding with a dense post-TDE disc with a vertically concentrated density profile.}}

{This work extends the star--disc collision model by systematically exploring how the main local system parameters shape the outflows and lightcurves. In agreement with previous studies, we find that the peak luminosity, flare duration, and strong and weak flare patterns are sensitive to different combinations of physical parameters, making QPE observations a potential probe of the collision geometry and disc structure. We also find that specific regions of parameter space can reproduce the main observed flare properties of GSN~069. Future work should extend the simulations to more realistic stellar structures, include disc rotation and shear, and compute viewing angle-dependent luminosities. These improvements will be needed to determine whether repeated star--disc collisions can quantitatively reproduce the diversity of observed QPE sources.}

\section*{Data availability}

The scripts used to construct a local section of an accretion disc and to simulate star-disc collisions for different system parameters will be made publicly available as a part of \textsc{Phantom} at \href{https://github.com/danieljprice/phantom}{https://github.com/danieljprice/phantom}. The software used to analyze simulation snapshots is available at \href{https://github.com/tajjankovic/Radiation-hydrodynamics-of-star-disc-collisions/}{https://github.com/tajjankovic/Radiation-hydrodynamics-of-star-disc-collisions/}. \footnote{Movies made from the simulations are available online at \href{https://www.youtube.com/playlist?list=PLH8qhWjKWQ92nPx_tPaPYnobRCPUjlfdF}{https://www.youtube.com/playlist?list=PLH8qhWjKWQ92nPx\_tPaPY\\nobRCPUjlfdF}.}
\begin{acknowledgements}
Researcher T. J. conducts his research under the Marie Skłodowska-Curie Actions – COFUND project, which is co-funded by the European Union (Physics for Future – Grant Agreement No. 101081515).  
This work was co-funded by the European Union and supported by the Czech Ministry of Education, Youth and Sports (MEYS) (Project No. CZ.02.01.01/00/22\_008/0004632 -- FORTE).
T. J. acknowledges the use of HPC cluster Phoebe of the Central European Institute of Cosmology (CEICO) at the Institute of Physics of the Czech Academy of Sciences where the computations were performed. MZ is grateful for the support of the GA\v{C}R Junior Star grant no. GM24-10599M ``Stars in galactic nuclei: interrelation with massive black holes''.
MS acknowledges the Czech Science Foundation (GA\v{C}R) grant no. 26-23342I.
We thank C. Bonnerot for the useful discussions.

The following software was used in this work: 
     \textit{Matplotlib} \citep{Hunter:2007},
       \textit{NumPy} \citep{harris2020array}, 
    \textit{SciPy} \citep{2020SciPy-NMeth}.
    
\end{acknowledgements}

\bibliography{bibliography}

\onecolumn
\begin{appendix}

\section{All simulations}\label{app:all_sims}

\begin{figure*}[h] %  figure placement: here, top, bottom, or page
   \centering
   \includegraphics[width=0.8\textwidth]{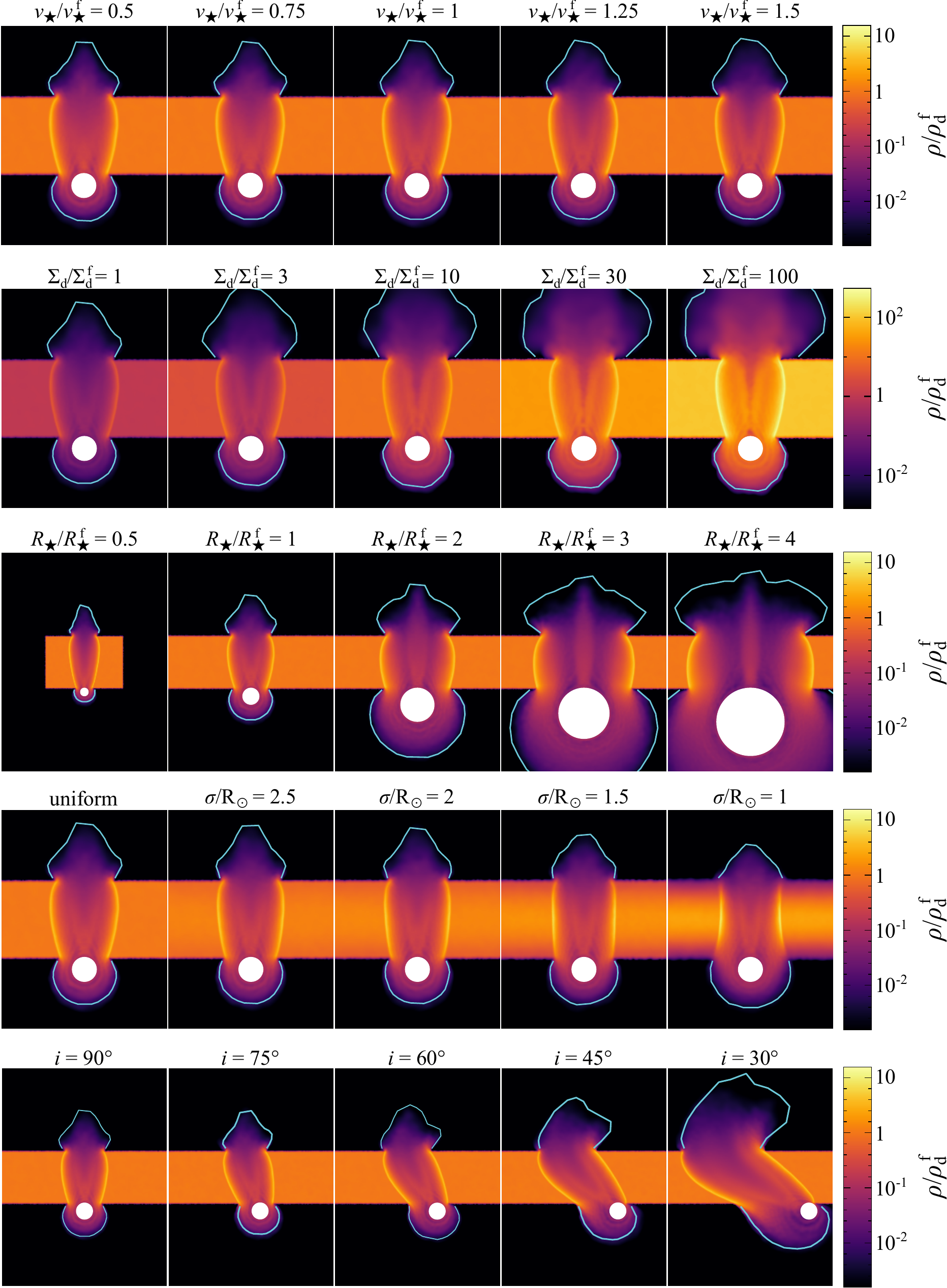}
   \caption{Gas density slices in the $yz$-plane at $x=0$ at $t/t_{\rm cr}=1$ for all the simulations varying $v_\star$ (first row), $\Sigma_{\rm d}$ (second row), $R_\star$  (third row), $\sigma/R_\odot$  (fourth row), and $i$  (fifth row). The white circle denotes the star.}
   	\label{fig:qpe_sim_density_all}
\end{figure*}

\begin{figure*}[h] %  figure placement: here, top, bottom, or page
   \centering
   \includegraphics[width=0.9\textwidth]{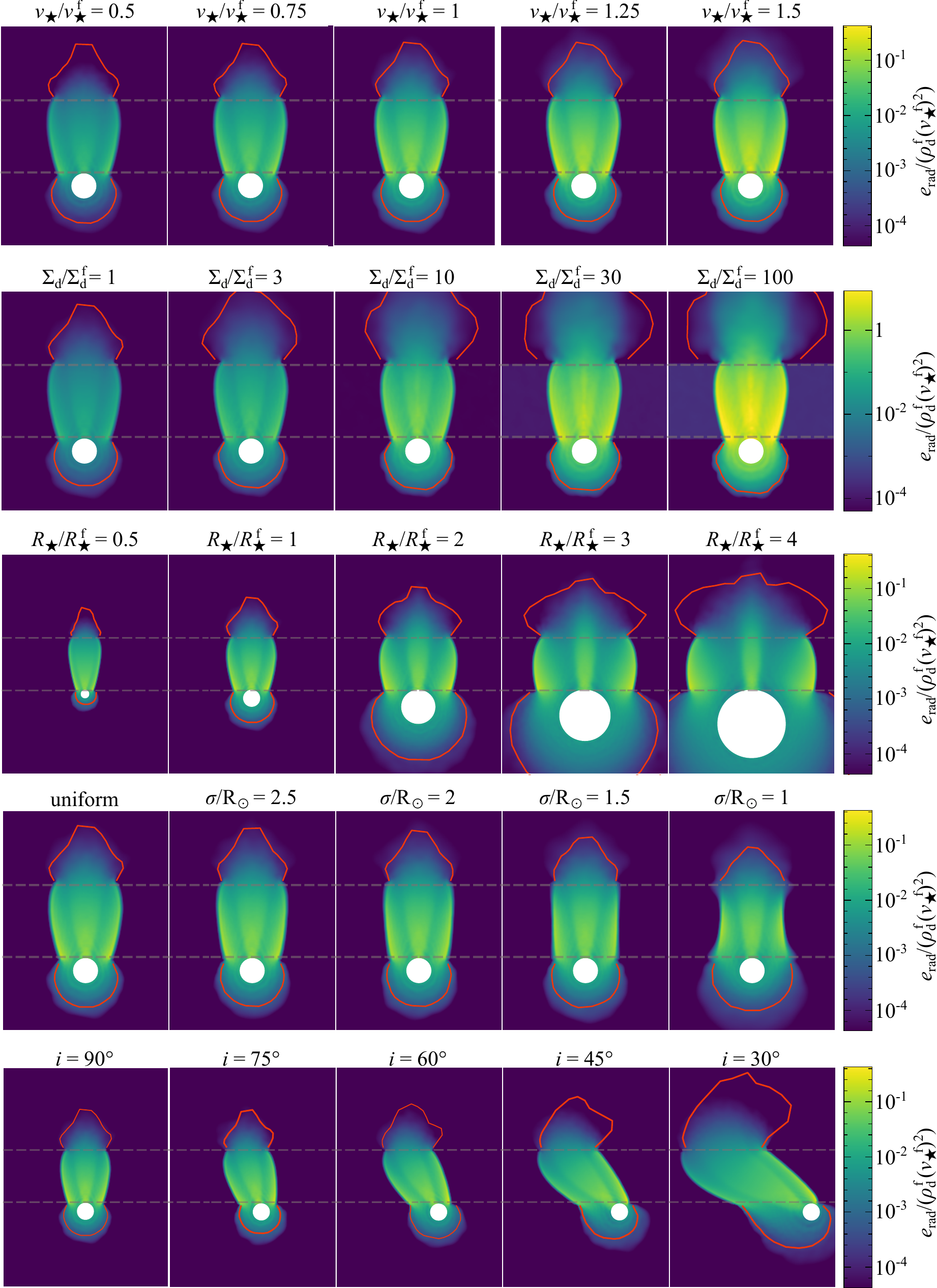}
  \caption{Radiation energy density slices in the $yz$-plane at $x=0$ at $t/t_{\rm cr}=1$ for all the simulations varying $v_\star$ (first row), $\Sigma_{\rm d}$ (second row), $R_\star$  (third row), $\sigma/R_\odot$  (fourth row), and $i$  (fifth row). The white circle denotes the star.}
  
   	\label{fig:qpe_sim_erad_all}
\end{figure*}

\begin{figure*}[h]
	\centering  
        \includegraphics[width=0.33\textwidth]{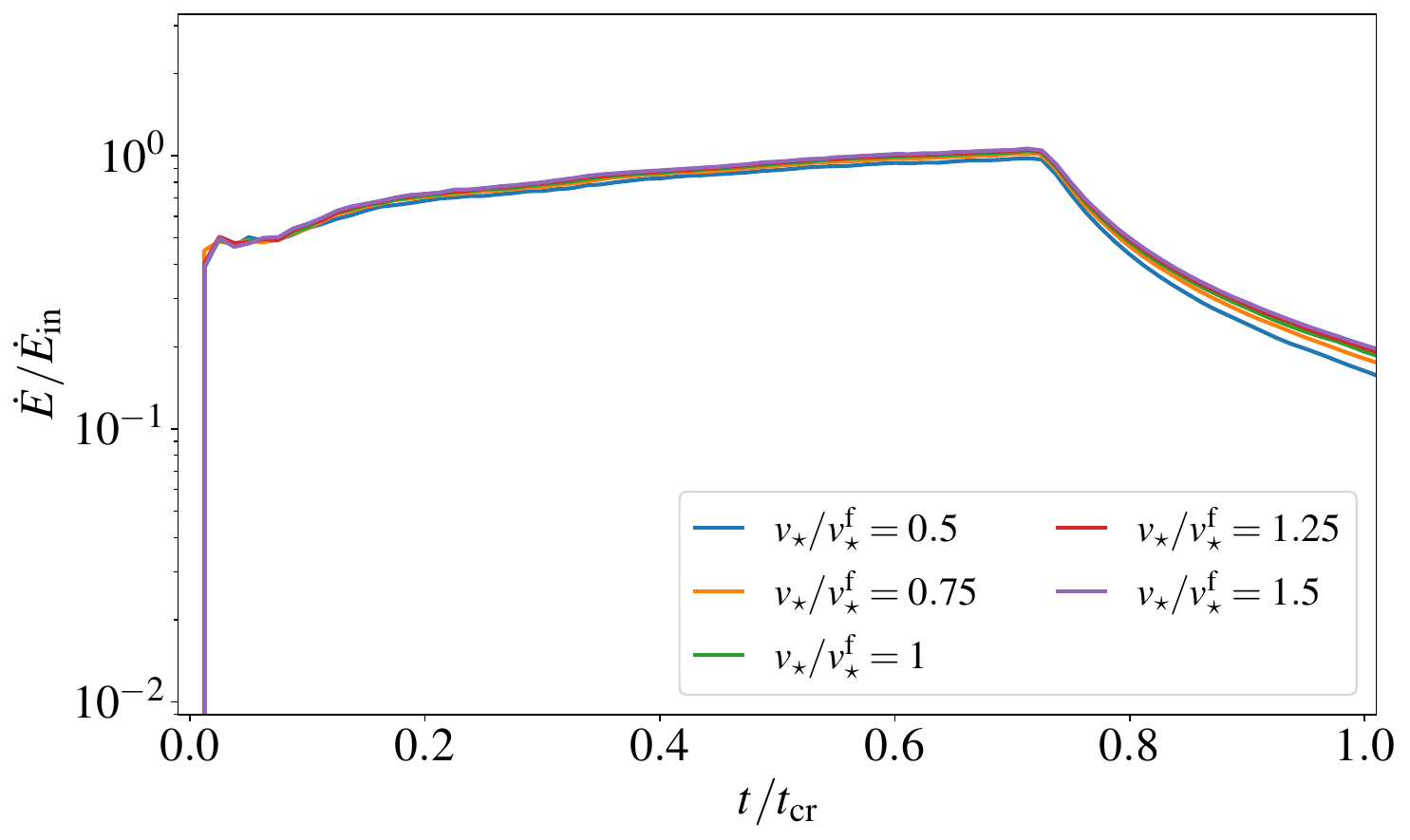}
        \includegraphics[width=0.33\textwidth]{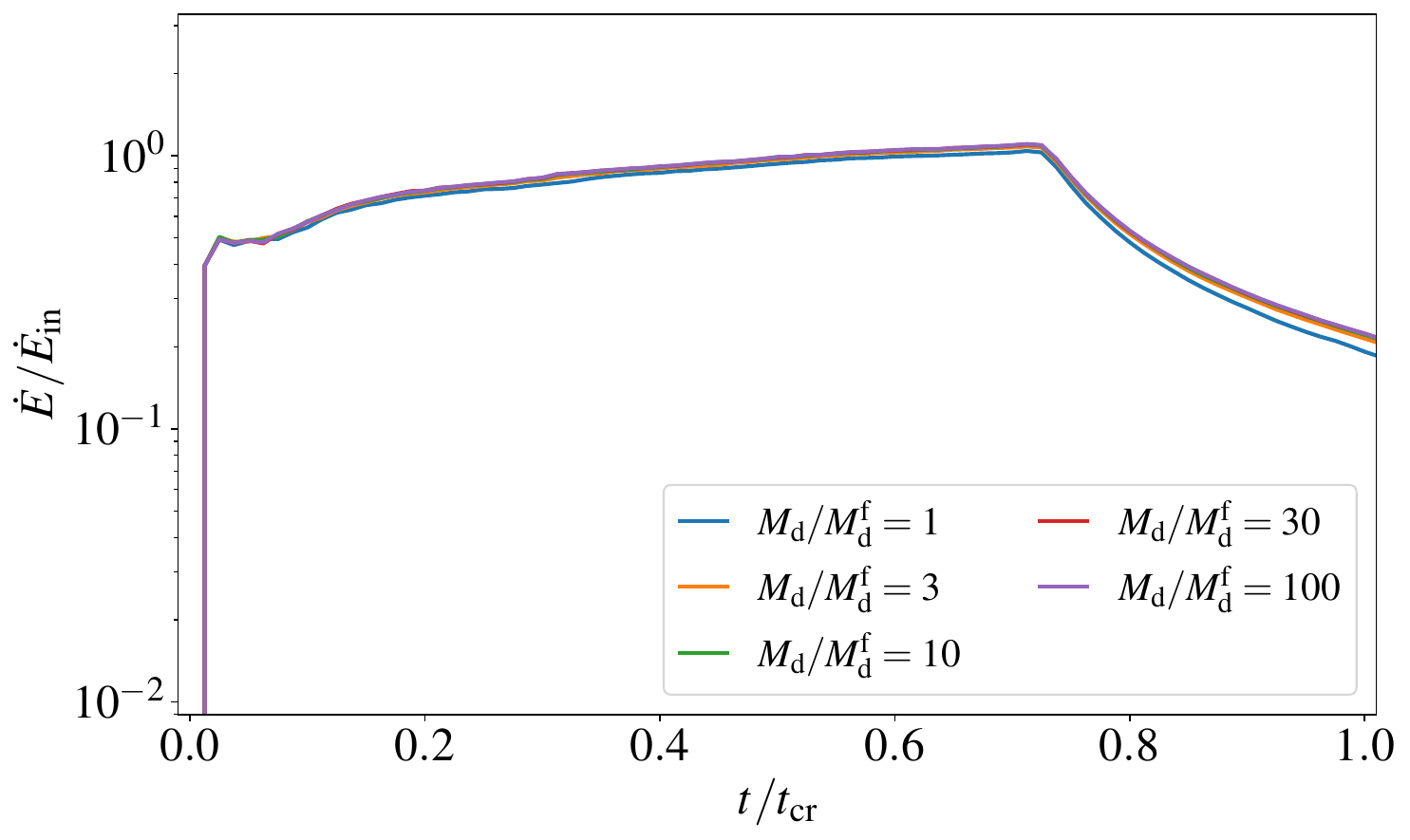}
        \includegraphics[width=0.33\textwidth]{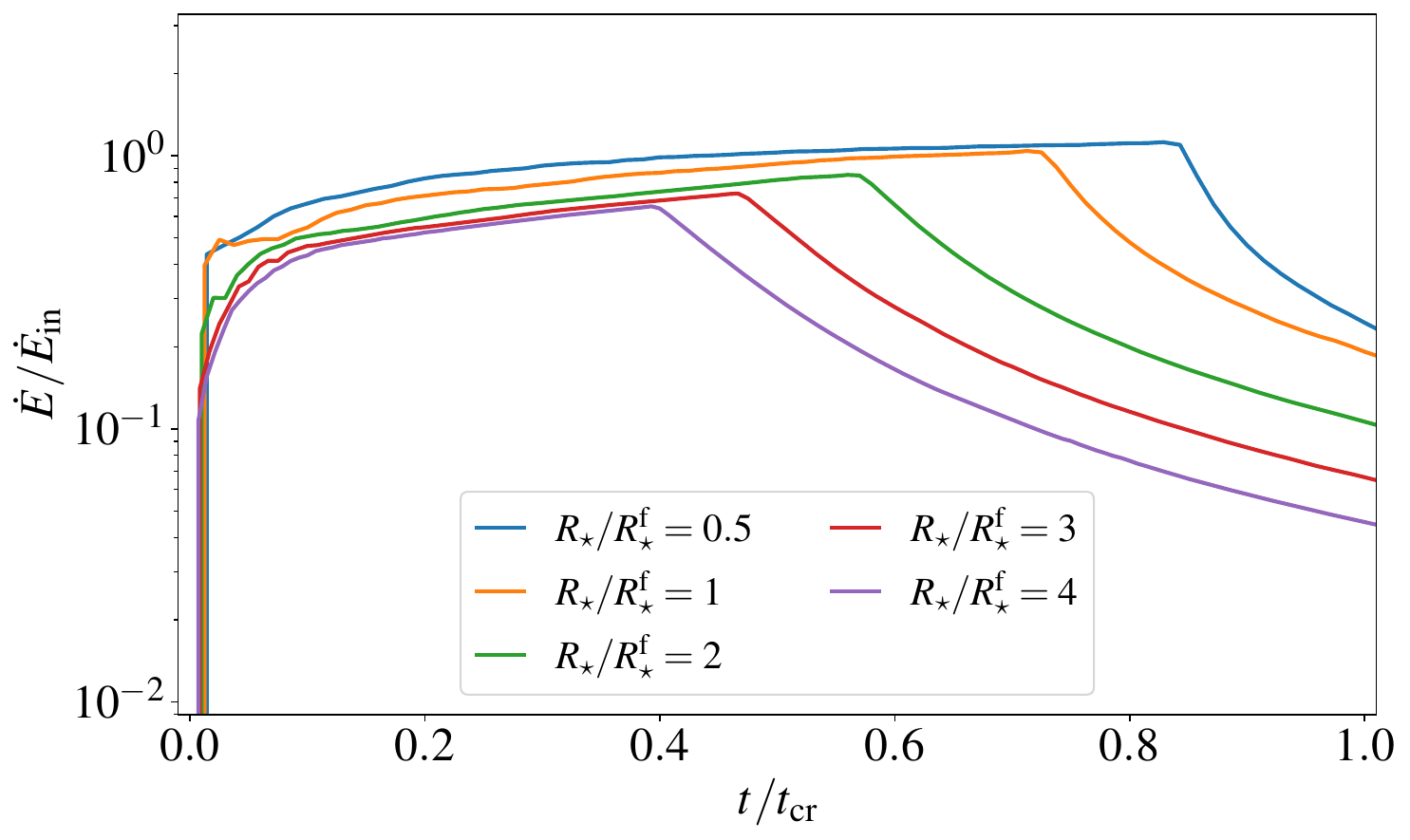}
        \includegraphics[width=0.33\textwidth]{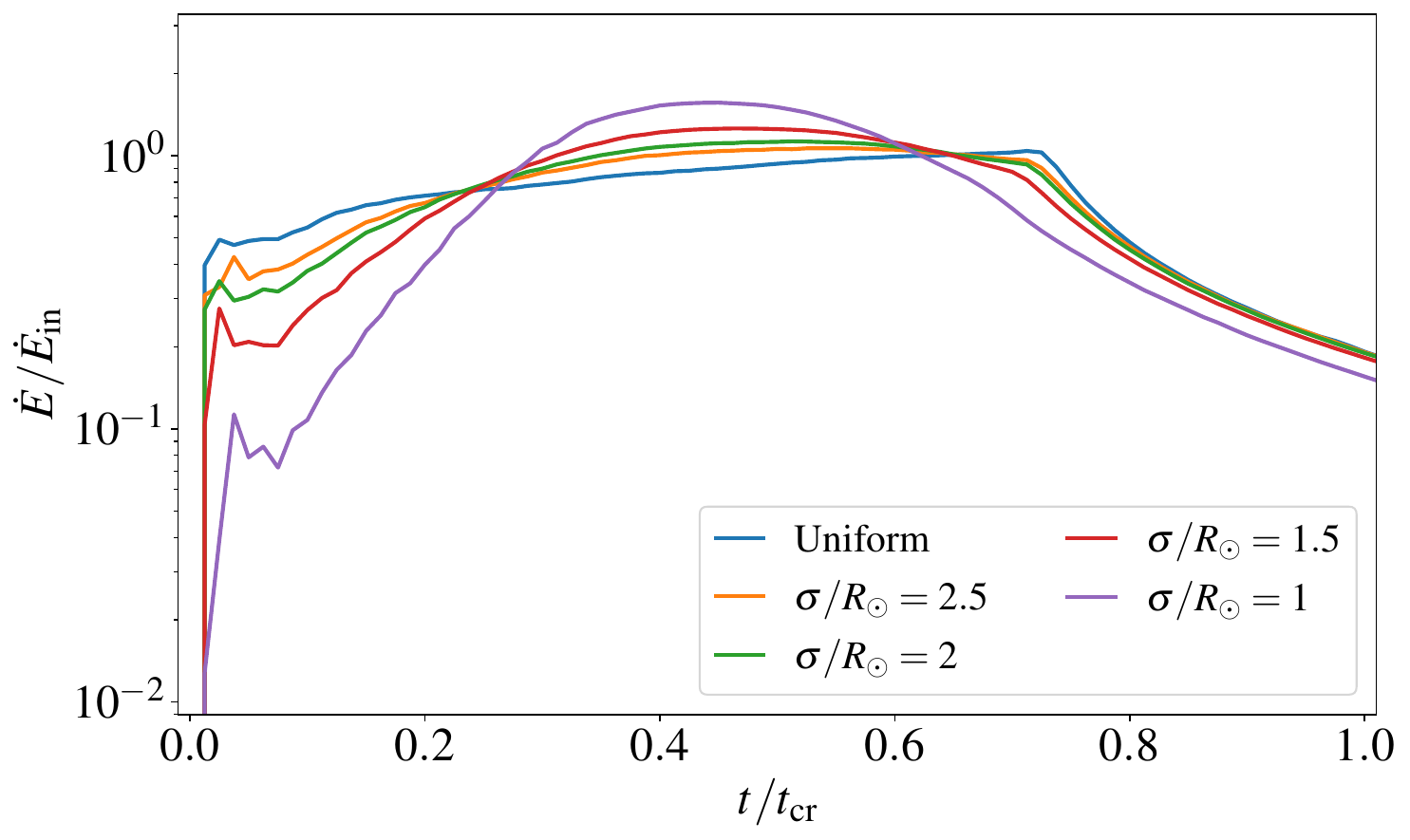}
        \includegraphics[width=0.33\textwidth]{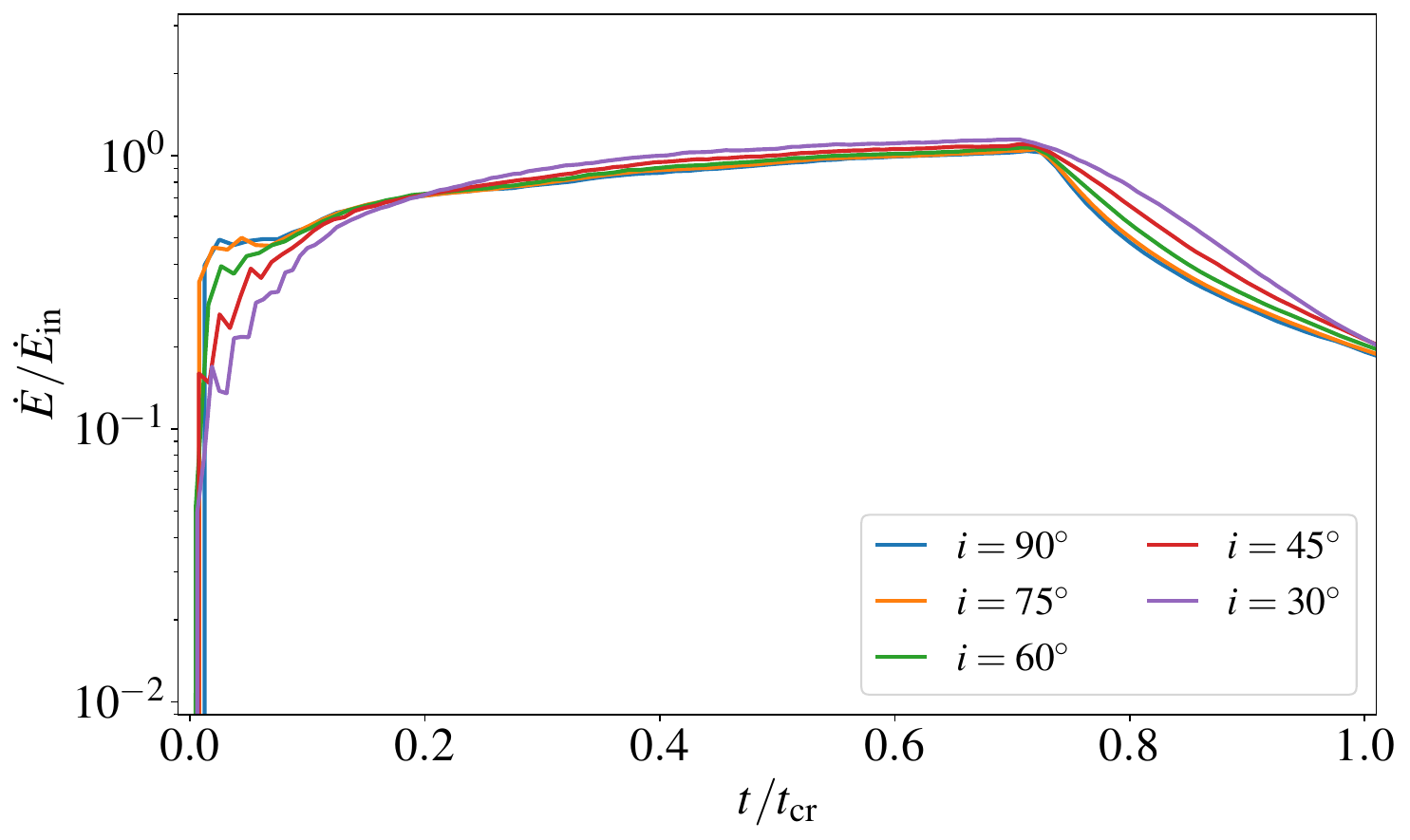}
\caption{Shock heating rate $\dot{E}$ as a function of time for simulations varying $v_\star$, $\Sigma_{\rm d}$, $R_\star$, $\sigma$, and $i$, shown respectively in the first to fifth panels, counting from left to right and then from top to bottom.}
	\label{fig:dotE_all}
\end{figure*}

\begin{figure*}[h]
\centering
       \includegraphics[width=0.33\textwidth]{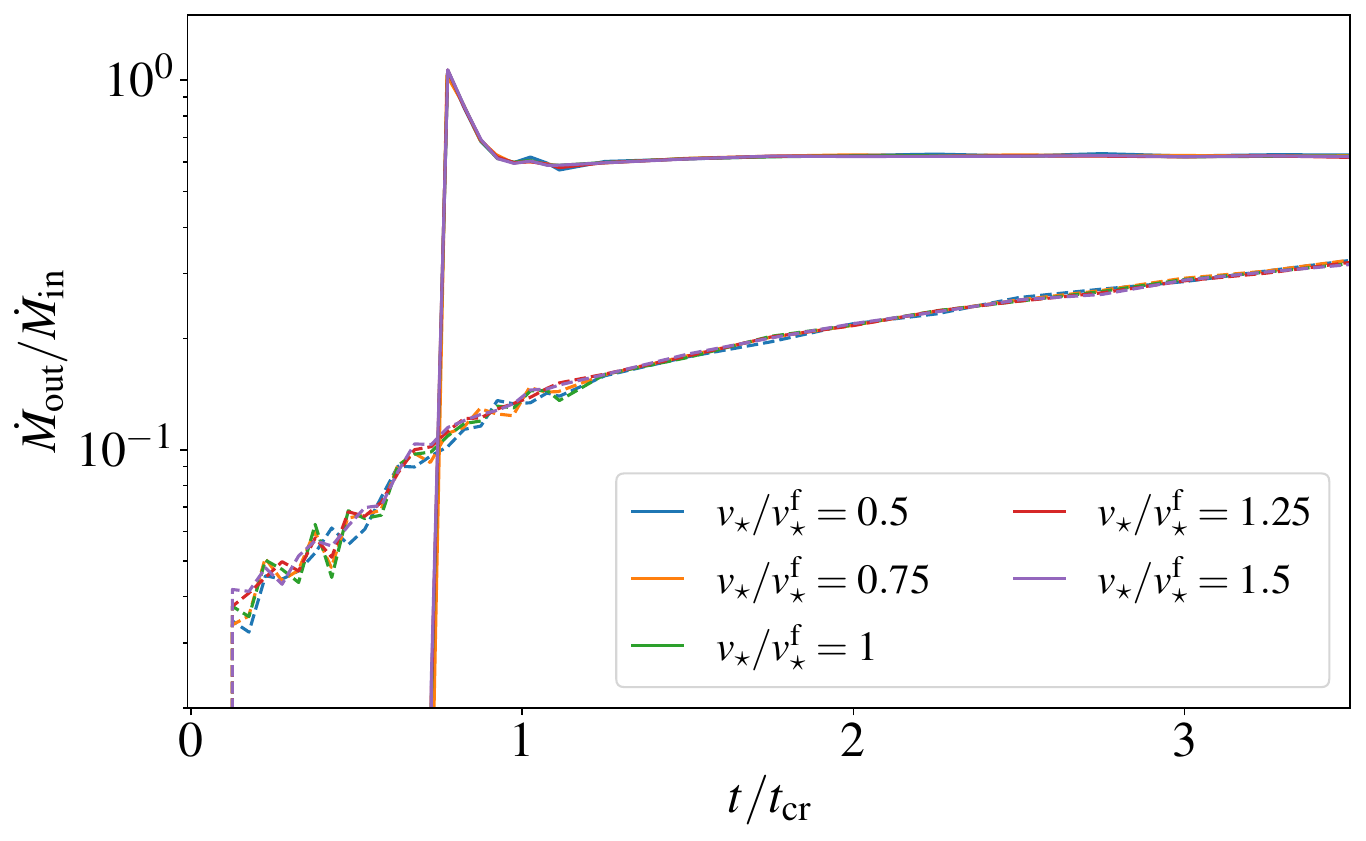}
        \includegraphics[width=0.33\textwidth]{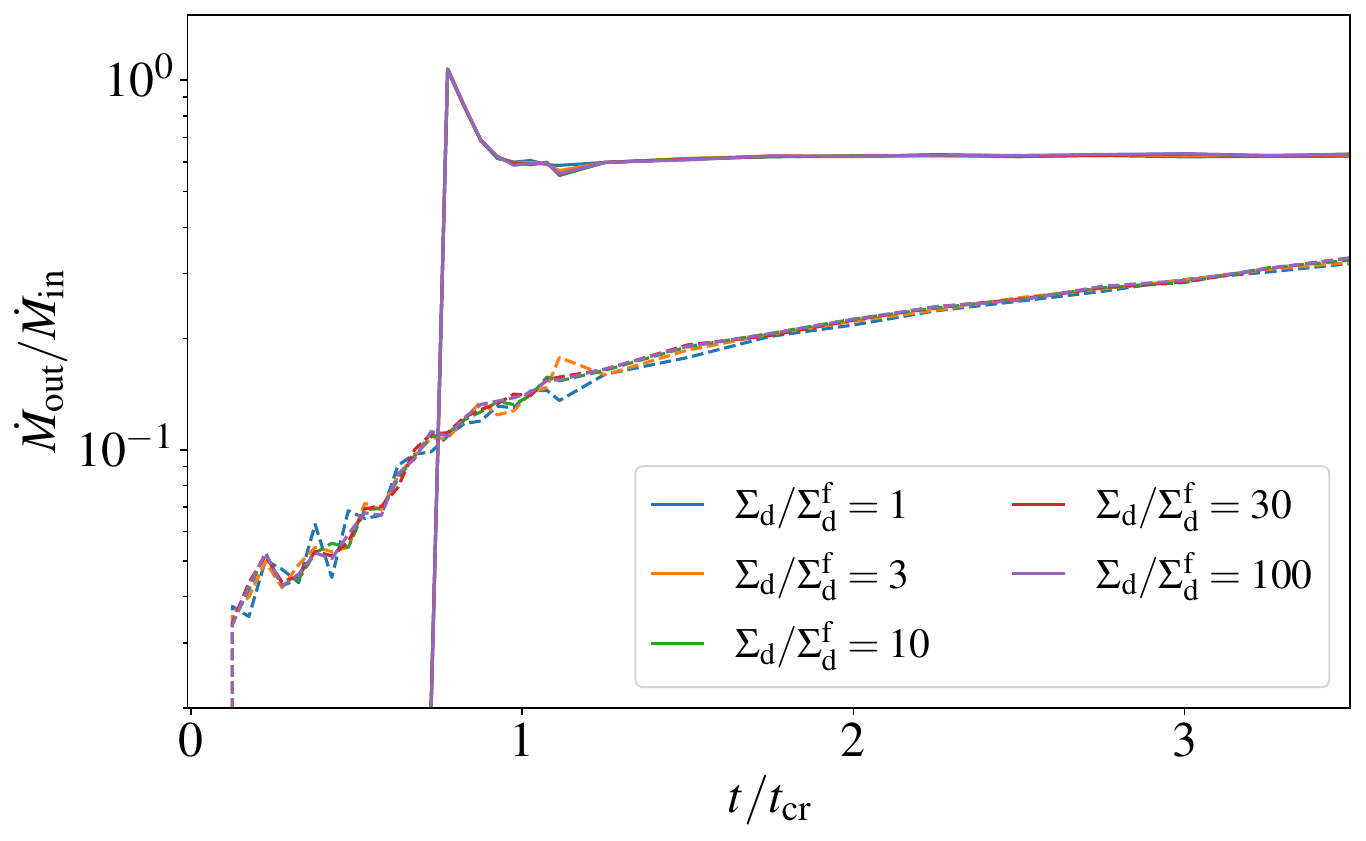}
        \includegraphics[width=0.33\textwidth]{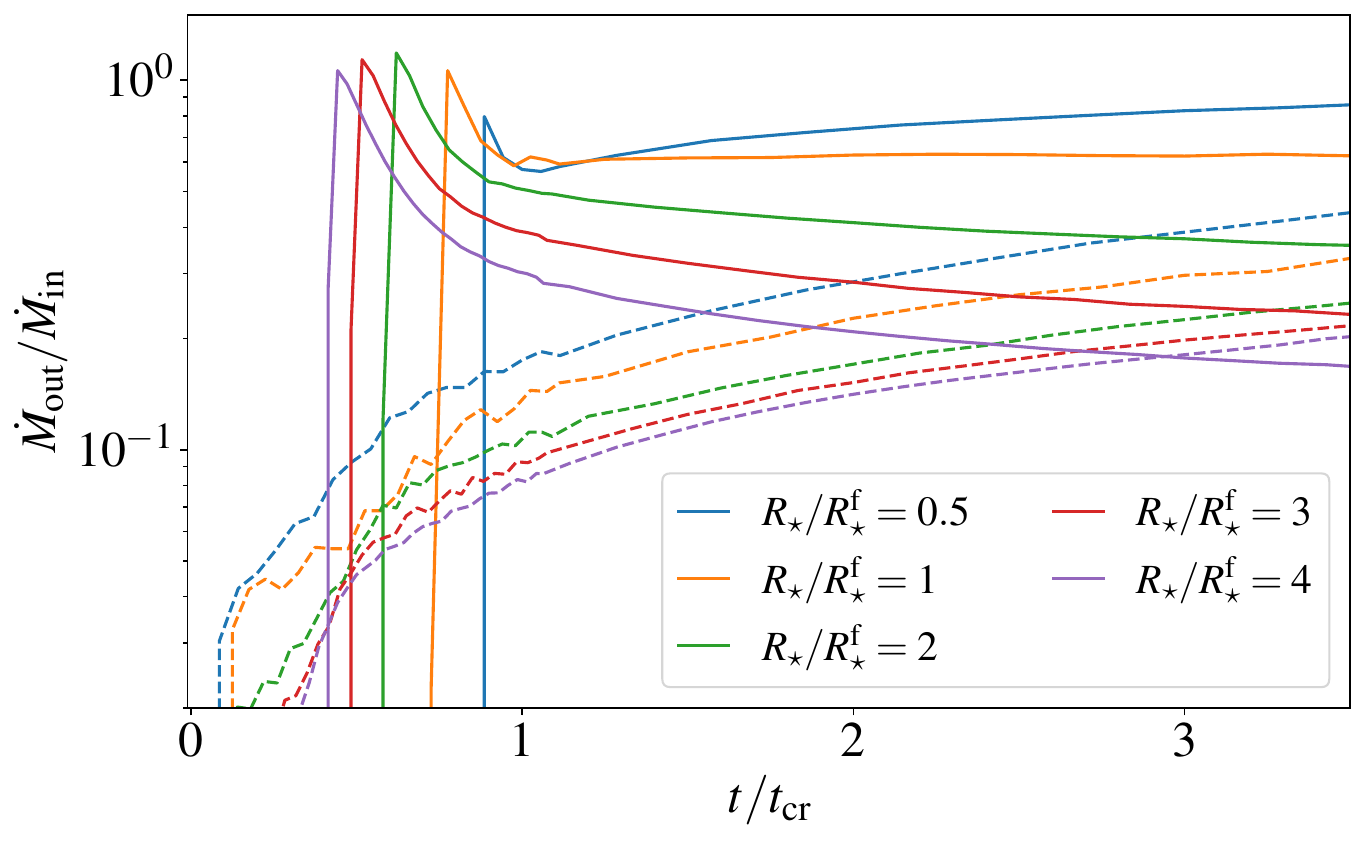}
        \includegraphics[width=0.33\textwidth]{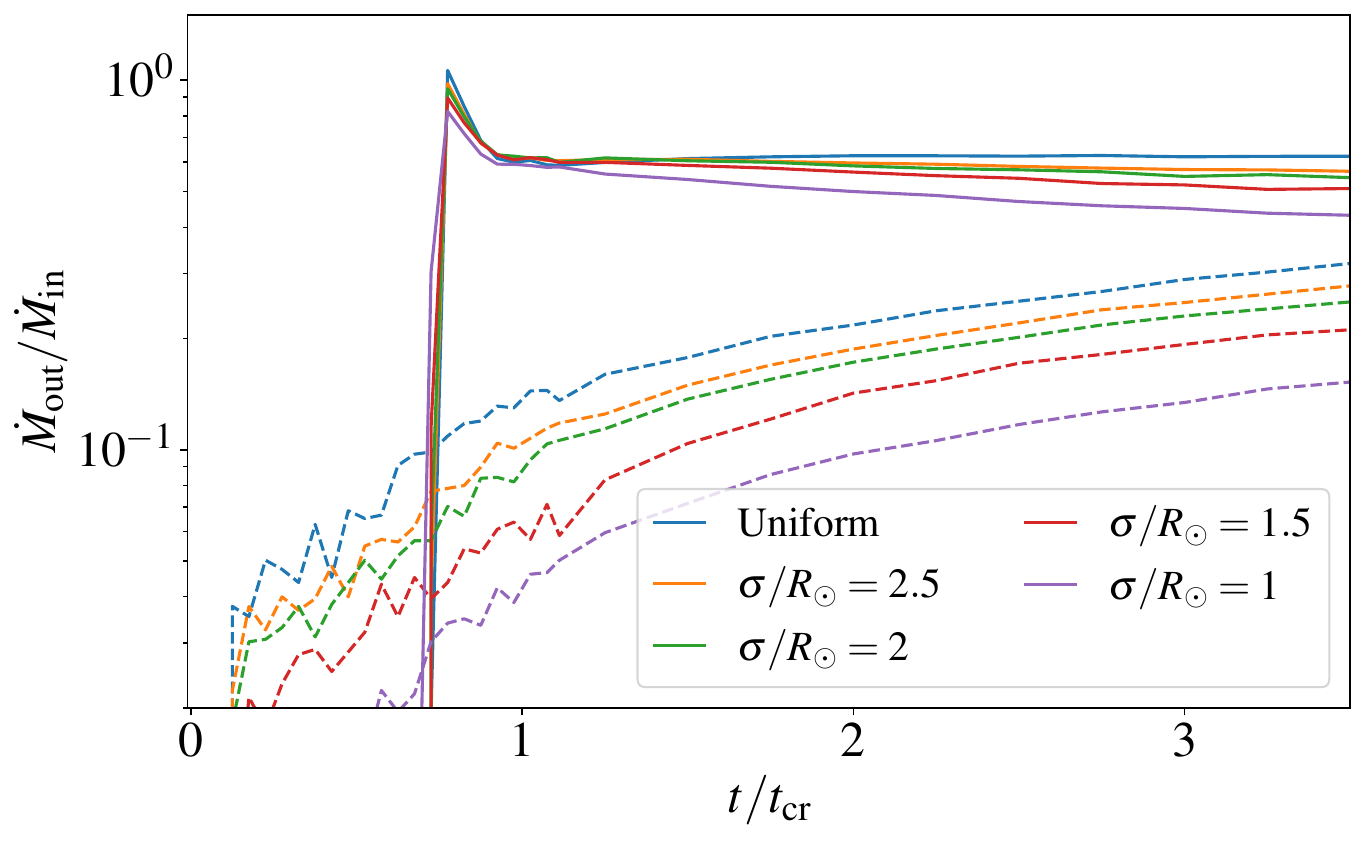}
        \includegraphics[width=0.33\textwidth]{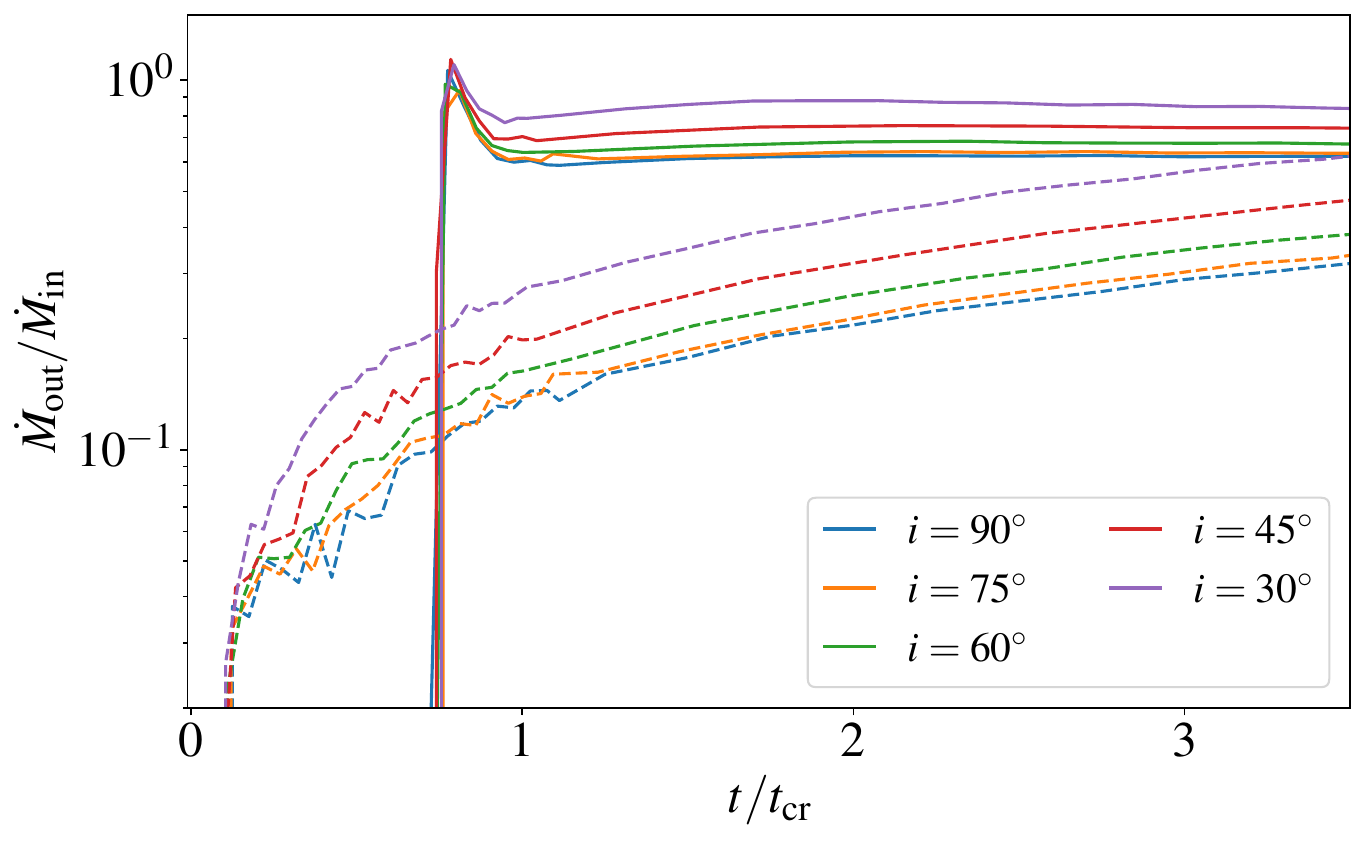}
    \caption{The mass outflow rate $\dot{M}_{\rm out}$ ejected along the forward (solid lines) and backward (dashed lines) direction for simulations varying $v_\star$, $\Sigma_{\rm d}$, $R_\star$, $\sigma$, and $i$, shown respectively in the first to fifth panels, counting from left to right and then from top to bottom.}
	\label{fig:dotM_Mej_all}
\end{figure*}

\begin{figure*}[h]
   \centering
   \includegraphics[width=0.33\textwidth]{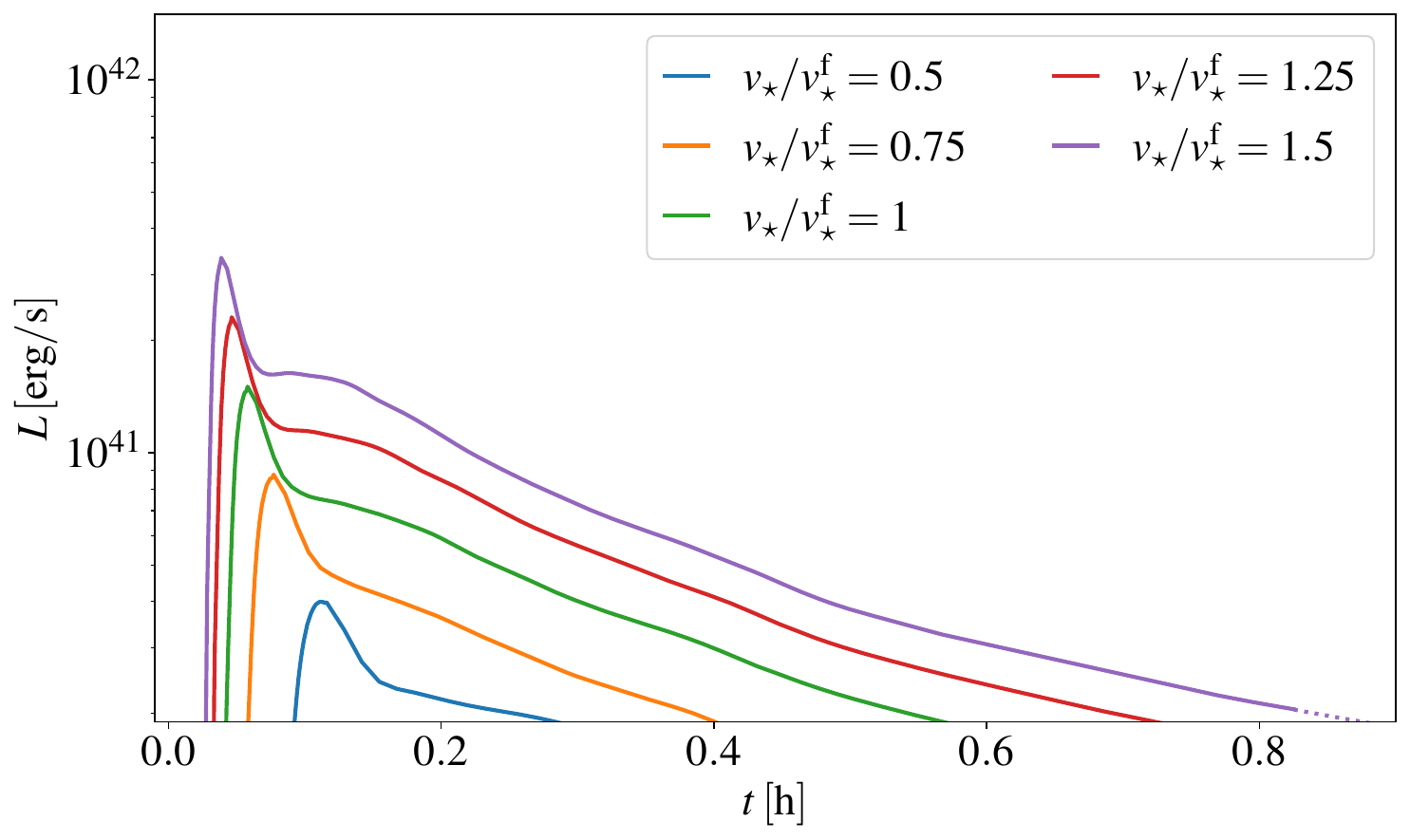}
   \includegraphics[width=0.33\textwidth]{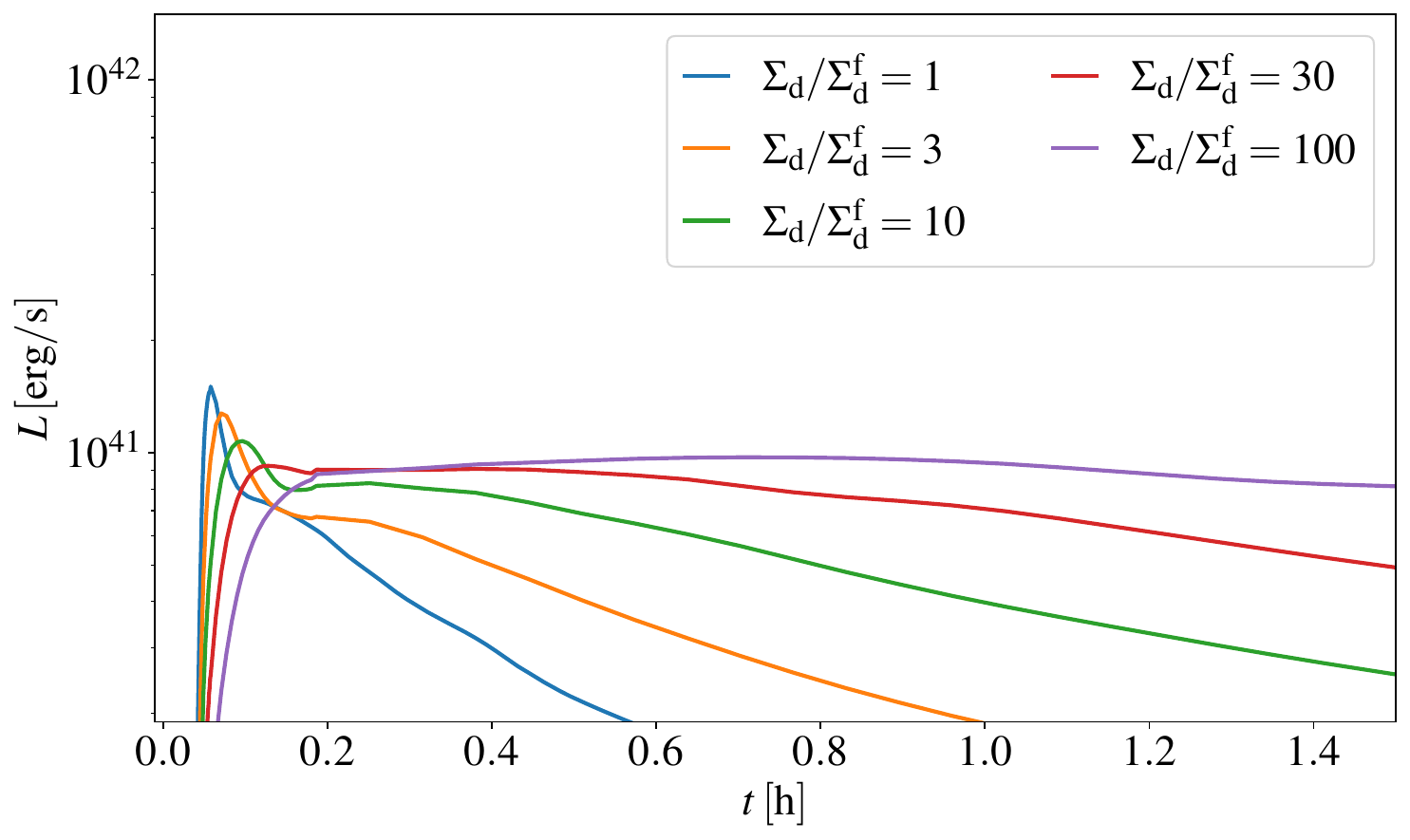}
   \includegraphics[width=0.33\textwidth]{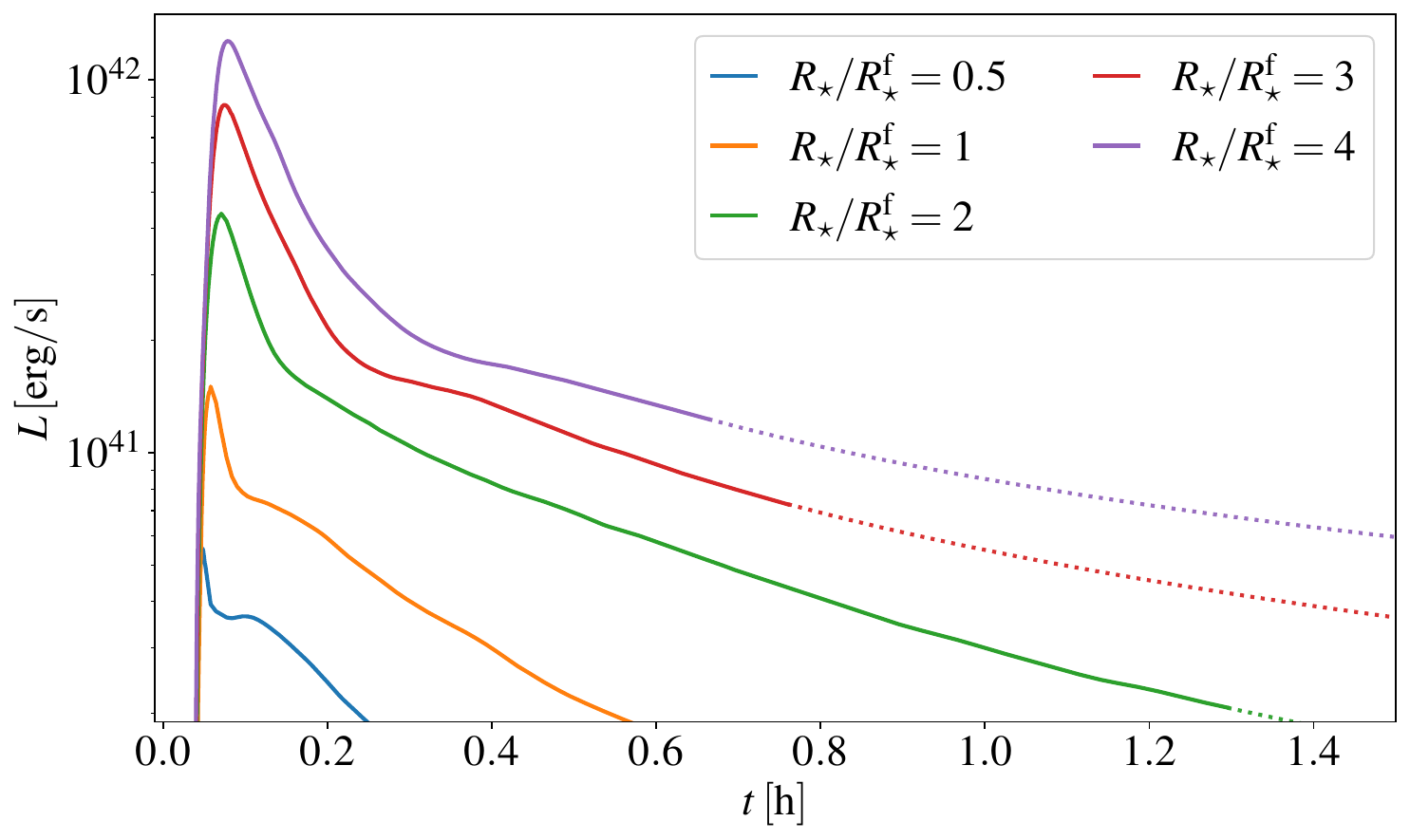}
   \includegraphics[width=0.33\textwidth]{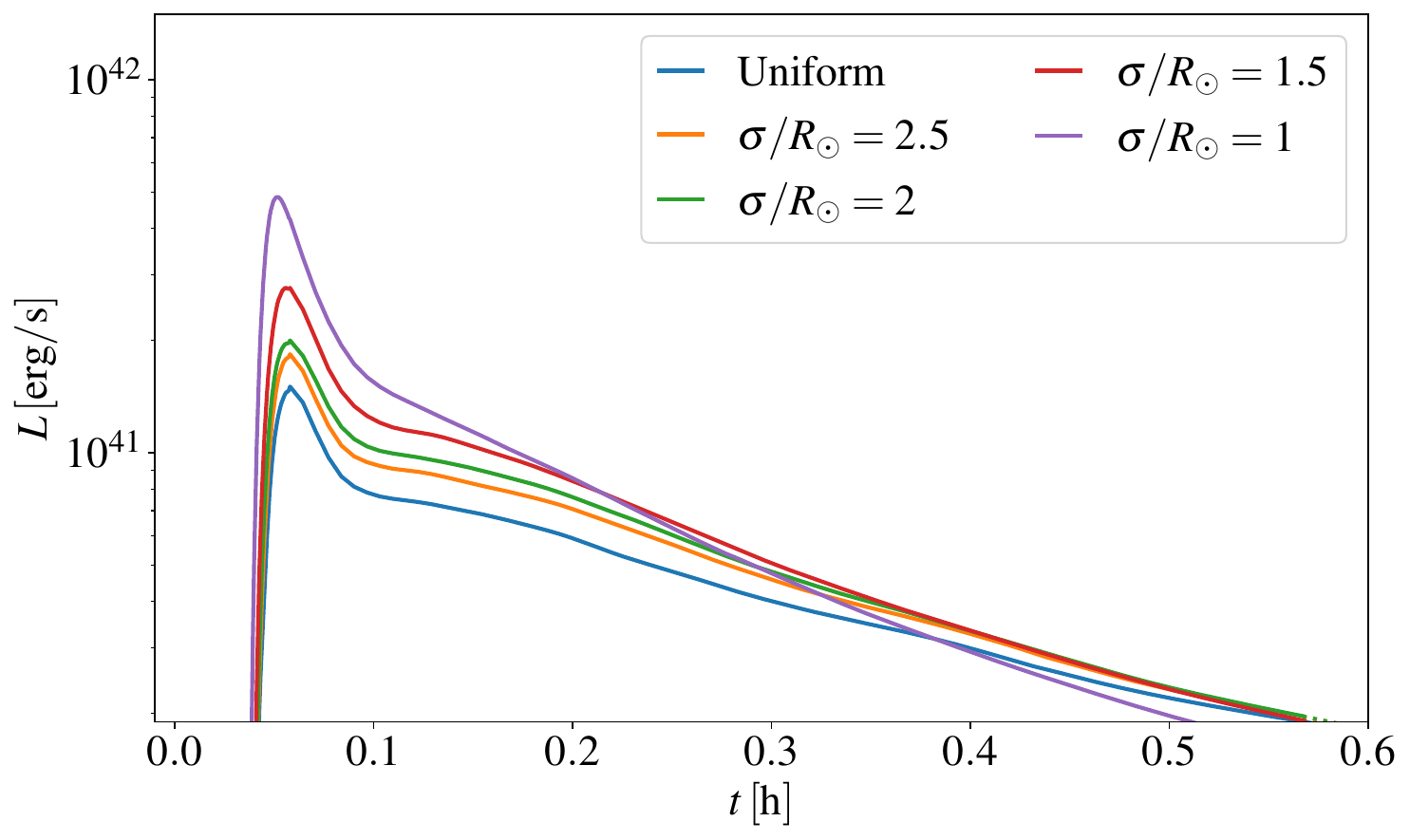}
   \includegraphics[width=0.33\textwidth]{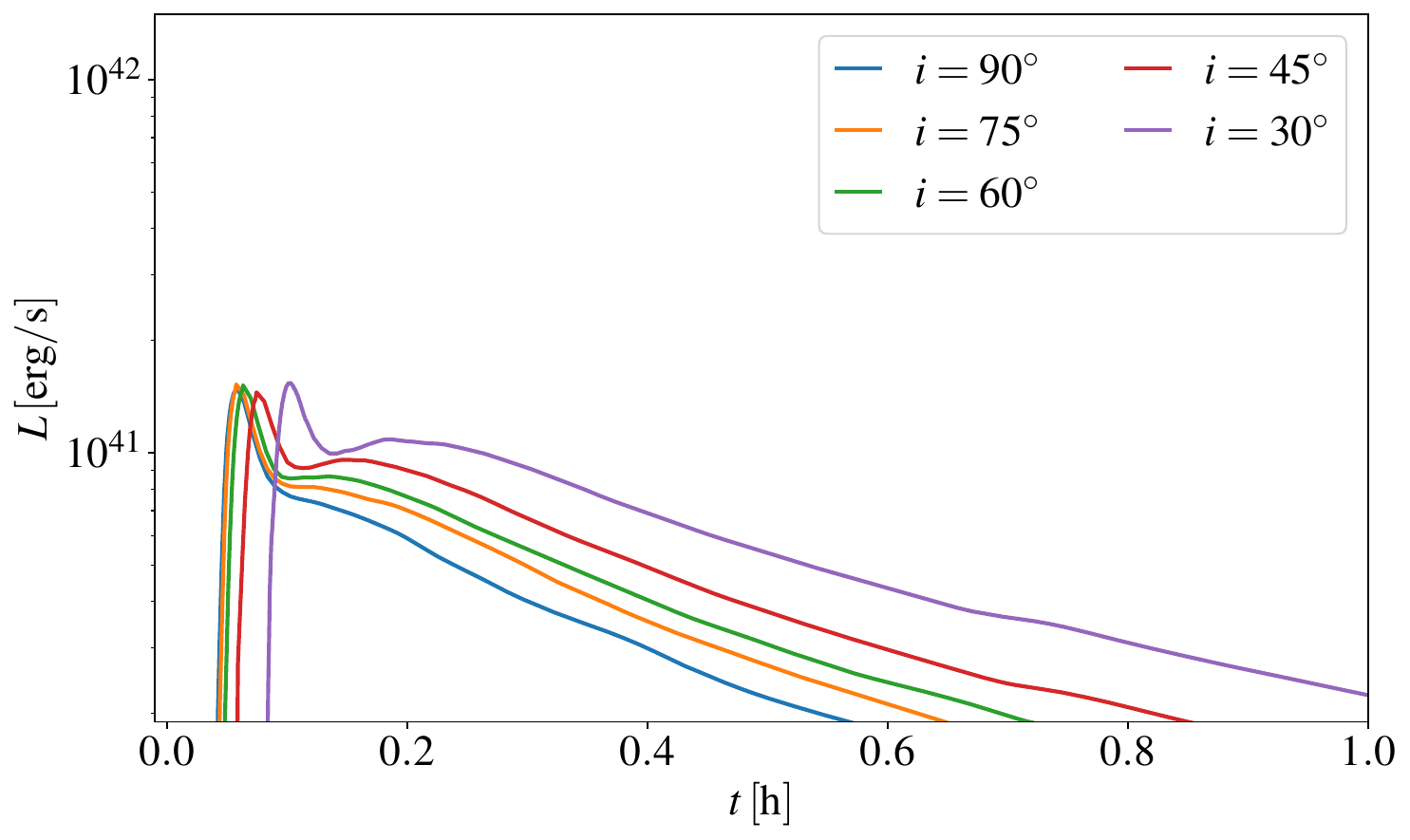}
   \caption{The time evolution of $L$ of the forward outflow for simulations varying $v_\star$, $\Sigma_{\rm d}$, $R_\star$, $\sigma$, and $i$, shown respectively in the first to fifth panels, counting from left to right and then from top to bottom. The dotted lines denote the extrapolated late-time decay.}
\label{fig:L_plot_forw_all}
\end{figure*}

\begin{figure*}[h]
   \centering
   \includegraphics[width=0.33\textwidth]{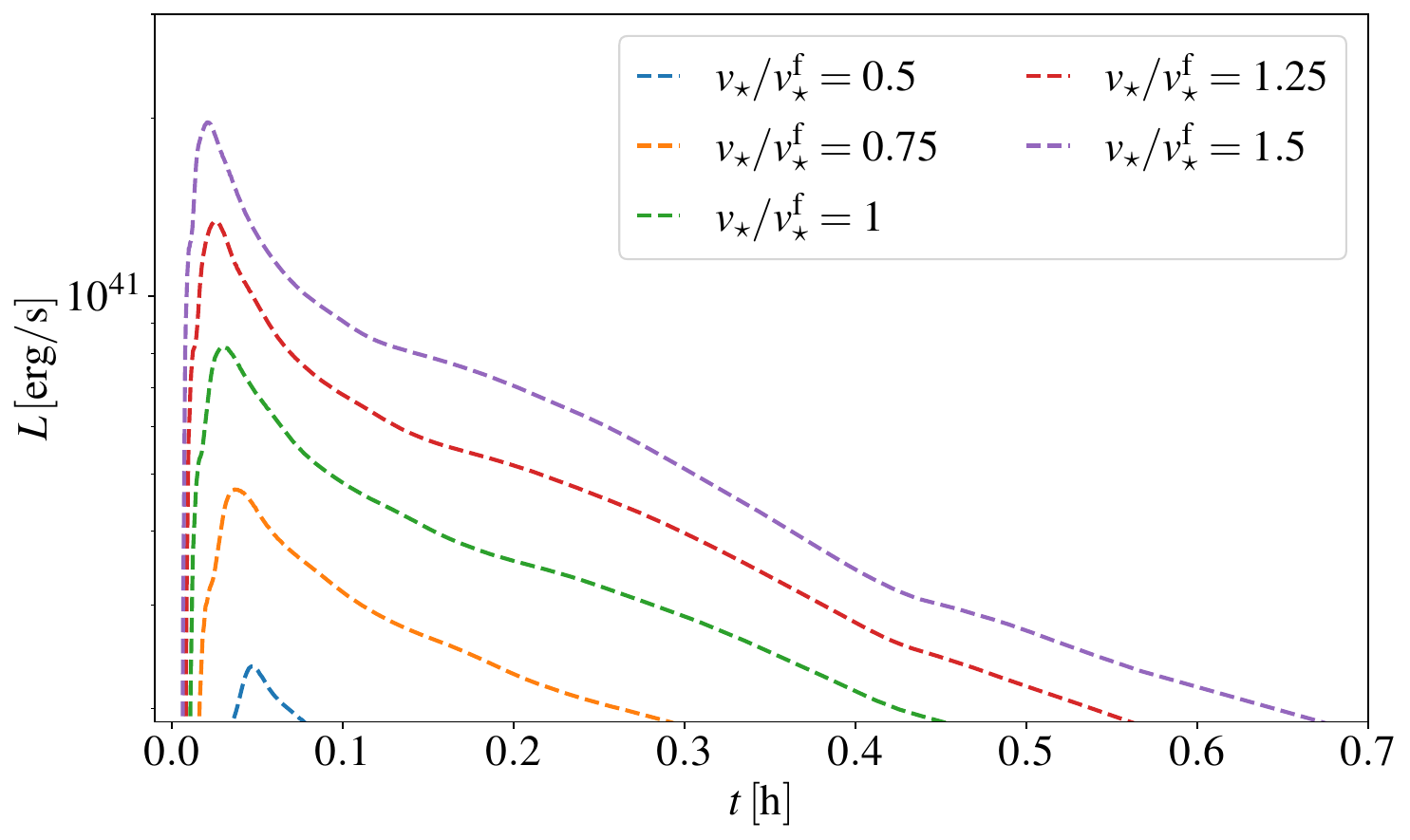}
   \includegraphics[width=0.33\textwidth]{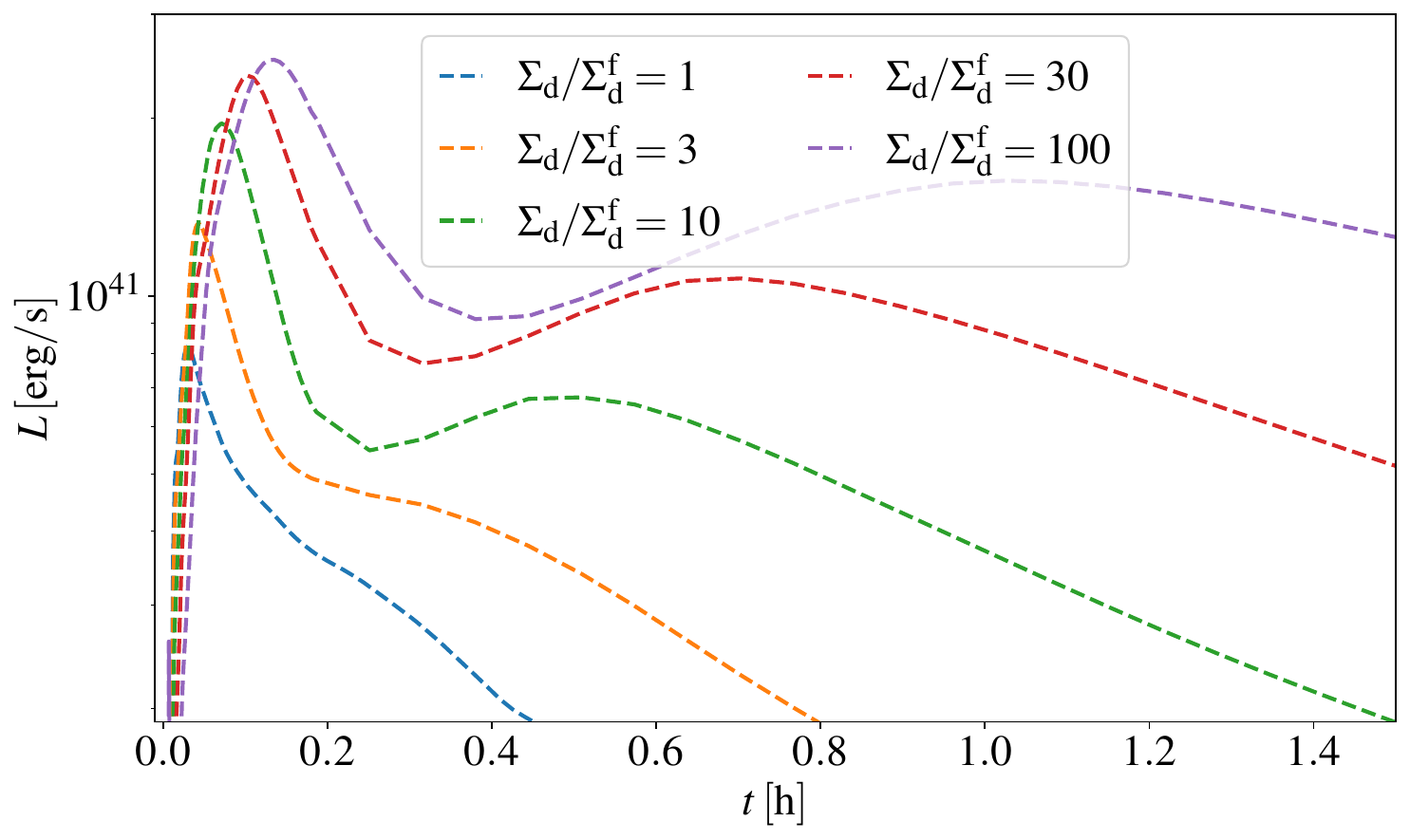}
   \includegraphics[width=0.33\textwidth]{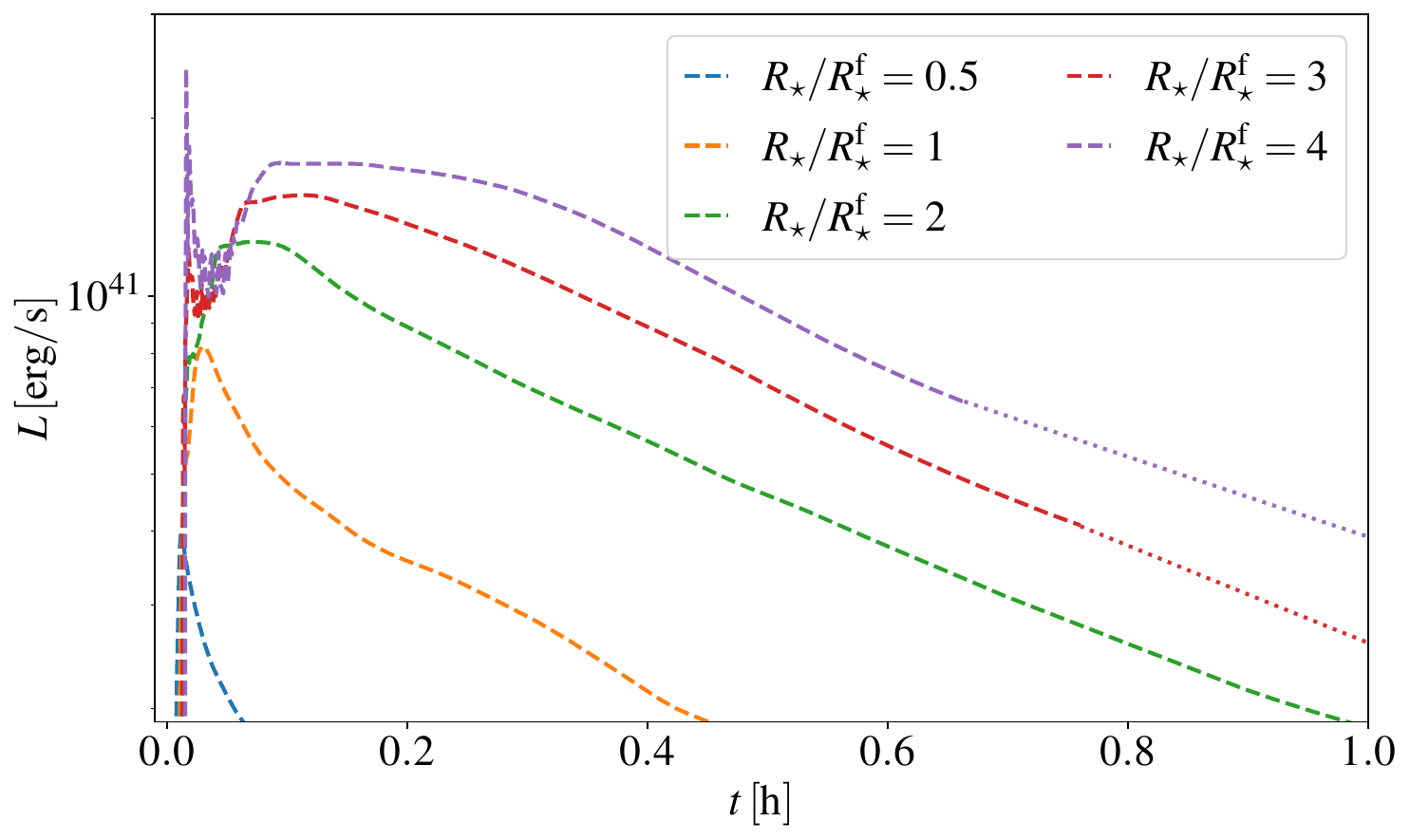}
   \includegraphics[width=0.33\textwidth]{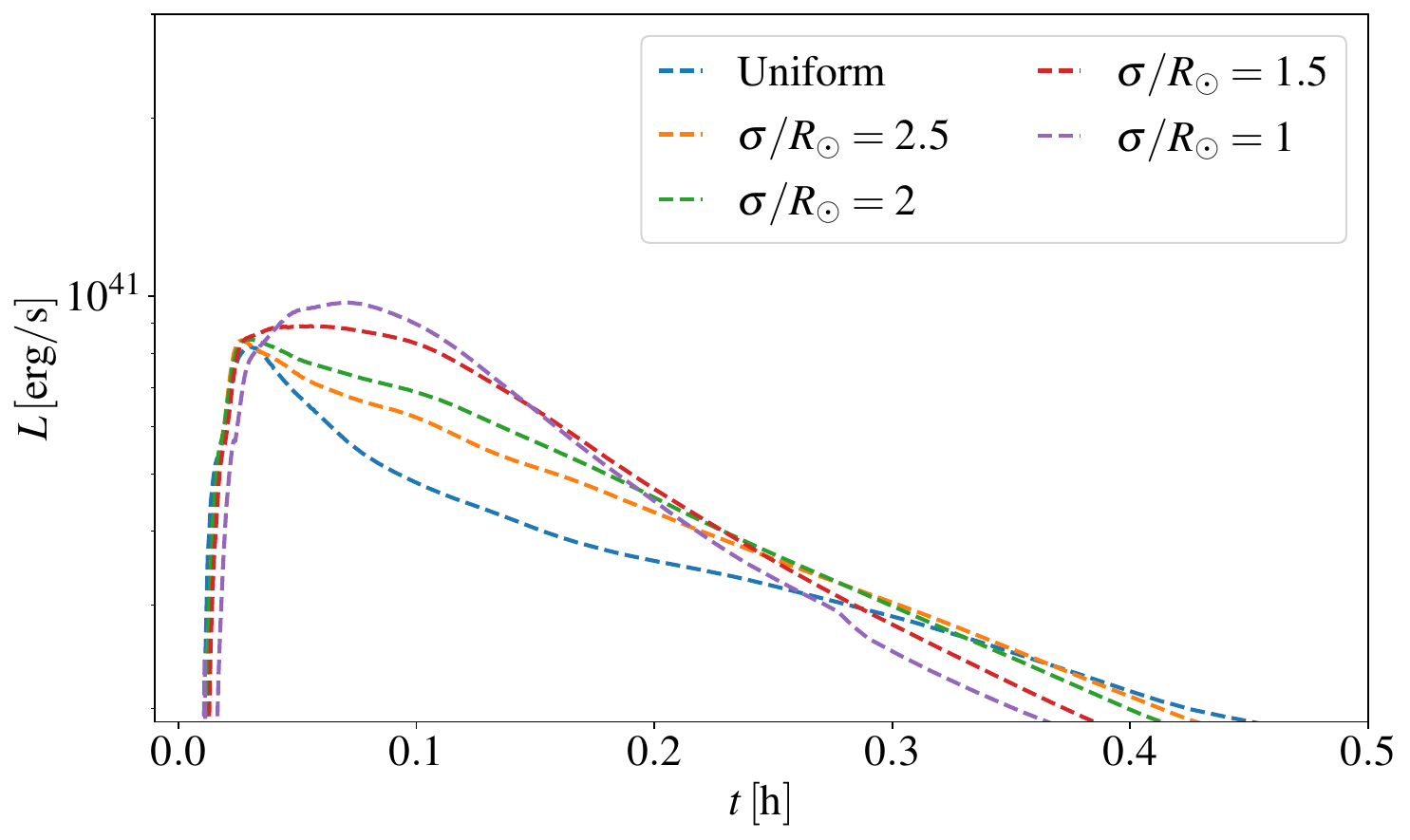}
   \includegraphics[width=0.33\textwidth]{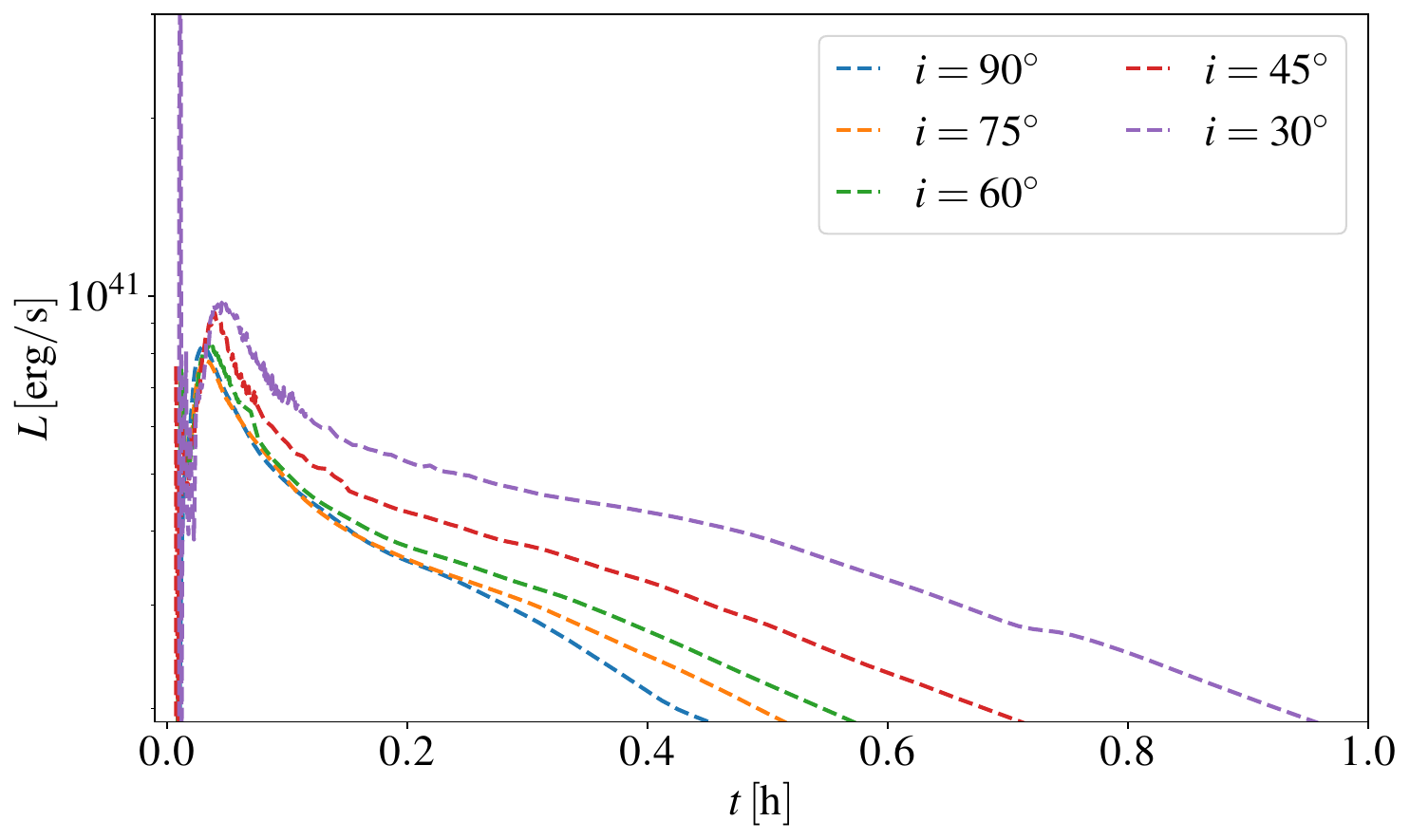}
 \caption{The time evolution of $L$ for the backward outflow for simulations varying $v_\star$, $\Sigma_{\rm d}$, $R_\star$, $\sigma$, and $i$, shown respectively in the first to fifth panels, counting from left to right and then from top to bottom. The dotted lines denote the extrapolated late-time decay.}
\label{fig:L_plot_back_all}
\end{figure*}

\begin{figure*}[h] %  figure placement: here, top, bottom, or page
   \centering
   \includegraphics[width=\textwidth]{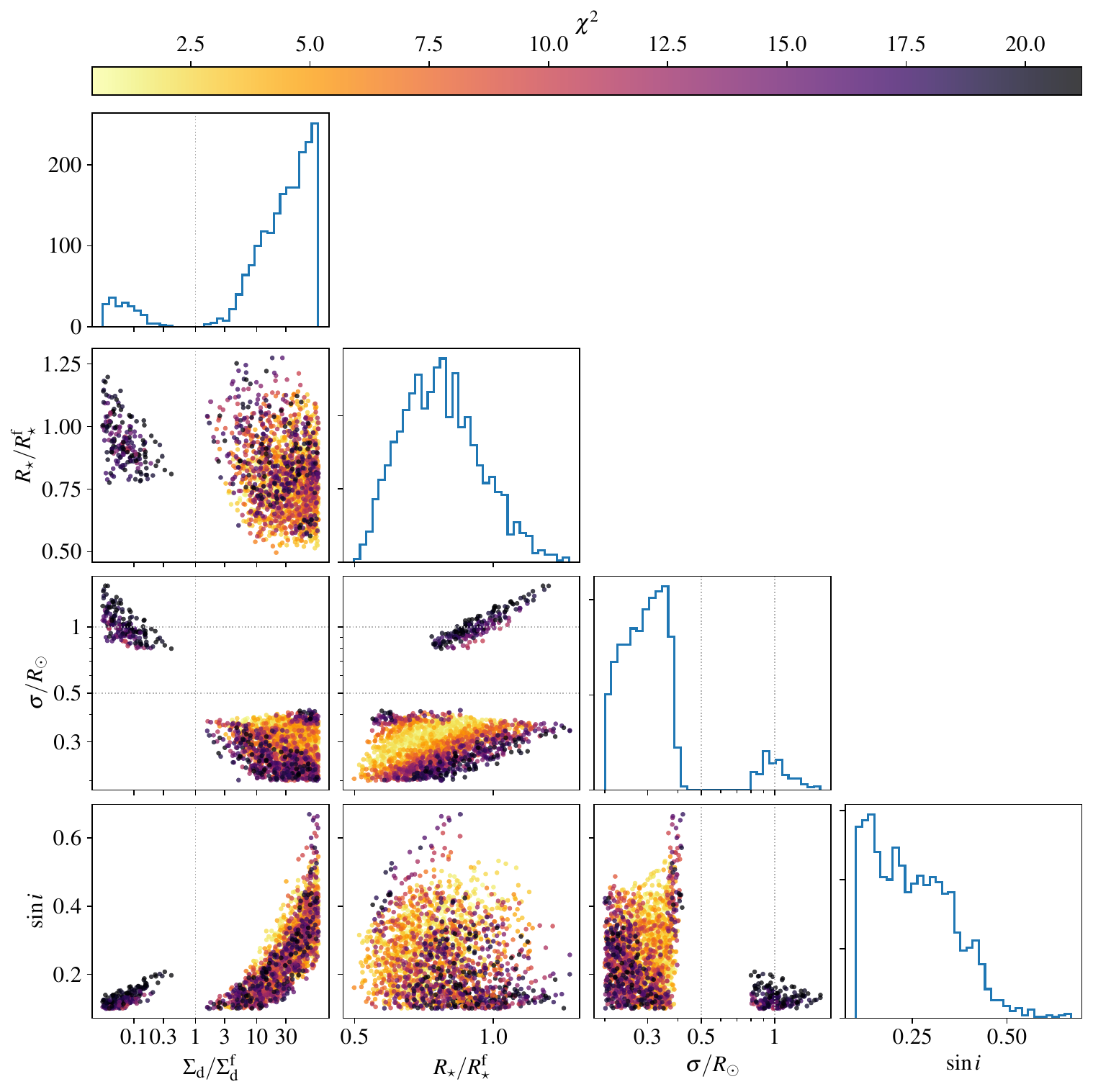}
   \caption{{Corner plot showing the regions of parameter space that reproduce the adopted GSN~069 flare properties within a factor of three. Each plotted point corresponds to a specific combination of $\Sigma_{\rm d}/\Sigma_{\rm d}^{\rm f}$, $R_\star/R_\star^{\rm f}$, $\sigma/R_\odot$, $\sin i$, and is coloured according to its $\chi^2$ value. Grey dotted lines mark values of $\Sigma_{\rm d}/\Sigma_{\rm d}^{\rm f}=1$ and $\sigma/R_\odot=0.5,1$. The candidate solutions separate into two broad regions: i) a high surface density, vertically concentrated branch with $\Sigma_{\rm d}/\Sigma_{\rm d}^{\rm f}\gtrsim1$ and $\sigma/R_\odot\lesssim0.5$, and ii) a lower surface density branch with $\Sigma_{\rm d}/\Sigma_{\rm d}^{\rm f}\lesssim1$, $\sigma/R_\odot\sim1$, and very grazing collisions with $\sin i\sim0.1$.}}
   	\label{fig:corner}
\end{figure*}

In this Section, we show the full set of Figures for all simulations discussed in Section~\ref{sec:results} and \ref{sec:discussion}. Figures~\ref{fig:qpe_sim_density_all} and \ref{fig:qpe_sim_erad_all} show the density and radiation-energy-density slices for all simulations at $t/t_{\rm cr}=1$. Figure~\ref{fig:dotE_all} shows the shock heating rate evolution for all simulations. 
Figure~\ref{fig:dotM_Mej_all} shows the mass outflow rates for all simulations. Figures~\ref{fig:L_plot_forw_all} and \ref{fig:L_plot_back_all} show the luminosity evolution of the forward and backward outflows for all simulations, respectively. Figure \ref{fig:corner} shows corner plot that indicates the regions of parameter space that reproduce the adopted GSN~069 flare properties within a factor of three.
 
\end{appendix}

\end{document}